\documentclass[10pt,twoside,epsf]{article}
\usepackage{amsmath}
\usepackage{amsfonts}
\usepackage{amssymb}
\usepackage{epsfig}
\usepackage{color}
\font\tenbf=cmbx9

\font\fourbf=cmbx14

\font\tenrm=cmr9

\font\tenit=cmti9

\font\sc=cmr12

\usepackage{graphics}
\newcommand{\lle}{\mbox{$\langle$}}
\newcommand{\rle}{\mbox{$\rangle$}}

\newcommand{\bfsi}{\mbox{\boldmath$\sigma$}}

\newcommand{\bfep}{\mbox{\boldmath$\varepsilon$}}

\newcommand{\bfze}{\mbox{\boldmath$\zeta$}}

\newcommand{\bfchi}{\mbox{\boldmath$\chi$}}
\newcommand{\bfzeta}{\mbox{\boldmath$\zeta$}}

\newcommand{\bfcK}{\mbox{\boldmath$\cal K$}}
\newcommand{\bfcL}{\mbox{\boldmath$\cal L$}}
\newcommand{\bfcG}{\mbox{\boldmath$\cal G$}}

\newcommand{\bfb}{\mbox{\boldmath$\bf b$}}

\newcommand{\bfe}{\mbox{\boldmath$\bf e$}}

\newcommand{\bfm}{\mbox{\boldmath$\bf m$}}
\newcommand{\bff}{\mbox{\boldmath$\bf f$}}
\newcommand{\bfh}{\mbox{\boldmath$\bf h$}}
\newcommand{\bfg}{\mbox{\boldmath$\bf g$}}

\newcommand{\bfn}{\mbox{\boldmath$\bf n$}}
\newcommand{\bfp}{\mbox{\boldmath$\bf p$}}
\newcommand{\bfq}{\mbox{\boldmath$\bf q$}}
\newcommand{\bft}{\mbox{\boldmath$\bf t$}}
\newcommand{\bfs}{\mbox{\boldmath$\bf s$}}

\newcommand{\bfu}{\mbox{\boldmath$\bf u$}}

\newcommand{\bfw}{\mbox{\boldmath$\bf w$}}
\newcommand{\bfx}{\mbox{\boldmath$\bf x$}}
\newcommand{\bfy}{\mbox{\boldmath$\bf y$}}
\newcommand{\bfz}{\mbox{\boldmath$\bf z$}}
\newcommand{\bfA}{\mbox{\boldmath$\bf A$}}
\newcommand{\bfB}{\mbox{\boldmath$\bf B$}}
\newcommand{\bfC}{\mbox{\boldmath$\bf C$}}
\newcommand{\bfD}{\mbox{\boldmath$\bf D$}}

\newcommand{\bfI}{\mbox{\boldmath$\bf I$}}
\newcommand{\bfJ}{\mbox{\boldmath$\bf J$}}
\newcommand{\bfL}{\mbox{\boldmath$\bf L$}}
\newcommand{\bfN}{\mbox{\boldmath$\bf N$}}

\newcommand{\bfX}{\mbox{\boldmath$\bf X$}}
\newcommand{\bfY}{\mbox{\boldmath$\bf Y$}}

\newcommand{\bfU}{\mbox{\boldmath$\bf U$}}
\newcommand{\bfF}{\mbox{\boldmath$\bf F$}}
\newcommand{\bfR}{\mbox{\boldmath$\bf R$}}
\newcommand{\bfK}{\mbox{\boldmath$\bf K$}}
\newcommand{\bfM}{\mbox{\boldmath$\bf M$}}
\newcommand{\bfT}{\mbox{\boldmath$\bf T$}}

\newcommand{\bfdelta}{\mbox{\boldmath$\delta$}}
\newcommand{\bfdel}{\mbox{\boldmath$\delta$}}
\newcommand{\bfxi}{\mbox{\boldmath$\xi$}}

\newcommand{\cH}{\mbox{$\cal H$}}
\newcommand{\cL}{\mbox{$\cal L$}}

\newcommand{\bfbK}{\mbox{$\mathbb{K}$}}

\newcommand{\bfbD}{\mbox{$\mathbb{D}$}}

\newcommand{\bfLa}{\mbox{\boldmath$\Lambda$}}
\newcommand{\bfcD}{\mbox{\boldmath$\cal D$}}

\newcommand{\bfccl}{\mbox{\boldmath$\mathfrak l$}}

\newcommand{\bfcR}{\mbox{\boldmath$\cal R$}}
\newcommand{\bfcA}{\mbox{\boldmath$\cal A$}}

\newcommand{\bfcJ}{\mbox{\boldmath$\cal J$}}

\newcommand{\cV}{\mbox{$\cal V$}}

\newcommand{\bfal}{\mbox{\boldmath$\alpha$}}
\newcommand{\bfbe}{\mbox{\boldmath$\beta$}}
\newcommand{\bfga}{\mbox{\boldmath$\gamma$}}
\newcommand{\bfGa}{\mbox{\boldmath$\Gamma$}}

\newcommand{\bfet}{\mbox{\boldmath$\eta$}}
\newcommand{\bftau}{\mbox{\boldmath$\tau$}}

\newcommand{\bfthe}{\mbox{\boldmath$\vartheta$}}
\newcommand{\bfeta}{\mbox{\boldmath$\eta$}}

\newcommand{\BB}{\begin{equation}}
\newcommand{\EE}{\end{equation}}

\newcommand{\BBEQ}{\begin{eqnarray}}
\newcommand{\EEEQ}{\end{eqnarray}}
\begin{document}

\centerline{\fourbf
Critical analyses of RVE concepts in local and
} 

\centerline{\fourbf peridynamic micromechanics of composites 
}




\vspace{6pt}
\vspace{6pt}
\centerline{\bf Valeriy A. Buryachenko \footnote{This is a different paper (not a version of the paper Buryachenko V. A. (2024)  Peridynamic micromechanics of composites: a review. JPER, 531–601(71pp, 378refs). Extension:  arxiv.org/abs/2402.13908v4 (109pp., 466 refs.))}}
\vspace{6pt}
\vspace{6pt}

\centerline{\it Micromechanics \& Composites LLC, Cincinnati, Ohio 45202, USA }




\begin{abstract}
\noindent A static peridynamic (proposed by Silling, see J. Mech. Phys. Solids 2000; 48:175--209) composite materials (CMs) of the random and periodic structures are considered. In the framework of the second background of micromechanics (also called computational analytical micromechanics, CAM), one proved that local micromechanics (LM) and peridynamic micromechanics (PM) are formally similar to each other for CM of both random and periodic structures. It allows for the straightforward generalization of LM methods to their PM counterparts. The representative volume element (RVE) playing a central role in the LM is generalized to PM. For the inhomogeneous body force with compact support, CAM is developed for the estimation of the effective behavior of CM which is used as a new compressed dataset. The preparation of these datasets is selected by the use of a fundamentally new RVE concept that directly depends on neither constitutive laws of phases nor the form of the predicted effective (surrogate) operator.  The mentioned datasets with intrinsically incorporated new RVE concept as a necessary ingredient can be  implemented into any known machine learning (ML) and neural network (NN) technique  used for the prediction of nonlocal surrogate operators. Exploitation of the new RVE concept eliminates any potential incorrectness related to both size scale, boundary layer,  and edge effects. 
\end{abstract}


\section{Introduction}
\label{intro}
The key concepts of micromechanics involve predicting the mechanical behavior of composite materials (CM) based on the mechanical properties of their phases and microstructure. The process often requires analyzing the stress field in the constituent phases of the material, typically by using micromechanical models that account for inclusions within a matrix subjected to some homogeneous effective field. The prediction of the stress field relies on solving a basic problem that represents a single ellipsoidal inclusion within an infinite, homogeneous matrix subjected to a uniform effective field; 
 this solution has an analytical representation in the form of Eshelby \cite{Eshelby`1957} tensor (see for references
\cite{Parnell`2016}, \cite{Zhou`et`2013}).
For more complex inclusion shapes, analytical solutions like Eshelby’s tensor are not feasible. In such cases, finite element analysis (FEA) is widely used in combination with the truncation method modeling  the infinite medium by expanding the sample size.
Boundary integral equations  (see e.g., \cite{Liu`et`2011}, Chapter 11 in \cite{Buryachenko`2022a}) and volume integral equations
 (see for references, e.g., \cite{Buryachenko`2007}) methods can handle infinite domains by converting the problem into a set of integral equations over the boundary and volume of the inclusions, respectively.
Notable references include comprehensive texts and books in works such as those by authors
\cite{Buryachenko`2007}, \cite{Buryachenko`2022a}, \cite{Dvorak`2013}, \cite{Kachanov`S`2018}, \cite{Mura`1987},\cite{Torquato`2002}.

Nonlocal mechanics was initially developed in the 1960s by several researchers such as Kr\"oner, Eringen, Kunin, Ba$\check{\tenrm z}$ant and others (see for references \cite{Buryachenko`2022a}). These early contributions explored the idea that the behavior of materials cannot always be fully described by classical local theories, which consider only immediate neighbors and spatial derivatives of displacement.
A breakthrough in nonlocal mechanics came with the proposal of peridynamics by Silling \cite{Silling`2000} (see also
\cite{Bobaru`et`2016}, \cite{Dorduncu`et`2024}, \cite{Madenci`O`2014}, \cite{Oterkus`O`2024}, \cite{Silling`L`2008}, \cite{Silling`2010}), which fundamentally changes the classical formulation of solid mechanics. Peridynamics replaces the local partial differential equations (such as the balance of linear momentum) with integral equations, which are free of spatial derivatives of displacement. 
In this framework, the equilibrium of a material point is governed by the sum of internal forces exerted by the surrounding points within a finite distance (referred to as the horizon). This is a key departure from the classical theory, where interactions are only local and exerted by adjacent points through contact forces. Peridynamics uses a state-based approach, meaning that the deformation of a material at a given point depends not only on the forces exerted by neighboring points but on the deformation of all the bonds within the horizon
(\cite{Silling`et`2007}, \cite{Silling`2010})
via a response function, which completely describes the interactions. 
This allows the theory to naturally accommodate discontinuities such as cracks or material interfaces, making it suitable for modeling failure, fracture, and damage propagation. 
The original formulation of peridynamics is known as the bond-based (BB) approach, where the interactions are modeled between pairs of points within a finite horizon. A notable consequence of the bond-based approach is that for isotropic linear materials, the Poisson's ratio is fixed at a value of $\nu = 1/4$ (3D and 2D plane strain) or $\nu = 1/3$ (2D plane stress) \cite{Silling`2000}.
The major advantages of the state-based approach
include a material response depending on collective quantities (like volume change or shear angle), which
allows constitutive models from the local theory of solid mechanics to be incorporated directly within
the peridynamic approach (called {\it correspondence model}; see \cite{Aguiar`F`2014}, \cite{Silling`et`2007})
 The state-based models can be subdivided into ordinary and non-ordinary PDs, in
which the force is aligned and unaligned to the bond direction, respectively. 
The correspondence model in peridynamics allows the theory to mimic all components of the classical elasticity tensor for a fully anisotropic material. The non-ordinary state-based models can achieve greater flexibility but often suffer from instability. These models require careful numerical treatment to stabilize them and obtain reasonable results (see for references and details \cite{Wang`et`2024a}). The ordinary state-based model \cite{Scabbia`et`2024}
overcomes
these limitations, since it allows to accurate reproduce any component of the CCM elasticity tensor both in 2D and 3D cases.
Tensor-involved peridynamics builds on the BB PD framework to simulate both isotropic and anisotropic materials, further expanding the flexibility of the peridynamic approach.  (see \cite{Tian`2024}).

{\color{black} The versatility and effectiveness of peridynamic models has been demonstrated in various practical applications,
including damage accumulation, corrosion, the fracture
and failure of composites of deterministic structure, crack instability, impact, the fracture of
polycrystals, and nanofiber networks (see, e.g., \cite{Askari`et`2006}, \cite{Askari`et`2008},
\cite{Basoglu`et`2022},
\cite{Chen`et`2021}, \cite{Diehl`et`2019}, \cite{Isiet`et`2021}, \cite{Javili`et`2019}, \cite{Li`et`2020}, \cite{Li`et`2022b}, \cite{Madenci`O`2014}, \cite{Nowak`et`2023}, \cite{Ren`et`2022}, \cite{Silling`et`2023}, \cite{Zhou`W`2021}).}

For statistically homogeneous thermoperidynamic media subjected to homogeneous volumetric macro boundary loading,
one proposed the background principles (see \cite{Buryachenko`2014a}, \cite{Buryachenko`2017})
in the framework of the bond-based approach considered.
A key result from the bond-based approach is that the effective behavior of the thermoperidynamic media can be governed by a local effective constitutive equation, similar to what is seen in classical thermoelasticity \cite{Buryachenko`2007}.
The relationship between the effective properties and the mechanical or transformation influence functions 
(do not miss the influence functions in peridynamics) describe how the behavior of one part of the material affects other parts.
It is obtained by the use of decomposition of the local fields into load fields (those directly induced by external loads) and residual fields (those arising from the internal material properties, such as the matrix behavior).
The energetic definition of effective elastic moduli is obtained. 
A similarity is identified between the behavior of locally elastic composites and peridynamic composites. This similarity arises because both approaches rely on similar underlying principles, such as Hill’s condition and the self-adjointness of the peridynamic operator.
The similarity between peridynamic and classical elastic models, especially in terms of effective elastic moduli and thermal expansion, opens the door for a generalization of classical solutions to nonlocal peridynamic systems
 (instead of the simplified methods such as, e.g. the mixture theory and scale separation hypothesis, see \cite{Askari`et`2006}, \cite{Askari`et`2008},
\cite{Askari`et`2015}, \cite{Cheng`et`2024}, \cite{Diyaroglu`et`2016}, \cite{Frank`et`2023}, \cite{Hu`et`2012a}, 
 \cite{Hu`et`2014},  \cite{LiF`et`2023}, \cite{Madenci`O`2014}, \cite{Madenci`et`2021}, \cite{Madenci`et`2023}, 
\cite{Mehrmashhadi `et`2019}, \cite{Ren`et`2022},
\cite{Wu`C`2023}, \cite{Wu`et`2021}, \cite{Xu`et`2008},  \cite{Yang`et`2024}).

The first background of LM  is based on the so-called effective field hypothesis, EFH, proposed by Faraday, Poisson, Mossotti,
Clausius, Lorenz, and Maxwell (1830-1880), see for Refs. \cite{Buryachenko`2022a}).
Wide expansion of the methods of the LM (corresponding to classical continuum mechanics, CCM) into nonlocal phenomena (see \cite{Buryachenko`2014a}, \cite{Buryachenko`2017}), \cite{Buryachenko`2022a})
was supported by a critical generalization of the LM such as 
 the General Integral Equation (GIE) of microinhomogeneous media. 
This approach is considered as the second background of micromechanics (sometimes referred to as computational analytical micromechanics, or CAM, see \cite{Buryachenko`2010a}, \cite{Buryachenko`2010b}, \cite{Buryachenko`2014b}, \cite{Buryachenko`2015a}). 
The GIE does not rely directly on the Green function or constitutive laws (either local or nonlocal), which were previously central to the LM approach. This makes the GIE more flexible and generalizable to a broader range of material behaviors
providing a fundamental jump in multiscale and multiphysics research with drastically improved accuracy
of local field estimations (even to the point of correction of a sign, see \cite{Buryachenko`2022a}). 

 It should be mentioned the stochastic methods where
the material property is modeled by a random variable, see for references \cite{Desai`2024}, \cite{Fan`et`2022a}, \cite{Fan`et`2022b},
\cite{Fan`et`2024} (see also \cite{Chen`et`2021}, \cite{Mehrmashhadi `et`2019}, \cite{Song`et`2023}, \cite{Wu`C`2023}, \cite{Wu`et`2021})). The probability density function (PDF) of a state variable is used to describe the statistical distribution of that variable across the material. 
Stochastic approaches often face limitations, such as not accounting for practical engineering constraints like the non-overlapping of randomly located inclusions; 
the mixture theory also does not account for the ratio of the horizon size (in peridynamics) to the inclusion size, which can be significant in nonlocal models  \cite{Chen`et`2021}, \cite{Frank`et`2023}, \cite{Hobbs`et`2024}, \cite{Mehrmashhadi `et`2019}, \cite{Pan`et`2024}, \cite{Wu`C`2023}, \cite{Wu`et`2021}. {\color{black} 
Another limiting case associated with ``randomly" inhomogeneous structure is the modeling of a critical stretch of each bond
by the Weibull distribution function, see \cite{Song`et`2023} and \cite{Xu`et`2024} (see also \cite{Zhao`et`2024}). However, it doesn’t map directly onto classical methods used in local micromechanics for heterogeneous materials.

However, even for locally elastic CMs subjected to inhomogeneous loading, the effective deformations are described by a nonlocal (either the differential or integral) operator (see for references, e.g., \cite{Buryachenko`2007}, \cite{Buryachenko`2022a}, \cite{Du`et`2020},
\cite{Silling`2014}, {\color{black}\cite{Wang`et`2020}}) relating a statistical average of fields in the point being considered with a statistical average of fields in the vicinity of this point. The use of nonlocal operators enables more accurate modeling of materials, especially when considering inhomogeneous media, nonlinear or nonlocal constitutive laws, and coupled physical phenomena. CAM’s ability to handle these factors makes it a valuable method for multiscale and multiphysics problems, and its use of homogenized surrogate models allows for effective and efficient simulations of small-scale effects. By considering scales of both material and applied fields, CAM provides a comprehensive framework for studying composite materials.
The schemes of these approaches are considered in the current paper.

Another area of the LM is periodic structure composites (see, e.g., \cite{Fish`2014}, \cite{Ghosh`2011}, \cite{Sejnoha`Z`2013},
\cite{Zohdi`W`2008}) which can be easier to analyze because of the regularity in the microstructure allows for the application of specific homogenization techniques. So, the method of
{\it asymptotic homogenization} (or two-scale expansion) 
 introduced by Babuska \cite{Babuska`1976} focuses on the behavior of the composite when the ratio of the size of the unit cell (UC), which represents the smallest repeating structure in the material, to the overall size of the material becomes very small. In this case, the unit cell behaves like a local feature, and its influence on the global behavior can be approximated using asymptotic methods
(see e.g., \cite{Bakhvalov`P`1984}, \cite{Fish`2014}).
In {\it computational homogenization}, the macroscopic variables, such as stress and deformation, are derived by calculating the response at the microscopic level. This approach is especially useful for capturing complex material behaviors (such as nonlinearity or inelasticity) performed in the framework of the same general scheme
 (see
e.g., \cite{Kouznetsova`et`2001}, \cite{Miehe`K`2002}, \cite{Matous`et`2017}, \cite{Terada`K`2001}).

So, Alali and Lipton \cite{Allali`L`2012} (see also the related works \cite{Du`et`2020}, \cite{Du`et`2016}, \cite{Scott`M`2020}) developed a two-scale expansion approach for solving problems in periodic peridynamic CMs. However, this approach did not include numerical results or a formal definition of effective moduli.
Madenci and coauthors \cite{Madenci`et`2017} (see also \cite{Madenci`et`2018}, \cite{Diyaroglu`et`2019a}, \cite{Diyaroglu`et`2019b}, \cite{Hu`et`2022}, 
\cite{Li`et`2022b}, and \cite{WangQ`et`2024}) introduced the peridynamic unit cell model for computational homogenization. This model adapts the traditional methods of homogenization to the peridynamic context, allowing for the estimation of effective material properties in periodic structures by the use of the classical periodic boundary conditions, PBC. 
The introduction of the volumetric periodic boundary conditions (VPBC) in \cite{Buryachenko`2018}, \cite{Buryachenko`2018b},
\cite{Buryachenko`2022a} (see also \cite{Galadima`et`2023}, \cite{Galadima`et`2023b},
\cite{Galadima`et`2023c}, \cite{Hu`et`2022}, \cite{Qi`et`2024}, \cite{Xia`et`2020}, \cite{Xia`et`2019}, \cite{Xia`et`2021a}, \cite{Xia`et`2021b}) permits generalizing of classical computational homogenization approaches to their peridynamic counterpart.
 Another fundamental step in peridynamic computational homogenization is the estimation of effective moduli by the use of the averages of traction and displacements at the UC boundary \cite{Buryachenko`2018}, \cite{Buryachenko`2018b}, \cite{Buryachenko`2022a} (this evaluation is most simple) rather than 
estimation of the volume averages of the stresses and strains inside the UC
volume in a less general and more cumbersome method in 
\cite{Galadima`et`2023}, \cite{Galadima`et`2023b}, \cite{Galadima`et`2024}, \cite{Hu`et`2022} (because they require differentiating displacement fields that are not smooth).

The representative volume element (RVE) is essential for predicting the macroscopic (effective) behavior of heterogeneous materials. The correct RVE size ensures that the material’s microstructural heterogeneity is captured while avoiding artificial size scale and edge effects and ensuring that the material's response is independent of boundary conditions. Classical definition by Hill \cite{Hill`1963} implies that these boundary conditions are macroscopically homogeneous and the effective behavior is described by the tensor of effective moduli. 
 The mentioned sample response is estimated from direct numerical simulation (DNS) of microstructural
volume elements (MVEs) simulated or extracted, e.g. from microcomputer tomography (micro-CT).
 Determining the correct size of the RVE is a delicate complex task that requires careful consideration of several factors.
So, RVE should be statistically
large for all microstructural heterogeneities representation and at
the same time remain small enough so that the principle of scale
separation is not violated. A separation of
scales is given by $a\ll \Lambda\ll L$, where $ a$ represents the typical length scale
characteristic of the microstructural heterogeneity, $\Lambda$ is an applied field scale,  and $L$ the macroscopic
length scale.  If scale separation holds, classical homogenization
gives an adequate estimate of the average macroscopic properties.
 Furthermore, the RVE response should
be independent of the applied boundary conditions.
One way to determine the RVE size is to conduct a convergence study, where the effective properties of the material are calculated for increasing RVE sizes. The smallest RVE that yields stable (converged) results for these properties can be considered appropriate. 
Determining the RVE’s size is related to the notion of {\it statistically equivalent
representative volume element} (SERVE) based on image-based (data-driven)
modeling for obtaining detailed three-dimensional data sets from imaging methods,
for example, micro-CT, and constructing optimal
computational domains.  A detailed discussion 
and relevant techniques to determine the RVE size is given in \cite{Bargmann`et`2018}
 \cite{Francqueville`et`2019},  \cite{Harper`et`2012},  \cite{Kanit`et`2003},  \cite{Matous`et`2017},  \cite{Moumen`et`2021},  \cite{Ostoja`et`2016},  \cite{Sab`N`2005}.

Violation of the scale separation hypothesis leads to the eventual abandonment of the hypothesis of 
statistically homogeneous fields yielding a nonlocal coupling between statistical averages
of the stress and strain tensors which are weighted
by a tensorial kernel, reflecting the nonlocal interactions between different material
points. This approach requires the use of an effective elastic operator that
incorporates the nonlocal coupling between stress and strain, and it leads to
an integral formulation of the constitutive law. There are known the strongly
nonlocal (strain type and displacement type, peridynamics) and weakly nonlocal
(strain-gradient, stress-gradient, and higher-order models) forms of these
nonlocal operators. Perhaps the most challenging issue is how micromechanics
can contribute to the understanding of the bridging mechanism between
the coupled scales, which is described by the nonlocal constitutive equations
involving the parameters of a relevant effective nonlocal operator. Thus, instead of the classical effective moduli Hill \cite{Hill`1963} we deal with the effective nonlocal operators (either the integral or differential forms) even for CMs with locally elastic properties of phases. It leads to the definition of a conceptually new RVE associated with a prescribed effective nonlocal operator (which is usually second order differential operator) for either random (\cite{Drugan`2000}, \cite{Drugan`2003}, \cite{Drugan`W`1996}) or periodic (\cite{Ameen`et`2018}, \cite{Kouznetsova `et`2004a}, \cite{Kouznetsova `et`2004b}, \cite{Smyshlyaev`C`2000})) structure CMs. 
The significance of RVE concept is drastically increased at the simultaneous analysis of three sorts of nonlocal effects generated by both inhomogeneous applied fields, material nonlocality (horizon), and
interactions between inclusions. The effect of interactions of these
phenomena increases owing to the synergism effect. 

The evolution style step-by-step development of effective nonlocal operator theory was blown up by the tool of absolutely different levels of opportunities, generality, and flexibility. This is the machine learning (ML) and neural network (NN) technique. 
The prospective direction of data-driven ML technique in the modeling of CMs was initiated by Silling \cite{Silling`2020}, \cite{You`et`2020} (see also  \cite{You`et`2024}). Construction of a surrogate integral operator of prescribed form was performed from a dataset obtained by 
DNS. Recently, the nonlocal neural operator was proposed as a way to learn a surrogate mapping between function spaces, see \cite{Lanthaler`et`2024} and \cite{Li`et`2003}. A new integral neural operator architecture called
the Peridynamic Neural Operator (PNO) \cite{Jafarzadeh`et`2024} provides a surrogate operator for physically consistent predictions (without predefined constitutive laws) of the overall behavior of highly nonlinear, anisotropic, and heterogeneous materials.  The heterogeneous PNO (HeteroPNO) approach is proposed in \cite{Jafarzadeh`et`2024b} for data-driven constitutive modeling of heterogeneous fiber orientation field in anisotropic materials. Physics-informed neural networks (PINN, see \cite{Cuomo`et`et`2022}, \cite{Haghighata`et`2021}, \cite{Harandi`et`2024},\cite{Hu`et`2024}, \cite{Karniadakis`et`2021}, \cite{Kim`L`2024}, \cite{Raissi`et`2019}, and \cite{Ren`L`2024}) have attracted considerable attention
because this framework embeds physics equations into the NN as
constraints, ensuring the training results correspond to physical laws. A neural operator can be combined with the PINN methods (see \cite{Faroughi `et`2024}, \cite{Gosmani`et`2022}, \cite{Wang`Y`2024}) to train a model that can learn complex nonlinearity, multi-material heterogeneity, and nonlocality in physical systems with extremely high generalization accuracy. 

However, powerful tools such as ML and NN techniques often ignore the background concepts of micromechanics (of both LM and PM) such as size scale and edge effects, and RVE. To eliminate this disadvantage, the proposed CAM methods generate fundamentally new compressed datasets for either random or periodic structure CMs.  The preparation of these datasets is selected by the use of a fundamentally new RVE concept that directly depends on neither constitutive laws of phases nor the form of the predicted surrogate operator (depending, in opposite to known surrogate operators, on field concentration factors in the phases of CM). 
The mentioned
datasets with intrinsically incorporated new RVE concept as a necessary
the ingredient can be  implemented into any known machine learning (ML) and neural
network (NN) tecnique 
 used for the prediction of nonlocal surrogate operators. 
Exploitation of new RVE concept eliminates any potential incorrectness related with
both size scale, boundary layer,  and edge effects.

The paper is organized as follows. In Section 2, a short introduction to the peridynamic
theory adapted for a subsequent presentation is presented. Statistical description of the composite
microstructures is described in accompanied
by the volumetric homogeneous displacement loading conditions. Some field
averages are considered. In Section 3,  nonlinear GIEs are proposed
which contain either a statistical
average field or the field produced in the infinite matrix by the body force with compact
support. Solutions of the GIEs are presented in Section 4. Closing assumptions for solutions of two sorts of the GIEs mentioned are considered. Effective moduli are obtained in the framework of some additional
simplified hypotheses. 
By the use of the body force with compact support as a training parameter, a set of surrogate
effective operators is constructed. A few sorts of RVE concepts are presented in Section 5. 
Estimation of a set of surrogate operators is proposed in Section 6.

\section{Preliminaries. }
\vspace{-2.mm}
\subsection{Basic equations of peridynamics}
\vspace{-2.mm}
\setcounter{equation}{0}
\renewcommand{\theequation}{2.\arabic{equation}}
One considers a linear elastic medium occupying an open simply connected bounded domain $w\subset \mathbb{R}^d$
with a smooth boundary $\Gamma_0$ and with an indicator function $W$ and space dimensionality
$d$ ($d=1,2,3$).
The domain $w$ {\color{black} with the boundary $\Gamma^0$} consists from a homogeneous matrix $v^{(0)}$ and a statistically homogeneous
{\color{black} field} $X=(v_i)$ of heterogeneity $v_i$ with indicator functions, $V_i$ and bounded by the closed
smooth surfaces $\Gamma_i$ $(i=1,2,\ldots)$.
It is presumed that the heterogeneities can be grouped into phases $v^{(q)} \quad (q=1,2,\ldots,N)$ with identical mechanical and geometrical properties.
The basic equations of local thermoelasticity of composites are considered
\BBEQ
\label{2.1}
\nabla\cdot \bfsi(\bfx)&=&-\bfb(\bfx), \\ 
\label{2.2}
\bfsi(\bfx)&=&\bfL(\bfx)\bfep(\bfx)+\bfal(\bfx), \ \ {\rm or}\ \ \
\bfep(\bfx)=\bfM(\bfx)\bfsi(\bfx)+\bfbe(\bfx), \\ 
\label{2.3}
\bfep(\bfx)&=&[\nabla {\otimes}{\bf u}+(\nabla{\otimes}{\bf u})^{\top}]/2, \ \
\nabla\times\bfep(\bfx)\times\nabla={\bf 0}, 
\EEEQ
where $\otimes$ and
and $\times$ are the tensor and vector products, respectively, and $(.)^{\top}$ denotes a matrix transposition.
It is presumed that the body force density function $\bfb(\bfx)$ is self-equilibrated
and vanished outside some loading region $B^b$:
\BB
\label{2.4}
\int\bfb(\bfx)={\bf 0},\ \ \ \ \bfy\not\in  b({\bf 0}, B^b):=\{\bfy|\ |\bfy|\leq B^b\},
\EE
where l$b(\bfx_i,B^{b})$ is the ball of radius $B^{b}$
centered at ${\bf 0}$.  
${\bf L(x)}$ and ${\bf M(x) \equiv L(x)}^{-1}$ are the known phase
stiffness and compliance tensors.
$\bfbe(\bfx)$ and $\bfal(\bfx)=-\bfL(\bfx)\bfbe(\bfx)$ are second-order tensors of local eigenstrains and eigenstresses.
In particular, for isotropic
phases, the local stiffness tensor $\bfL(\bfx)$ is presented in
terms of the local bulk $k(\bfx)$ and shear $\mu(\bfx)$
moduli and:
\BB
\nonumber
\bfL(\bfx)=(dk,2\mu)\equiv dk(\bfx)\bfN_1+2\mu(\bfx)\bfN_2, \ \ \bfbe(\bfx)=\beta^{t}\theta\bfdel,
\EE
${\bf N}_1=\bfdelta\otimes\bfdelta/d, \ {\bf N}_2={\bf I-N}_1$ $(d=2\ {\rm or}\ 3$) whereas
$\bfdelta$ and $\bfI$ are the unit second-order and fourth-order tensors; $\theta=T-T_0$ denotes the temperature changes with respect to a reference temperature $T_0$ and $\beta^{t}$ is a thermal expansion.
For all material tensors $\bfg$ (e.g., $\bfL, \bfM,\bfbe,\bfal)$ the notation $\bfg_1(\bfx)\equiv \bfg(\bfx)-\bfg^{(0)}=\bfg^{(m)}_1(\bfx)$ $(\bfx\in v^{(m)},\ m=0,1$) is exploited.
The upper index $^{(m)}$ indicates the
components, and the lower index $i$ shows the individual
heterogeneities; $v^{(0)}=w\backslash v$, $ v\equiv \cup v_i,
\ V(\bfx)=\sum V_i(\bfx)$, and $V_i(\bfx)$ are the
indicator functions of $v_i$, equals 1 at
$\bfx\in v_i$ and 0 otherwise, $(i=1,2,\ldots)$.
Substitution of Eqs. (\ref{2.2}) and (\ref{2.3}) into Eq. (\ref{2.1}) leads to a representation of the equilibrium equation (\ref{2.1}) in the form
\BB
\label{2.5}
^L\widetilde{\bfcL}(\bfu)(\bfx)+\bfb(\bfx)={\bf 0},\ \ \ ^L\widetilde{\bfcL}(\bfu)(\bfx):=\nabla[\bfL\nabla\bfu(\bfx)+\bfal(\bfx)],
\EE
where $^L\widetilde{\bfcL}(\bfu)(\bfx)$ is an elliptic differential operator of the second order.

In this section, a summary of the linear peridynamic model introduced
by Silling \cite{Silling`2000} (see also 
\cite{Bobaru`H`2011}, \cite{Javili`et`2019}, \cite{Lehoucq`S`2008}, \cite{Silling`A`2005}).
An equilibrium equation is free of any derivatives of displacement (contrary to Eq. (\ref{2.5})) and presented in the following form
\BB
\label{2.6}
\widetilde{\bfcL}(\bfu)(\bfx)+\bfb(\bfx)={\bf 0}, \ \ \ \widetilde{\bfcL}(\bfu)(\bfx):=\int \bff(\bfu(\hat {\bfx})-\bfu(\bfx),\hat{\bfx}-\bfx,\bfx)d\hat {\bfx} ,
\EE
where $\bff$ is a {\it bond force density} whose value is the force vector
that the point located at $\hat {\bfx}$ (in the reference configuration) exerts on the point located at $\bfx$ (also in the reference configuration); the third argument $\bfx$ of
$\bff$ (\ref{2.6}) can be dropped for the homogeneous media.
Equations (\ref{2.5}$_1$) and (\ref{2.6}$_1$) have the same form for both local and peridynamic formulation with
the different operators (\ref{2.5}$_2$) and (\ref{2.6}$_2$). Because of this, the superscripts $^L(\cdot)$
will correspond to the local case.

The relative position of two material points in the reference configuration $\bfxi=\hat{\bfx}-\bfx$ and their relative displacement $\bfeta
=\bfu(\hat {\bfx})-\bfu(\bfx)$ are connected with the relative position vector between the two points in the deformed
(or current) configuration $\bfeta+\bfxi$.
Only points $\hat{\bfx}$ inside some neighborhood (horizon region) ${\cal H}_{\bf x}$ of $\bfx$ interact with $\bfx$:
\BB
\label{2.7}
\bff(\bfeta,\bfxi,\bfx)\equiv {\bf 0}\ \ \ \forall \hat{\bfx}\not \in {\cal H}_{\bf x}.
\EE
The vector $\bfxi=\hat{\bfx}-\bfx$ ($\hat{\bfx}\in {\cal H}_{\bf x}$) is called a {\it bond} to $\bfx$, and the collection of all bonds to $\bfx$ form the horizon region ${\cal H}_{\bf x}$.
Without a loss of generality, it is assumed that a shape of ${\cal H}_{\bf x}$ is spherical: ${\cal H}_{\bf x}=\{\hat{\bfx}:\ |\hat{\bfx}-\bfx|\leq l_{\delta}\}$ and a number $l_{\delta}$, called the {\it horizon}, does not depend on $\bfx$. The properties of
$\bff(\bfeta,\bfxi,\bfx)$ are concidered in \cite{Silling`2000}.

Peridynamic states introduced by Silling \cite{Silling`et`2007} 
(for a more detailed summary, see \cite{Silling`L`2010})) 
are the functions acting on bounds.
There are scalar, vector, and modulus states producing the scalars, vectors, and
2nd order tensors, respectively. $\underline{\bfT}[\bfx]\lle\bfxi\rle$ and $\underline{\bfT}[\hat \bfx]\lle-\bfxi\rle$ are the {\it force vector states} at $\bfx$ and
$\hat\bfx$, which return the force densities associated with $\bfxi$ and $-\bfxi$, respectively.
In the ordinary state-based peridynamics being considered, $\underline{\bfT}[\bfx]\lle\bfxi\rle$ is parallel (in contrast with the non-ordinary model) to the deformation vector state and the equilibrium Eq. (\ref{2.6}) is expressed through the
force vector states as (see for details \cite{Silling`et`2007}, \cite{Silling`L`2010})
\BB
\label{2.8}
\widetilde{\bfcL}(\bfu)(\bfx)+\bfb(\bfx)={\bf 0}, \ \ \ \widetilde{\bfcL}(\bfu)(\bfx):=\int
\{\underline{\bfT}[\bfx]\lle\bfxi\rle-\underline{\bfT}[\hat\bfx]\lle-\bfxi\rle
\}
d\hat {\bfx}.
\EE

A small displacement state $\bfu$ when
\BB
\label{2.9}
\bfccl:=\sup_{|\bf \xi|\leq {\it l}_{\delta}} |\bfeta(\bfxi)\rle|\ll l_{\delta}.
\EE
is considered.
Force vector state
\BBEQ
\label{2.10}
\underline{\bfT}&=&\underline{\bfT}^0+
\underline{\bfbK}\bullet\underline{\bfU}.
\EEEQ
is expressed through the {\it modulus state} $\bfbK$.
Here, the operation of {\it dot product} $\bullet$ of two vector states $\underline{\bfA}$, $\underline{\bfB}$ and a double state
$\underline{\bfbD}$ are introduced as
\BBEQ
\label{2.11}
\underline{\bfA}\bullet \underline{\bfB}&=&\big\langle\underline{\bfA}\lle\bfxi\rle\cdot
\underline{\bfB}\lle\bfxi\rle \big\rangle ^{\cal H_{\bf x}}= \int_{\cal H_{\bf x}} \underline{\bfA}\lle\bfxi\rle\cdot
\underline{\bfB}\lle\bfxi\rle ~d\bfxi,\nonumber\\
(\underline{\bfbD}\bullet \underline{\bfB})_i\lle\bfxi\rle&=&\int_{\cal H_{\bf x}} \underline{\bfbD}_{ij}\lle\bfxi,\bfze\rle\cdot
\underline{\bfB}_j\lle\bfze\rle ~d\bfze.
\EEEQ
Hereafter $\lle(\cdot)\rle^{\cal H_{\bf x}}(\bfx)$ denotes the average over the horizon region ${\cal H_{\bf x}}$ with the center $\bfx$.

A linearized model for pure mechanical loading ($\bfbe\equiv{\bf 0}$) can be written from (\ref{2.8}$_2$) as described by Silling \cite{Silling`2010}
\BBEQ
\label{2.12}
\widetilde\bfcL(\bfC,\bfu)(\bfx)+\bfb(\bfx)&=&{\bf 0}, \\
\label{2.13}
\widetilde\bfcL(\bfC,\bfu)(\bfx):&=&\int_w\bfC^{}(\bfx,\bfq)(\bfu(\bfq)-\bfu(\bfx))~dV_q,
\EEEQ
where the integrand in Eq. (\ref{2.10}) vanishing at
$|\bfx-\bfq|\geq 2l_{\delta}$ may be non-null at $l_{\delta}<|\bfx-\bfq|< 2l_{\delta}$.
The kernel with the following symmetries holds for any $\bfx$ and $\bfq$:
\BBEQ
\label{2.14}
\bfC^{\top}(\bfx,\bfq)=\bfC^{}(\bfq,\bfx).
\EEEQ
Thermoelastic cases ($\bfbe\not \equiv{\bf 0}$) were considered in
\cite{Madenci`O`2014}, \cite{Beckmann`et`2013}, \cite{Kilic`M`2010}, \cite{Madenci`O`2016}.
Fully coupled thermo-mechanical PD theory was proposed in \cite{Oterqus`et`2014}, \cite {YangC`et`2024}.

For subsequent convenience, one introduces a vector-valued function $\widetilde{\bff}: \mathbb{R}^d\times \mathbb{R}^d\to \mathbb{R}^d$ by
\BBEQ
\label{2.15}
\!\!\!\!\!\widetilde{\bff}(\bfp,\bfq)=\begin{cases}
\bff(\bfu(\bfp)-\bfu(\bfq),\bfp-\bfq,\bfq), & {\rm if} \ \bfp,\bfq\in w,\\
{\bf 0}, & {\rm otherwise},
\end{cases}
\EEEQ
which is presumed to be Riemann-integrable.
Then, one can
define the ``peridynamic stress" $\bfsi(\bfz)$ at the point $\bfz$
as the total force that all material points $\hat {\bfx}$ to the right of $\bfz$
exert on all material points to its left
(see e.g., \cite{Buryachenko`2022a}, \cite{Lehoucq`S`2008}, \cite{Silling`et`2003}, \cite{Weckner`A`2005}).
For $dD$ cases ($d=1,2,3$)
\BBEQ
\label{2.16}
\!\!\!\!\!\!\!\!\!\!\!\!\!\!\!\!\!\!\!\!\!\bfsi(\bfx)&=&\bfcL^{\sigma}(\bfu),\\
\label{2.17}
\!\!\!\!\!\!\!\!\!\!\!\!\!\!\!\!\!\!\!\!\!\bfcL^{\sigma}(\bfu)&:=&\frac {1}{2} \int_S\int_0^{\infty}\int_0^{\infty} (y+z)^{d-1}\widetilde{\bff}(\bfx+y\bfm,\bfx-z\bfm)\otimes\bfm dzdyd\Omega_{\bf m}.
\EEEQ
Here, $S$ stands for the unit sphere, and $d\Omega_{\bf m}$ denotes a differential solid angle on $S$ in the direction of any unite vector $\bfm$. It was proved \cite{Li`et`2022a} that the peridynamic stress is the same as the
first Piola-Kirchhoff static
Virial stress which offers a simple and clear expression for numerical calculations of
peridynamic stress.

The equilibrium Eqs. (\ref{2.8}), (\ref{2.21}) and (\ref{2.16}) of ordinary state-based PD are considered.
When the interactions between material points are only pairwise, the equilibrium equations are reduced to the bond-based PD equations. One of the simplest nonlinear is
the proportional microelastic material model \cite{Silling`A`2005}
\BBEQ
\label{2.18}
\bff^{\rm bond}(\bfeta,\bfxi,\bfx)\!\!&=&\!\!f(|\bfeta+\bfxi|,\bfxi)\bfe, \ \ \ f(|\bfeta+\bfxi|,\bfxi)=c(\bfxi)s,
\\
\label{2.19}
\bfe\!\!&:=&\!\!
\frac {\bfeta+\bfxi}{|\bfeta+\bfxi|},\ \ \ s:=\frac {|\bfeta+\bfxi|-|\bfxi|}{|\bfxi|},
\EEEQ
where $s$ denotes the bond stretch (also called bond strain) which is the relative change of the length of a bond, and $c$ is referenced as the ``bond constant". 
Although this model is linear in terms of the bond stretches, it is nonlinear in terms of displacements;
thermoelastic cases ($\bfbe\not \equiv{\bf 0}$) were considered in
\cite{Beckmann`et`2013}, \cite{Kilic`M`2010}, \cite{Madenci`O`2014}, \cite{Madenci`O`2016}. A nonlinear model in terms of
bond stretches 
(at $\bfbe\equiv {\bf 0}$)
\BB
\label{2.20}
f(|\bfeta+\bfxi|,\bfxi)=c(\bfxi)[s+3s^2/2+s^3/2]
\EE
is considered in \cite{Jafarzadeh`et`2022}. The potential role of employing a nonlinear peridynamic kernel
in predicting the onset of fractures has been explored in \cite{Dimola`et`2022} (see also \cite{Coclite`et`2022a},
\cite{Coclite`et`2022b}).

A linearized version of the theory (for small displacement) for a microelastic homogeneous
material (\ref{2.21}) takes the form ($\forall \bfeta,\bfxi)$
\BB
\label{2.21}
\bff^{\rm bond}(\bfeta,\bfxi,\bfx)=\bff_{\rm lin}^{\rm bond}(\bfeta,\bfxi,\bfx)=\bfC^{\rm bond}(\bfxi,\bfx)\bfet,
\EE
Here, the material's {\it micromodulus} function $\bfC$
contains all constitutive information and its
value is a second-order tensor given by
\BB
\label{2.22}
\bfC^{\rm bond}(\bfxi,\bfx)=\frac {\partial \bff({\bf 0},\bfxi,\bfx)}{\partial \bfeta }\ \ \ \forall \bfxi.
\EE
Substitution of Eq. (\ref{2.21}) into Eq. (\ref{2.6}) leads to the equilibrium equation
\BBEQ
\label{2.23}
\widetilde{\bfcL}(\bfC^{\rm bond},\bfu)(\bfx)&+&\bfb(\bfx)={\bf 0}, \\
\label{2.24}
\widetilde{\bfcL}(\bfC^{\rm bond},\bfu)(\bfx)&:=&\int \bfC^{\rm bond}(\bfx,\bfq)(\bfu(\bfq)-
\bfu(\bfx))~dV_q.
\EEEQ

For consistency with Newton's third law,
the micromodulus function $\bfC$ for the homogeneous materials must be symmetric
to its tensor structure as well as to arguments
\BB
\label{2.25}
\bfC^{\rm bond}(\bfx,\bfq)=\bfC^{\rm bond}(\bfq,\bfx)=(\bfC^{\rm bond})^{\top}(\bfx,\bfq) \ \ \ \forall \bfx,\bfq,
\EE
where the properties of $\bfC^{\rm bond}$ are discussed in detail in Silling [54]
(see also [10, 33]). 
For example, for the micromodulus functions with the step-function and triangular profiles,
\BB
\label{2.26}
\bfC(\bfxi)= \bfC V({\cal H}_{\bf x}), \ \
\bfC(\bfxi)=\bfC(1-|\bfxi|/l_{\delta})V({\cal H}_{\bf x}),
\EE
respectively, where $V({\cal H}_{\bf x})$ is the indicator function of ${\cal H}_{\bf x}$.
{\color{black} The peridynamic solution
of Eq. (\ref{2.23}) with $\bfC$ described by Eq. (\ref{2.26}) is investigated in detail by both
numerical and analytical methods in 1D (see \cite{Bobaru`et`2009}, \cite{Mikata`2012}, \cite{Silling`et`2003}, \cite{Weckner`A`2005}),
2D (see \cite{Hu`et`2012a}, \cite{Hu`et`2012b}), and 3D cases \cite{Mikata`2023}, \cite{Weckner`et`2009}.}

For bond-based peridynamics, the stress representation (\ref{2.17}) can be recast in terms of displacements
\BBEQ
\label{2.27}
\bfcL^{\sigma}(\bfC,\bfu):&=&\frac {1}{2} \int_S\int_0^{\infty}\int_0^{\infty} (y+z)^{d-1}
\bfC^{\rm bond}((y+z)\bfm,\bfx-z\bfm)
\nonumber\\
&\times&[\bfu(\bfx+y\bfm)-\bfu(\bfx-z\bfm)
]\otimes\bfm dzdyd\Omega_{\bf m}.
\EEEQ

It is interesting that equilibrium Eqs. (\ref{2.22}) and (\ref{2.23}) formally coincide, although the kernels
$\bfC(\bfx,\bfq)$ (\ref{2.23}) and $\bfC^{\rm bond}(\bfx,\bfq)$ (\ref{2.21}) are conceptually different with different symmetry properties (\ref{2.15}) and (\ref{2.25}), respectively. Moreover, in the state-based version, the maximum interaction distance between the points is $2l_{\delta}$ whereas this distance in bond-based peridynamics coincides with the horizon $l_{\delta}$.
However, this similarity provides a possibility to reformulate the results obtained before for the linear bond-based peridynamic micromechanics (see for details \cite{Buryachenko`2022a} and \cite{Buryachenko`2023k}) to their counterparts for the linear ordinary state-based ones.

\subsection {Statistical description of the composite microstructures}
Two material length scales (see, e.g., \cite{Torquato`2002}, \cite{Zaoui`2002}) are considered:
the macroscopic scale $L$, characterizing the extent of $w$, and the microscopic scale $a$, related with the
heterogeneities $v_i$. Moreover, one supposes that the applied field varies on a characteristic length scale $\Lambda$.
The limit of our interests for both the material
scales and field one are either
\BBEQ
\label{2.28}
L\geq\Lambda\geq a^{\rm int}\geq a\geq l_{\delta}\ \ \ &{\rm or}&\ \ \ L\gg \Lambda\gg a^{\rm int}\geq a\gg l_{\delta},\\
\label{2.29}
L\geq \Lambda\geq|\Omega_{00}| \geq l_{\delta}\ \ \ &{\rm or}&\ \ \ L\gg \Lambda\gg|\Omega_{00}|\gg l_{\delta},
\EEEQ
where the inequalities (\ref{2.28}$_2$) and (\ref{2.29}$_2$) are called a scale separation hypotheses.
The inequalities (\ref{2.28}) correspond to the case of random structure CMs, where $a^{\rm int}$ stands for the scale of long-range interactions of inclusions (e.g. $a^{\rm int}=6a$). The inequalities (\ref{2.29}) describe the scale representations for periodic structure CMs, where $\Omega_{00}$ is a unite cell (see for details Subsection 2.4), and, for shortening, we use $|\Omega_{00}|$ instead of linear $|\Omega_{00}|^{1/d}$.

All random quantities under discussion are described by statistically inhomogeneous random fields.
Let us introduce a conditional
probability density $\varphi (v_i,{\bf x}_i \vert v_1,{\bf x}_1,$ $ \ldots,v_n,{\bf x}_n)$
for finding a heterogeneity of type $i$ with the center $\bfx_i$ in the domain $v_i$, given
that the fixed heterogeneities $v_1,\ldots,v_n$ are centered at ${\bf x}_1,\ldots ,{\bf x}_n$.
The notation $\varphi (v_i , {\bf x}_i\vert ;v_1,{\bf x}_1,\ldots ,v_n,{\bf x}_n)$ denotes the case ${\bf x}_i\neq
{\bf x}_1,\ldots ,{\bf x}_n$.
A random field being considered is called statistically homogeneous its multi-point statistical moments of
any order are shift-invariant functions of spatial variables.
Of course, $\varphi(v_i, {\bf x}_i\vert ;v_1,{\bf x}_1,\ldots,v_n,{\bf x}_n)=0$ {\color{black} (since heterogeneities cannot overlap) for values of ${\bf x}_i$ lying inside the
some area $\cup v^0_{mi}$ ($m=1, \ldots ,n)$ called ``excluded volumes'',} where $v^0_{mi}\supset v_m$
with indicator function $V^0_{mi}$ is the ``excluded volumes'' of $\bfx_i$ with respect to $v_m$ (it is usually assumed that
$v^0_{mi}\equiv v^0_m$), and $\varphi (v_i, {\bf x}_i\vert ;v_1,{\bf x}_1,\ldots ,v_n,{\bf x}_n)\to \varphi(v_i, {\bf x}_i)$
as $\vert {\bf x}_i-{\bf x}_m\vert\to \infty$, $m=1,\ldots,n$ (since no long-range order is assumed).
$\varphi (v_i,{\bf x})$ is a number density, $n^{(k)}=n^{(k)}({\bf x})$ of component $v^{(k)}\ni v_i$
at the point ${\bf x}$ and $c^{(k)}=c^{(k)}({\bf x})$ is the concentration, i.e. volume fraction,
of the component $v_i\in v^{(k)}$ at the point ${\bf x}$:
$ c^{(k)}({\bf x})=\langle V^{(k)}\rangle ({\bf x})=\overline v_in^{(k)}({\bf x}),
\ \overline v_i={\rm mes} v_i\ \ (k=1,2,\ldots,N;\ i=1,2,\ldots),\quad
c^{(0)}({\bf x})=1-\langle V\rangle ({\bf x}).$

A very short description of the notations presented is immediately used in subsequent representations. For clear mathematical definitions of basic stochastic geometry conceptions (such as, e.g., random fields, probability space, probability density, statistically homogeneous and inhomogeneous fields, ergodicity,
clustered media, etc.), the interested readers are referred to \cite{Chiu`et`2013}; 
applications to a wide class of micromechanical problems and continuum physics can be found, e.g.,
in \cite{Buryachenko`2022a}, \cite{Malyarenko`O`2019}, \cite{Torquato`2002}.

\subsection{ Volumetric boundary conditions}
Owing to nonlocality, the equilibrium equation (\ref{2.6})
is combined with a ``boundary" condition, used as a volumetric constraint in the so-called interaction domain $w_{\Gamma}$ (in opposite to the local case where the boundary conditions are imposed directly at the bounding surface $\Gamma^{(0)}$, see for details
\cite{Silling`2000}, \cite{Kilic`2008}; 
i.e., the nonlocal boundary $w_{\Gamma}$ is a $d$-dimensional region
unlike its $(d-1)$-dimensional counterpart $\Gamma^0$ in local problems.
The interaction domain $w_{\Gamma}$
contains points $\bfy$ not in $w$ interacting with points $\bfx\in w$.
The most popular shape for $w_{\Gamma}$ with prescribed either the forces or displacements is a boundary layer
of thickness given by $2l_c$
(see \cite{Macek`S`2007}):
$w_{\Gamma}=\{w\oplus{\cal H}_{\bf 0}\}\backslash w$, where $w\oplus{\cal H}_{\bf 0}$ is the Minkowski sum $w$ (${\cal A}\oplus{\cal B}:=
\cup_{\bfx\in{\cal A},\bfy\in {\cal B}}\{\bfx+\bfy\}$); then $w$ is the internal region of
$\overline{\overline{w}}$ (see \cite{Silling`2000}).

It is presumed that $\overline{\overline{w}}$ is considered as a cutting out of a macrodomain (containing a statistically large number of inclusion's realization) from the random
heterogeneous medium covering the entire space $\mathbb{R}^d$.
Then, the Dirichlet, or Neumann volumetric boundary conditions (VBC, see \cite{Du`et`2013}), \cite{Mengesha`D`2014})
are called homogeneous volumetric loading conditions if there exist some symmetric constant tensors either
$\bfep^{w_{\Gamma}}$ or $\bfsi^{w_{\Gamma}}$ such that
\BBEQ
\label{2.30}
\bfu(\bfx)&=&\bfh(\bfy)= \bfep^{w_{\Gamma}}\bfy, \ \forall\bfy\in w_{\Gamma u}=w_{\Gamma},\\
\label{2.31}
\widetilde{\bfcL}(\bfx)&=&-\bfg(\bfy)=-\bfsi^{w_{\Gamma}}\bfn(\bfy), \ \forall\bfy\in w_{\Gamma \sigma}=w_{\Gamma},
\EEEQ
respectively.
There are no specific restrictions on the smoothness and shape of $\Gamma_0$, which is defined only by the convenience of representation.
It should be mentioned that in the LM, the analogs of the VBC (\ref{2.30}) and (\ref{2.31}) (at the nonlocality vanishing $l_{\delta}\to 0$) are the homogeneous boundary conditions (also called the
kinematic uniform boundary conditions (KUBC) and static uniform boundary
conditions (SUBC), respectively)
\BBEQ
\label{2.32}
\bfu(\bfy)&=& \bfep^{w_{\Gamma}}\bfy, \ \forall\bfy\in \Gamma_{0u}={\Gamma}_0,\\
\label{2.33}
\bft(\bfy)&=&\bfsi^{w_{\Gamma}}\bfn(\bfy), \ \forall\bfy\in \Gamma_{0\sigma}={\Gamma}_0,
\EEEQ
correspond
to the analyses of the equations for either strain or stresses, respectively, which are formally similar to each other. However,
in peridynamic micromechanics, a primary unknown variable is displacement (rather than stresses), and, because of this, the VBC
(\ref{2.30}) is assumed.

Seemingly to the volumetric boundary domain $w_{\Gamma i}$, we introduce a volumetric interface boundary (called also interaction interface, see \cite{Seleson`et`2013}, \cite{Allali`G`2015})
$v_{\Gamma i}=v_{\Gamma i}^{+}\cup v_{\Gamma i}^{-}$ where
$v_{\Gamma i}^{+}$ and $v_{\Gamma i}^{-}$ are the boundary layers (internal and external, respectively) divided by the geometric boundary $\Gamma_i$ and have a thickness expressed through the horizon as $2l_{\delta}$
{\color{black}The geometrical boundaries of the boundary layers
$v_{\Gamma i}^{+}$ and $v_{\Gamma i}^{-}$ are
$\Gamma_i^{+}$ and $\Gamma_i^{-}$, respectively.}
For a general form of the inclusion $v_i$ the external volumetric interface $\Gamma^-_i$ can be expressed through the Minkowski sum
$\Gamma^-_i=\{v_i\oplus2{\cal H}_{\bf 0}\}\backslash v_i$, where
$2{\cal H}_{\bf 0}:=\{\bfx|\ \bfx/2\in {\cal H}_0\}$.
A nonlocal closure of the inclusion $v^l_i:=v_i\oplus2{\cal H}_{\bf 0}$ (with an indicator function $V_i^l(\bfx)$) is called an {\it extended inclusion} while $v^l:=\cup v_i^l$ ($i=1,2,\ldots$) (with the indicator function $V^l(\bfx)=\sum_i V_i^l(\bfx)$) stands for the extended inclusion phase. In so doing $v^{l(0)}:=w\backslash v^l\subset v^{(0)}$ is called a {\it truncated matrix}.

{\color{black} In the simplest case, a micromodulus of interaction interface of the inclusion $v_i$ (see, e.g. \cite{Allali`G`2015}, \cite{Allali`L`2012},
\cite{Silling`A`2005}) is presented as an average value of the material properties at two material points connecting dissimilar materials
($V^{(0)}(\bfx)+V_i(\hat\bfx)=V^{(0)}(\hat\bfx)+V_i(\bfx)=1$)}
\BBEQ
\label{2.34}
\bfC^i(\bfx,\hat\bfx)=[\bfC^{(0)}(\bfx,\hat\bfx)+\bfC_i(\bfx,\hat\bfx)]/2
\EEEQ
{\color{black} although more sophisticated models of interaction interface properties (see, e.g., \cite{Bobaru`H`2011}, \cite{Madenci`et`2017},
\cite{Laurien`et`2023}, see for references \cite{Bie`et`2024}), can be incorporated into
the general subsequent representations. In particular, a variation of $\bfC^i(\bfx,\hat\bfx)$ was presented in \cite{Madenci`et`2017}
and \cite{Qi`et`2024}
in a spirit of the functionally graded materials theory described in, e.g. \cite{Buryachenko`2007}.}

\subsection{ Periodic structures and volumetric periodic boundary conditions}

For simplicity of notations for periodic media, we consider 2D cases
$w=\cup \Omega_{ij}$ ($i,j=0,\pm 1, \pm 2,\ldots$)
with the square
unit cells $\Omega_{ij}$ and the centers of the unite cells $\bfLa=\{\bfx_{ij}\}$. Let a representative unit cell $\Omega_{00}$
with the corner points $\bfx^c_{kl}$ ($k,l=\pm 1$) has the boundary
$\Gamma^0=\cup\Gamma^0_{ij}$ where the boundary partition $\bfx_{ij}^{0}\in \Gamma^0_{ij}$ separates the UCs $\Omega_{00}$ and $\Omega_{ij}$ ($i=0,\pm 1,\ j=\pm(1-|i|)$, see Fig. 19.2 in \cite{Buryachenko`2022a}).
A representative volume element (or the unit cell, UC) $\Omega_{00}$ is deformed in a repetitive way as its neighbors.
The position vectors $\bfx^0_{ij}\in\Gamma^0_{ij}$ or
$\bfx^0_{kl}\in\Gamma^0_{kl}$ are presented by the corner
points $\bfx^c_{mn}$ ($i=0,\pm 1,\ j=\pm(1-|i|),\ k=-i,\ l=-j,\ m,n=\pm 1)$; e.g.,
$\bfx^0_{ij}=\bfx^0_{kl}+\bfx^c_{1,-1}-\bfx^c_{-1,-1}$
for the $i=1$, $j=0$.
$N$ field points $\bfx_i$ ($i = 1,\ldots,N$) in the central
cube $\Omega_{00}=\{[- l^{\Omega}, l^{\Omega}]^d\}$ are periodically reflected as
$\bfx_i^{\bf \alpha}$ into the neighboring cubes $\Omega_{\bf \alpha}$, where
$\bfal = (\alpha_1, \ldots, \alpha_d)\in Z^+$, and $\alpha_i = 0,\pm 1$
$(i = 1,\ldots,d)$.
If the source point
$\bfx_p+\bfxi\in\Omega_{\bf \alpha}$ $(\bfxi\in{\cH}_p$) then
the peridynamic counterpart of the local
PBC called
new {\it volumetric periodic boundary conditions} (VPBC, see \cite{Buryachenko`2018}, \cite{Buryachenko`2022a}) represent periodic displacements and antiperiodic tractions
\BBEQ
\label{2.35}
\bfu(\bfx_p+\bfxi)&=&\bfu(\bfx_p+\bfxi-2\bfal^l)
+2\bfep^{w_\Gamma}\bfal^l,\nonumber\\
\label{2.35}
\bft(\bfx^0_{\bf\alpha})&=&-\bft(\bfx^0_{\bf\gamma}),
\EEEQ
respectively,
where $\bfga=-\bfal$ and $\bfx^0_{\bf\alpha}$ are defined analogously
to $\bfx^0_{ij}$. The VPBC (\ref{2.35}) coincide with the classical PBC
\BBEQ
\bfu(\bfx_p)&=&\bfu(\bfx_p-2\bfal^l)
+2\bfep^{w_\Gamma}\bfal^l,\nonumber\\
\label{2.36}
\bft(\bfx^0_{\bf\alpha})&=&-\bft(\bfx^0_{\bf\gamma})
\EEEQ
only for $l_{\delta}=0$.
The VPBC (\ref{2.35}) were proposed for CMic in \cite{Buryachenko`2018}, \cite{Buryachenko`2022a} for any peridynamic consitutive lows of phases; it was applied to CMs
with both the bond-based peridynamic properties of constituents \cite{Buryachenko`2018}, \cite{Buryachenko`2022a}
and non-ordinary state-based \cite{Galadima`et`2023} peridynamic properties of phases.

For periodic structure CMs, the probability density $\varphi (v_i,{\bf x}_i )$ and conditional probability density
$\varphi (v_i,{\bf x}_i \vert; v_j,{\bf x}_j)$ can be expressed through the $\delta$ functions ($\bfx_{\bf \alpha}\in \bfLa$)
\BBEQ
\label{2.37}
\varphi (v_i,{\bf x}_i )&=&\sum_{\bf \alpha}\delta (\bfx_i-\bfx_{\bf \alpha}), \nonumber\\
\varphi (v_i,{\bf x}_i\vert; v_j,\bfx_j)&=&\sum_{\bf \alpha}\delta (\bfx_i-\bfx_{\bf \alpha})-\delta(\bfx_i-\bfx_j).
\EEEQ
It would be interesting to analyze the correctness of the replacement of VPBC (\ref{2.35}) by
PBC (\ref{2.36}). In particular, 1D periodic structure bar was considered in \cite{Buryachenko`2018} and \cite{Buryachenko`2018b}
for the linear bond based model wiith $l^{(1)}_{\delta}=l_{\delta}^{(0)}$. For infinite bar, a discretized Eq. (\ref{2.23}) has a form
\BBEQ
\label{2.38}
(\bfbK\bfu_1)(\bfx)=\hat\bfb (\bfx),
\EEEQ
where $\bfbK$ is an infinite band matrix with the bandwids equal to $n_c$ ($n_c$ is number of nodes in the horizin $l_{\delta}$); $\bfu_1(\bfx):=\bfu(\bfx)-\bfep^{w\Gamma}\cdot\bfx$ and $\hat\bfb$ is a fictitious body force defined by $\bfep^{w\Gamma}$ (see the rigorous formalized representation in \cite{Buryachenko`2018}). For reduction of Eq. (\ref{2.35}) to UC in a visual graphical manner, we cut out $2 N^{\Omega}+1$ rows from the infinite matrix $\bfbK$, where $2 N^{\Omega}+1$ is a number of nods in the UC (see Fig. 1a). This cut out submatrix (containing $2N^{\Omega}+2n_c$ nonzero colomns) is partitioned on a square band matrix
$(2 N^{\Omega}+1)\times(2 N^{\Omega}+1)$ , upper
triangular submatrix ${\rm T^U}$ $(n_c\times n_c)$ and low triangular one ${\rm T^U}$ $(n_c\times n_c)$. The square matrix
$\bfbK^{\Omega}$ $(2 N^{\Omega}+1)\times(2 N^{\Omega}+1)$ for UC is formed by moving of the triangle matrices ${\rm T^U}$
and ${\rm T^L}$ into the matrix $\bfbK^{\Omega}$ as indicated in Fig. 1b. In so doing, a locally elastic counterpart of the band in
$\bfbK^{\Omega}$ is
presented by three diagonals 

\vspace{5.mm} \noindent \hspace{10mm} \parbox{6.2cm}{
\centering \epsfig{figure=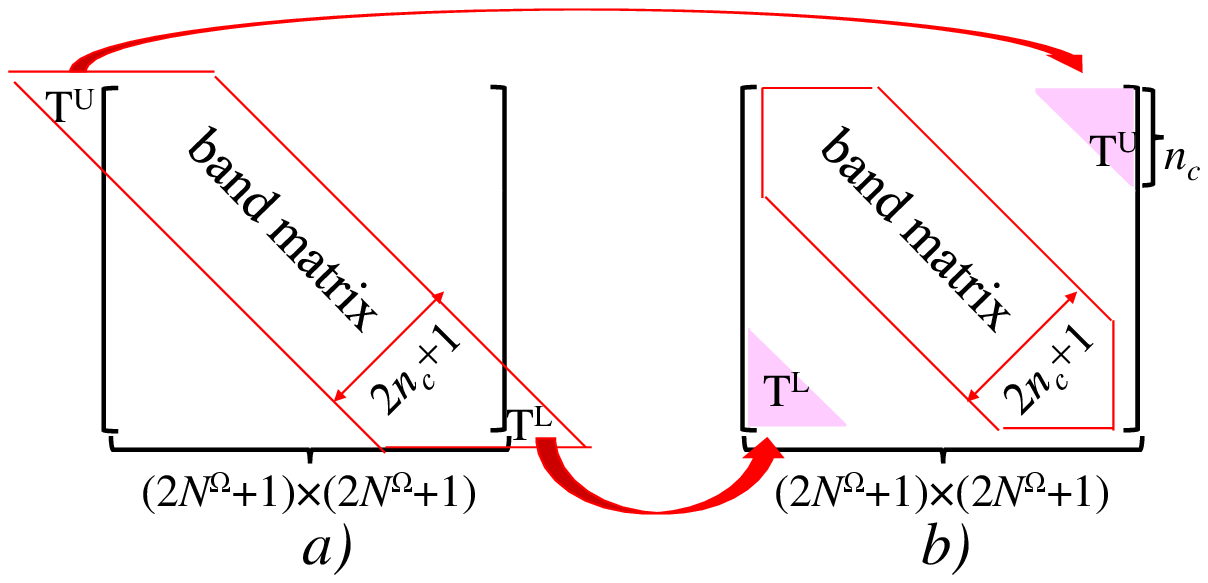, width=10.2cm}\\ \vspace{-22.mm}
\vspace{20.mm}
\vspace{-62.mm} \tenrm \baselineskip=8pt
{{} }}
\vspace{2.mm}

\vspace{-28.mm}
\hspace{-20mm} \parbox{12.8cm}{
\centering \epsfig{figure=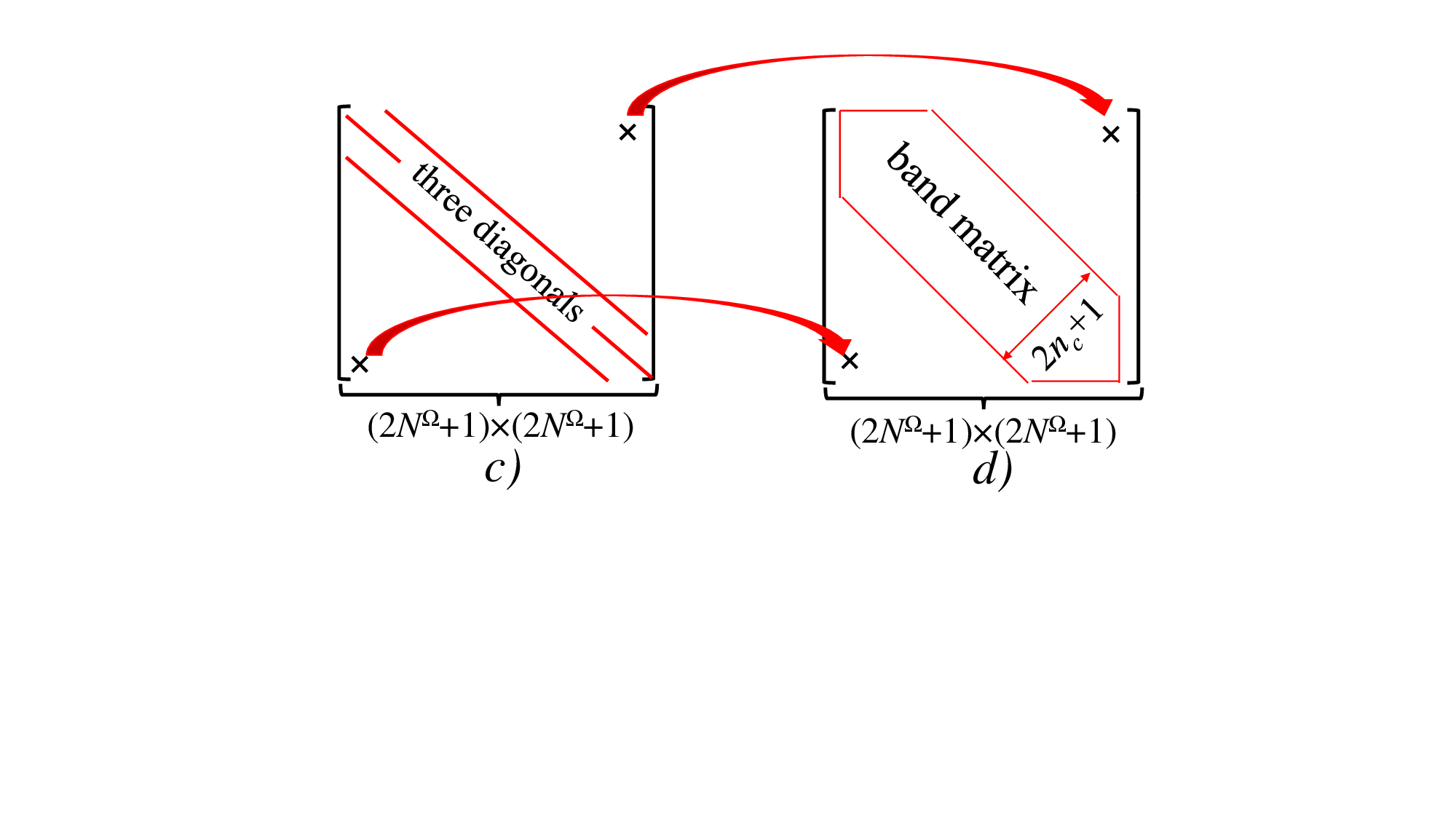, width=14.9cm}\\ \vspace{-22.mm}
\vspace{20.mm}
\vspace{-27.mm} \tenrm \baselineskip=8pt
{\hspace{12.mm}{\sc Fig. 1:} Schemes: a) $2N^{\Omega}+1$ rows of $\bfbK$;\ b) $\bfbK^{\Omega}$;\ c) three diagonal matrix $\bfbK^{\Omega}_3$,\
d) combining of $\bfbK$a) with $\bfbK^{\Omega}_3$c)}}
\vspace{1.mm}

\noindent (see Fig. 1c), while the PBC (\ref{2.36}) produces two
nonzero elements $\bfbK_{-N^{\Omega},N^{\Omega}}$ and $\bfbK_{N^{\Omega},-N^{\Omega}}$
(see two crosses in Fig. 1c). Thus,
in the limit of the horizon going to zero $l_{\delta}/a\to 0$ reducing the peridynamic
heterogeneous bar (\ref{2.23}) to their locally elastic counterpart (\ref{2.1})-(\ref{2.3}), two triangular submatrices
corresponding to the VPBC (\ref{2.35}) in Fig. 1b are shrunk
into two nonzero elements corresponding to the PBC (\ref{2.36}) which are marked by\label{2.35}
the crosses in Fig. 1c; i.e. the matrix in Fig. 1c corresponds to the combining of Eqs. (\ref{2.1})-(\ref{2.3}) with Eqs. (\ref{2.36}). Then logical combining of peridynamic Eq. (\ref{2.23}) with PBQ (\ref{2.36}) leads to the matrix in Fig. 1d differing from the correct matrix $\bfbK^{\Omega}$ in Fig. 1b.

Furthermore, Buryachenko \cite{Buryachenko`2018} considered 1D periodic PD bar where the VPBC (\ref{2.35}) holds. Two cases of CMs are analyzed; in the first case, CM has a periodic structure where the VPBC (\ref{2.35}) applies to each UC. In the second case, the same periodic structure is considered but the VPBC (\ref{2.35}) is only applied to the mesocell $\Omega^5_{00}$ containing five initial UC $\Omega_{00}$, i.e.
the original

PD equation (e.g. (\ref{2.35})) is solved in $\Omega^5_{00}$ without specific VPBC (\ref{2.35}) between UCs 
belonging to the same mesocell $\Omega^5_{00}$. As expected, the displacement distributions $\bfu(\bfx)$ are identical (with numerical tolerance) in both considered cases of CMs. It would be interesting to estimate the errors of replacement of the VPBC (\ref{2.35}) by PBC (\ref{2.36}) as the functions of different sizes of metho-cells (e.g. $\Omega^5_{00}$ and $\Omega^{10}_{00}$), and different scale ratios
$|\Omega_{00}|/a/l^{(1)}_{\delta}/l_{\delta}^{(0)}$. The limiting case of periodic structure CM is a homogeneous infinite medium
(i.e. $\bfC\equiv \bfC^{(0)}$ (\ref{2.13})) with an arbitrary partition of a full space $w=R^d$ over the unit cells $\Omega_{\bf \alpha}$.
The applying the VPBC (\ref{2.35}) at each UC leads to exact solution $\bfu(\bfx)=\bfep^{w_\Gamma}\cdot\bfx$. It would be interesting to estimate the errors of replacement of the VPBC (\ref{2.35}) by PBC (\ref{2.36}) (i.e. the volumetric BC are replaced by BC at the surface $\Gamma^0$, see for analysis \cite{Silling`2000}) as the functions of a single scale ratio
$|\Omega_{00}|/l_{\delta}$. At last, the VPBC (\ref{2.35}) correspond to the remote homogeneous volumetric BC
(\ref{2.30}) and separation of scales $ L\gg \Lambda\gg|\Omega_{00}|$ being  assumed. If the scales $\Lambda$ and $|\Omega_{00}|$
are comparable, the VPBC should be corrected (although the comparable scales $\Lambda$ and $|\Omega_{00}|$ were
analyzed in LM, see, e.g. \cite{Geers`et`2010}, \cite{Matous`et`2017}, \cite{Smyshlyaev`C`2000}).

The VPBC was proposed in \cite{Buryachenko`2018} and \cite{Buryachenko`2018b} following a formal similarity with random packing of spheres in the periodic UC. However, the process of random simulation containing two length scales ($a$ and $|\Omega_{00}|$)
can not be completely adequate to the PD phenomena in UC containing four length scales ($a, \ |\Omega_{00}|, \
l^{(0)}_{\delta}$, and $l^{(1)}_{\delta}$).
So, for ${\rm dist}(v_i(\bfx), \Gamma^0)\geq l_{\delta}^{(0)}$ (for $\forall \bfx\in v_i$ and $\forall v_i\subset \Omega_{00}$) and $l_{\delta}^{(1)}\leq l_{\delta}^{(0)}$, the VPBC (\ref{2.35}) holds (see Fig. 2a).
However, if $\exists \bfx\in v_i$ and $v_i\subset \Omega_{00}$ such that ${\rm dist}(v_i(\bfx), \Gamma^0)< l_{\delta}^{(0)}$
(see Fig. 2b) and $l_{\delta}^{(1)}> l_{\delta}^{(0)}$ then the VPBC (\ref{2.35}) should be corercted. Thought an equivalence of partitions presented in Figs. 2a and 2b was proved for 1D counterpart and $l_{\delta}^{(1)}=l_{\delta}^{(0)}$ \cite{Buryachenko`2018b}, the equivalence of partitious in Figs. 2a and 2b for 2D case and $l_{\delta}^{(1)}\not=l_{\delta}^{(0)}$ is questionable.

Thus, generalization of the VPBC (\ref{2.35}) for any
scale ratios $\Lambda/|\Omega_{00}|/a/l^{(1)}_{\delta}/l_{\delta}^{(0)}$ is desirable. Moreover, the most attractive tool of ML and NN techniques in PM of periodic structure CMs is using a general case of body force (\ref{2.4}) as a training parameter (see Subsection 4.7). Then solution periodicity is lost and PBC (\ref{2.36}) (and VPBC (\ref{2.35})) cannot be fulfilled. This problem of correction of PBC (\ref{2.36}) even in LM is not solved for the general case of body force (\ref{2.4}).

\vspace{-01.mm}
\hspace{-10mm} \parbox{12.8cm}{
\centering \epsfig{figure=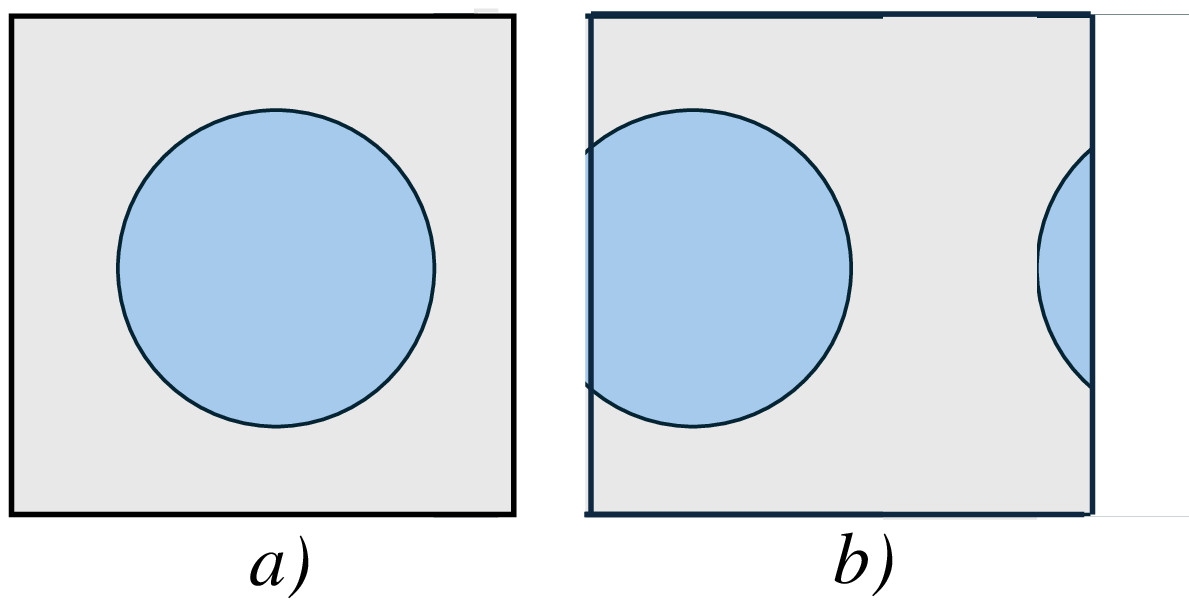, width=10.9cm}\\ \vspace{-22.mm}
\vspace{20.mm}
\vspace{-84.mm} \tenrm \baselineskip=8pt
{\hspace{10.mm}{\sc Fig. 2:} Unite cells }}
\vspace{1.mm}


\subsection {Some averages}
In the case of statistically homogeneous random functions $\bff (\bfx)$, the ergodicity condition is assumed when the spatial average is estimated over
one sufficiently large sample and statistical mean coincide for both the whole volume $w$ and the individual constituent $v^{(k)}$ ($k=0,1,\ldots,N$):
\BBEQ
\lle\bff\rle & =&
\{\bff\}\equiv \lim _{w\uparrow \mathbb{R}^d} |w|^{-1}\int _w
\bff(\bfx)(\bfx)d\bfx, \nonumber\\
\label{2.39}
\lle \bff\rle^{(k)} & =& \{\bff\}^{(k)}\equiv\lim _{w\uparrow \mathbb{R}^d} |w|^{-1}\int _{w}
\bff(\bfx)(\bfx)V^{(k)}d\bfx,
\EEEQ
were $|w|={\rm mes}\, w$.
Under the Gauss theorem, the volume averages by integrals over the corresponding boundary are expressed.
In particular, for the homogeneous boundary conditions either
(\ref{2.30}) or (\ref{2.31})
the mean value $\{\bfep\}$ or $\{\bfsi\}$ of $\bfep$ or $\bfsi$
coincide with
$\bfep^{w_{\Gamma}}$ and $\bfsi ^{w_{\Gamma}}$ (see, e.g. \cite{Buryachenko`2022a}).

The volume averages of the strains and stresses inside the extended inclusion $v^l_i$
can be presented by the averages over the external inclusion boundary
$\Gamma_i^-$
by the use of Gauss's theorem
\BBEQ
\label{2.40}
\lle\bfep V^l\rle=\bfep^{l(1)} & =& \bfep^{l\omega(1)}:= \frac {1}{\bar w}\sum_i\int_{\Gamma_i^-}
\bfu(\bfs){\,}{\,}^{\,^S}\!\!\!\!\!\!\otimes\bfn(\bfs)\,d\bfs,\\
\label{2.41}
\lle\bfsi V^l\rle=\bfsi^{l(1)} & =& \bfsi^{l\omega(1)}:= \frac {1}{\bar w}\sum_i\int_{\Gamma^-_i}
\bft(\bfs){\,}{\,}^{\,^S}\!\!\!\!\!\!\otimes\bfs\,d\bfs.
\EEEQ
The absence of numerical differentiation defines an advantage of the surface average
(\ref{2.40}) with respect to the volume average.
The surface average of stresses
(\ref{2.41}) has the well-known advantage of the
reduction of dimension by one that can be crucial for the analyst. In Eqs. (\ref{2.40}) and (\ref{2.41}) the averages over the modified phases (e.g., the extended inclusions or truncated matrix) are used
\BBEQ
\label{2.42}
\lle \bff\rle^{l(k)} = \{\bff\}^{l(k)}\equiv\lim _{w\uparrow \mathbb{R}^d} |w|^{-1}\int _{w}
\bff(\bfx)(\bfx)V^{l(k)}d\bfx,
\EEEQ
instead of the averages (\ref{2.38}$_2$) exploited in local micromechanics.

It should be mentioned that all averages(\ref{2.38})-(\ref{2.42}) and the equality
\BB
\label{2.43}
\{\bfg\}=c^{(0)}\{\bfg\}^{(0)}+c^{(1)}\{\bfg\}^{(1)},
\EE
(e.g., $\bfg=\bfep,\bfsi,\bftau,\bfeta$),
is fulfilled only for statistically homogeneous media subjected to the homogeneous boundary conditions.
If any of these conditions are broken, then it is necessary to consider two sorts of conditional averages (see for details
\cite{Buryachenko`2007}, \cite{Buryachenko`2022a}).
At first, $\lle\bfg \rle^{(q)}(\bfx)$ can be found as
$\lle\bfg \rle^{(q)}(\bfx)=\lle V^{(q)}\rle^{-1}(\bfx)\lle\bfg V^{(q)}\rle(\bfx)$.
Usually, it is simpler to estimate the conditional averages
of these tensors in the concrete point with the micro-coordinate $\bfz$ of the fixed inclusion $\bfz\in v_q$:
$\lle\bfg| v_q,\bfx_q\rle(\bfx,\bfz)\equiv \lle \bfg\rle_q(\bfx,\bfz)$.
Then a relation between the mentioned averages takes place[$\bfx=(x_1,\ldots,x_d)^{\top}$]:
{\color{black}\BBEQ
\label{2.44}
c^{l(q)}(\bfx)\lle \bfg\rle^{l(q)}(\bfx)=\int_{ v^{l1}_q({\bf x})}
n^{l(q)}(\bfy)\lle \bfg|v_q,\bfy\rle(\bfx,\bfy-\bfx) ~d\bfy.
\EEEQ
where $v_q^1(\bfx)$ is construct as a limit $v_{kq}^{l0}\to v_q^1({\bf x})$ if a fixed ellipsoid $v_k$ is shrinking to the point $\bfx$;
in a similar manner, $v_{kq}^{l0}\to v_q^{l1}({\bf x})$ is constructed for the peridynamic case.

For periodic structure CM, the CMl is constructed using the
building blocks or cells:
$w=\cup\Omega_{\bf m}$ containing the inclusions $ v_{\bf m}\subset \Omega_{\bf m}$.
Hereafter the notation
${\bf f}^{\small \Omega}(\bfx)$
will be used for the average of the function ${\bf f}$
over the cell $\bfx\in\Omega_i$ with the center
$\bfx^{\Omega}_i\in \Omega_i$:
\BB
{\bf f}^{\Omega}(\bfx)={\bf f}^{\Omega}(\bfx_i^{\Omega})\equiv
n(\bfx)\int_{\Omega_i}{\bf f}(\bfy)~d\bfy,\quad \bfx\in \Omega_i,
\label{2.45}
\EE
$n(\bfx)\equiv 1/\overline \Omega_i$
is the number density of inclusions in the cell $\Omega_i$.

Let $\cV_{\bf x}$ be a ``moving averaging" cell (or moving-window
\cite{GrahamBrady`et`2003}) with the center $\bfx$
and characteristic size $a_{{\cal V}}=({\overline{\cV}})^{1/d}$, and let
for the sake of definiteness
$\bfchi$ be a random vector uniformly distributed on $\cV_{\bf x}$
whose value at $\bfz\in \cV_{\bf x}$ is
$\varphi_{\small \bfchi}(\bfz)=1/\overline {\cV}_{\bf x}$ and
$\varphi_{\small \bfchi }(\bfz)\equiv 0$ otherwise. Then
we can define the average of the function ${\bf f}$ with respect to
translations of the vector $\bfchi:$
\BB
\langle {\bf f} \rangle _{\bf x}(\bfx-\bfy)={1\over\overline
{\cV}_{\bf x}}\int_{\cV_{\rm \bf X}}{\bf f}(\bfz -\bfy)~d\bfz,
\quad \bfx\in {\Omega} _i.
\label{2.46}
\EE
Among other things, ``moving averaging" cell $\cV_{\bf x}$
can be obtained by translation of a cell $\Omega_i$
and can vary in size and shape during motion from point
to point.
To make the exposition clear we will
assume that $\cV_{\bf x}$ results from $\Omega_i$ by translation of the vector
$\bfx-\bfx^{\Omega}_i$; it can be seen, however, that this assumption is not
mandatory.

\section {General integral equations (GIEs)}

\setcounter{equation}{0}
\renewcommand{\theequation}{3.\arabic{equation}}

\subsection {Analytical and computational local micromechanics }

The first background of local micromechanics was begun by
Poisson, Faraday, Mossotti, Clausius, Lorenz, and
Maxwell, see for references \cite{Buryachenko`2007}, \cite{Buryachenko`2022a}. They considered the different physical phenomena with an identical background concept (so-called effective field hypothesis, EFH, {\bf H1a}) as a local homogeneous
field acting on the inclusions and differing from the applied macroscopic one (at the infinity), e.g.,
\BBEQ
\label{3.1}
\overline\bfep_i(\bfx)={\rm const}, \ \ \ \overline\bfep(\bfx)\not=\bfep^{w\Gamma}\ \ \ (\bfx\in v_i).
\EEEQ
The concept of the EFH (even if this term is not mentioned) in combination with subsequent assumptions
totally predominates (and creates the fundamental limitations) in all four groups of {\it Analytical Micromechanics} (AMic, classification by
Willis \cite{Willis`1981}) of {\it random} random structure matrix CMs in physics and mechanics of heterogeneous media:
\vspace{-1mm}
\BBEQ
\label{3.2}
&&{\rm \underline{Gr1)}\ model\ methods, \ \ \ \ \ \ \ \ \ \underline{Gr2)}\ perturbation\ methods,} \nonumber \\
&& { \rm \underline{Gr3)}\ variational\ methods, \ \ \ \underline{Gr4)}\ self-consistent\ methods}
\EEEQ
of truncation of a hierarchy among which there are no rigorous boundaries (see for references and details
\cite{Buryachenko`2007}, \cite{Buryachenko`2022a}, \cite{Dvorak`2013}, \cite{Kachanov`S`2018}, \cite{Torquato`2002}).
The ultimate goal of AMic is to develop more cheap, fast, robust, and flexible methods (for making $\bfL^*$ estimations)
than direct numerical simulation (DNS), although it takes additional intellectual complexity to
the implementations.

In contrast, {\it Computational Micrmechanics} (CMic) for CM of {\it deterministic} structures is based on DNS which can be found by different numerical methods. Computational micromechanics can be classified into three broad categories (blocks):
\BBEQ
\label{3.3}
&& \!\!\! \!\!\! \!\!\! \!\!\! \! {\rm \underline{\rm Block\ 1)}\ Asymptotic \ \ \ \ \ \ \underline{B1ock \ 2)}\ Computational ,\ \ \ \underline{Block \
3)} \ Finite\ set \ of} \nonumber \\
&& \!\!\! \!\!\! \!\!\! \!\!\! \!\!\! { \rm \ \ \ \ \ \ \ \ \ \ \ \ homogenization, \ \ \ \ \ \ \ \ \ \ \ \ homogenization,\ \ \ \ \ \ \ \ \ \ \ \ \ inclusions.}
\EEEQ
Blocks 1 and 2 are applied to periodic structure composites (see Introduction). Block 3 corresponds to one and the finite set of inclusions for either the finite-size sample
or the infinite matrix with a finite set of inclusions (in such a case, the problem can be solved by either the volume integral equation methods or boundary integral equation one, see for references, e.g. \cite{Buryachenko`2022a}).
The fundamental difference between CMic and AMic lies in the use (or not use) of DNS, which distinguishes computational methods from analytical ones. While DNS-based methods provide detailed and accurate predictions of material behavior at a fine scale, analytical methods offer faster, more computationally efficient approximations that may sacrifice some detail for speed. This distinction doesn't stem from the traditional meanings of ``analytical" and ``computational," but rather from the specific methodologies used to model and estimate the material properties.

It should be mentioned that the classification of AMic (\ref{3.2}) and CMic (\ref{3.3}) hold also for PM (see for details Comments 4.5 and 4.6, as well as \cite{Buryachenko`2024b}).

\subsection {General integral equation (GIE)}

\label{sec:3}

Let us consider a wacrodomain $w$ with one inclusion $v_i$ subjected to the prescribed effective field loading $\overline\bfep(\bfx)$. Then, Eq.
(\ref{2.6}) for a general peridynamic operator $\widetilde\bfcL$ can be presented as
\BBEQ
\label{3.4}
\widetilde\bfcL(\bfu)(\bfx)=\widetilde\bfcL^{(0)}(\overline\bfep)(\bfx),
\EEEQ
where $\widetilde\bfcL^{(0)}(\overline\bfep)(\bfx)$ is a fictitious body force generated by the effective field $\overline\bfep(\bfx)$.
In particular, for the homogeneous $\overline{\bfep}=
\bfep^{w\Gamma}$, $\widetilde\bfcL^{(0)}(\overline\bfep)(\bfx)=\widetilde\bfcL^{(0)}(\bfep^{w\Gamma})(\bfx)$.
The main advantage of this decomposition (\ref{3.4}) (see also p. 774 in \cite{Buryachenko`2022a}) is that it avoids the challenges associated with the fuzzy boundaries that are characteristic of nonlocal theories. The decomposition in Eq. (\ref{3.4}) simplifies these challenges by eliminating the need for volumetric boundary conditions and sidestepping the difficulties of properly imposing surface effects (see for references and details, e.g. \cite{Scabbia`et`2023}; \cite{Yu`Z`2024}; and \cite{Bobaru`et`2016}, Chapter 14) in nonlocal models. This approach contributes to a more tractable and efficient formulation for analyzing materials under nonlocal influences.

Hereafter for the contraction a solution of Eq. (\ref{3.4}) for nonlinear elastic case
the {\it perturbators} and field concentrations are defined in the reduced form
\BBEQ
\label{3.5}
\bfthe(\bfz)-\overline{\bfthe}(\bfz)=\bfcL^{\theta\zeta}_i(\bfz,\overline{\bfze}), \ \ \
\bfthe ({\bf z}) = \bfcA^{\theta\zeta}_i (\overline{\bfzeta}) ({\bf z}),
\EEEQ
where the duplet substitutions
\BB
\label{3.6}
(\bfu,\bfeta)\leftrightarrow \bfthe,\ \ \ (\bfu,\bfep)
\leftrightarrow\bfzeta, \ \ \
[\bfx,( \hat\bfx, \bfx)]\leftrightarrow \bfz
\EE
are introduced.
Strictly speaking, the perturbators $\bfcL^{\theta\theta}_i(\bfz,\theta)$
are just the notations of some problem that is destined to be solved in the equation
$\bfthe-\overline{\bfthe}=\bfcL^{\theta\theta}_i(\bfz,\bfthe)$
while the perturbators $\bfcL^{\theta\zeta}_i(\bfz,\overline{\bfze})$ are the solutions of this problem.
The elements of the doublet $\bfthe$ correspond to the variables in the left-hand side of the Eq. (\ref{3.5});
the elements of the doublet $\bfze$ correspond to the effective fields on the right-hand sides of the definitions, whereas
the elements $\bfx$ and $( \hat\bfx, \bfx)$ of the doublet $\bfz$ corresponds to the parameters $\bfu$ and $\bfeta$, respectively.
The superindices $^{\theta\zeta}$ of the perturbators $\bfcL^{\theta\zeta}$ correspond to the variables in the left-hand side $\bfthe$ and right-hand side $\bfzeta$,
respectively. In particular, the perturbator $\bfcL^{\theta\theta}_i(\bfz,\bfthe)$ was expressed in \cite{Buryachenko`2022a} through the Green function (see \cite{Wang`et`2017} and \cite{Weckner`et`2009}) as in Eq. (\ref{3.4}), although it is not necessary.

Estimation of the perturabator $\bfcL^{\theta\zeta}_i(\bfz,\overline{\bfze})$ (\ref{3.5})
is, in fact, a basic problem of micromechanics (see Introduction) for one inclusion inside the infinite homogeneous
matrix.
In the PM, estimation of the perturabator $\bfcL^{\theta\zeta}_i(\bfz,\overline{\bfze})$ (\ref{3.5}) was considered by four different methods (see
\cite{Buryachenko`2022a}, \cite{Buryachenko`2019b}) for the linear body-force medium with the same horizon $l_{\delta}$ in both the inclusion and matrix.
The generalization to the linear state-based model as well as to the multiphysics coupled problem
(see the LM applications in \cite{Buryachenko`2015a}) are straightforward.
The popular discretization methods for the solution of PD equations (see for references \cite{DElia`et`2020}, \cite{DElia`et`2017},
\cite{Littlewood`et`2024}) are the meshfree method with one-point Gaussian quadrature referring
to it as “meshfree PD” \cite{Silling`A`2005} (see also the solutions for 1D case
\cite{Bobaru`et`2009}, \cite{Emmrich`W`2007a},
\cite{Emmrich`W`2007b}, \cite{Eriksson`S`2021}, \cite{Silling`et`2007}, \cite{Weckner`A`2005}, \cite{Weckner`E`2005}
and 2D case
\cite{Bobaru`H`2011}, \cite{Hu`et`2012b}, \cite{Le`et`2014}, \cite{Sarego`et`2016}), \cite{Selezon`et`2024},
finite element methods (FEM, \cite{Bode`et`2022},, \cite{Chen`G`2011}, \cite{Macek`S`2007}, \cite{Ren`et`2017},
\cite{Sun`F`2021},
\cite{Sun`et`2020}, \cite{Tian`D`2015}), \cite{Wildman`et`2017}),
quadrature and collocation approaches (\cite{Lu`N`2022}, \cite{Lu`et`2022}, \cite{Seleson`et`2016}, \cite{Zhang`et`2016a}, \cite{Zhang`et`2016b}, \cite{Zhang`N`2023}) and the boundary element method \cite{Liang`et`2021}.
Adaptive algorithm (see, e.g., \cite{Bobaru`H`2011}, \cite{Bobaru`et`2009}, \cite{Buryachenko`2020}) using a multi-grid approach with fine grid spacing only in critical regions
is designed in multi-adaptive approach (\cite{Ongaro`et`2023})
to dynamically switch both the discretization scheme
and the grid spacing of the regions.

Let us consider two inclusions $v_i$ and $v_j$ placed in an infinite homogeneous matrix and subjected to the inhomogeneous field $\widetilde{\bfze}_{i,j}(\bfx)$ ($\bfu,\bfeta =\bfthe;\ \ \bfu,\bfep=\bfze;\ \ [\bfx, ( \hat\bfx, \bfx)]=\bfz;\ \ \bfx\in \mathbb{R}^d$).
We can transform Eq. (\ref{3.5}) into the following ones ($\bfz\in v_i^l$)
\BB
\label{3.7}
\bfthe(\bfz)-\widetilde {\bfthe}_{i,j}(\bfz) -\bfcL^{\theta\zeta}_i(\bfz-\bfx_i,{\widetilde {\bfze}}_{i,j}):=
\bfcL^{\theta\zeta}_{i,j}(\bfz ,{\widetilde {\bfze}}_{i,j})
\EE
defining the perturbator $\bfcL^{\theta\zeta}_{i,j}(\bfz ,{\widetilde {\bfze}}_{i,j})$ which can be found
by any numerical method analogously to the operator $\bfcL^{\theta\zeta}_i(\bfz-\bfx_i,{\overline {\bfze}}_{i})$ (\ref{3.5}).
It should be mentioned that the operators $\bfcL^{\theta\zeta}_i(\bfz-\bfx_i,{\widetilde {\bfze}}_{i,j})$ and
$\bfcL^{\theta\zeta}_{i,j}(\bfz ,{\widetilde {\bfze}}_{i,j})$ (\ref{3.7}) act on the effective fields $\widetilde{\bfze}_{i,j}(\bfx)$ at $\bfx\in v_i$ and $\bfx\in v_i,v_j$, respectively, and the kernel of the operator $\bfcL^{\theta\zeta}_{i,j}$ can be decomposed
($K=I,J)$:
\BB
\label{3.8}
\bfcL^{\theta\zeta}_{i,j}(\bfz,\bfy)=\bfcL^{I\theta\zeta }_{i,j}(\bfz,\bfy)+\bfcL^{J\theta\zeta }_{i,j}(\bfz,\bfy),\ \ \
\bfcL^{K\theta \zeta }_{i,j}(\bfz,\bfy)=\bfcL^{\theta\zeta}_{i,j}(\bfz,\bfy)V_k(\bfy),
\EE
where one follows Mura's tensor indicial notation (see for details \cite{Buryachenko`2022a}).
The double superindices $^{\theta\zeta}$
is used analogously to the double superindices $^{uu}$ and
$^{u \varepsilon}$
in Eqs. (\ref{3.5}).

Similarly, the effective field perturbators $\bfcJ^{\theta \zeta}_{i,j}$
and $\bfcJ^{\theta\zeta\infty }_{i,j}$
can be defined;
they describe the perturbation of the effective field
$\overline{\bfthe}_i(\bfz)-\widetilde {\bfthe}_{i,j}(\bfz)$ introduced by both the heterogeneity $v_j$ (interacting with $v_i$) and the fictitious inclusion
with the response operator $\bfcL^{(0)}$ and eigenfield $\bfbe_1^{\rm fict}(\bfy)$ corresponding to the field in the remote inclusion $v_j$ (without interaction with $v_i$) ($\bfy\in v_j,\ \bfx\in v_i,\
\bfz \in \mathbb{R}^d$)
\BBEQ
\label{3.9}
\overline{\bfthe}_i(\bfz)-\widetilde {\bfthe}_{i,j}(\bfz) &=&
\bfcJ_{i,j}^{\theta \zeta}(\widetilde{\bfze}_{i,j})(\bfz)\\
\label{3.10}
\overline{\bfthe}_i(\bfz)-\widetilde {\bfthe}_{i,j}(\bfz) &=&
\bfcJ_{i,j}^{\theta\zeta\infty }(\widetilde{\bfze}_{i,j})(\bfz),
\EEEQ
(see for details \cite{Buryachenko`2022a}).


For the loading by the body force with a compact support
$\bfb(\bfx)$ (\ref{2.4}), the direct summations of all surrounding perturbators $\bfcL^{\theta\zeta}_j(\bfz,\overline{\bfze})$
(\ref{3.5}) exerting on the fixed inclusion $v_i$ is described by the GIE ($\bfz\in v_i$) (see for details \cite{Buryachenko`2023k})
\BBEQ
\label{3.11}
\langle {\bfthe} \rle_i (\bfz)= \bfthe^{b(0)} ({\bf z})+
\int \bfcL^{\theta\zeta}_j(\bfz-\bfx_j,\overline{\bfze})
\varphi (v_j,{\bf x}_j\vert v_1,{\bf x}_1)d{\bf x}_j.
\EEEQ
with the deterministic fields $\bfthe^{b(0)} ({\bf z})$ produced by the body force $\bfb(\bfx)$ in the infinite homogeneous matrix.
A centering of Eqs. (\ref{3.11}) is considered which performs a subtraction from both sides of Eq. (\ref{6.2}) of their statistical averages. It leads to the GIE
\BBEQ
\label{3.12}
\langle {\bfthe} \rle_i (\bfz)&=& \lle \bfthe\rle ({\bf z})+
\int \big[\bfcL^{\theta\zeta}_j(\bfz-\bfx_j,\overline{\bfze})
\varphi (v_j,{\bf x}_j\vert v_i,{\bf x}_i)\nonumber\\
&-& \lle\bfcL^{\theta\zeta}_j(\bfz-\bfx_j,\overline{\bfze})\rle(\bfx_j)\big]d{\bf x}_j,
\EEEQ
which is more general and valid for any inhomogeneous $\langle \bfthe\rangle ({\bf z})$ while Eqs. (\ref{3.11}) are only correct for the the body force $\bfb(\bfx)$ with a compact support. Owing to the centering of Eq. (\ref{3.11}), Eq. (\ref{3.12})
contains the renormalizing term $\lle \bfcL^{\theta\zeta}_j(\bfz-\bfx_j,\overline{\bfze})\rle(\bfx_j)$ providing an absolute convergence
of the integrals involved in Eqs. (\ref{3.12}).

Afrer statistical average of Eqs. (\ref{3.11}) and (\ref{3.12}), the conditional perturbator $\lle\bfcL^{\theta\zeta}_j(\bfz ,\overline{\bfze}) \vert ; v_i,{\bf x}_i\rle_j$ can be expressed through the
explicit perturbator for two interacting heterogeneities subjected to the field $\widetilde{\bfze}_{i,j}$
\BBEQ
\langle \overline{\bfthe}_i\rangle(\bfz) \!\!\!& =\!\!\!\!& \langle \bfthe\rangle ({\bf z})
+\int \bigl\{\bfcJ^{\theta\zeta}_{i,j}(\lle\widetilde{\bfze}_{i,j}\rle)(\bfz)\varphi (v_j,{\bf x}_j\vert; v_i,{\bf x}_i)
-\bfcJ^{\theta\zeta\infty}_{i,j}(\lle\widetilde{\bfze}_{i,j}\rle)(\bfz)
\bigl\}d{\bf x}_j,\nonumber\\
\label{3.13}
\!\!\!\! \langle \overline{\bfthe}_i\rangle(\bfz) \!\!\!\!& =&\!\!\!\! \bfthe^{b(0)} ({\bf z})
+\int \bfcJ^{\theta\zeta}_{i,j}(\lle\widetilde{\bfze}_{i,j}\rle)(\bfz)\varphi (v_j,{\bf x}_j\vert; v_i,{\bf x}_i)
\,d{\bf x}_j,
\EEEQ
Equations (\ref{3.13}) are central to the overall formulation, capturing the material’s response under the applied effective field loading. These equations are constructed based on numerical solutions for one or two inclusions, and they incorporate statistically averaged fields $\lle\overline{\bfze}_{i}\rle(\bfx)$ and $\lle\widetilde{\bfze}_{i,j}\rle(\bfx)$
to represent the overall macroscopic behavior of the composite or heterogeneous material.
These effective fields are determined based on a closing assumption, which simplifies the complex interactions between inclusions in the matrix and helps make the model solvable. This approach provides a powerful tool for modeling composite materials and offers an exact formulation of their effective behavior under loading.

The fact that Eqs. (\ref{3.11}) and (\ref{3.12}) are applicable to both locally elastic and peridynamic CMs demonstrates the generality and versatility of this formulation. By not relying on the EFH (\ref{3.1}) or a specific constitutive law, and by avoiding the need for a Green's function, the equations are able to describe a broad range of material behaviors, from classical linear elasticity of CMs (see \cite{Buryachenko`2015a}) to nonlocal interactions. This approach offers a powerful tool for modeling both traditional and more complex, nonlocal materials without requiring separate formulations for each.
The passage highlights a significant advancement in the field of micromechanics by extending the methods from peridynamics (traditionally used for nonlocal interactions) to create nonlinear generalized integral equations (GIEs)
for CM (statistically homogeneous and inhomogeneous structures) with
the phases described by any peridynamic model (e.g., bond-based \cite{Buryachenko`2017}, \cite{Buryachenko`2023h} and state-based \cite{Buryachenko`2023b},
linear and nonlinear, ordinary and non-ordinary). These new equations can treat both the local thermoelastic models (see, e.g., \cite{Buryachenko`2007}, \cite{Buryachenko`2010a}, \cite{Buryachenko`2011b}, \cite{Buryachenko`2013}, \cite{Buryachenko`2014b}, \cite{Buryachenko`2015a},
\cite{Buryachenko`2019a}, \cite{Buryachenko`2022b}) and peridynamic models
(\cite{Buryachenko`2014a}, \cite{Buryachenko`2014c}, \cite{Buryachenko`2015b}, \cite{Buryachenko`2017},
\cite{Buryachenko`2018}, \cite{Buryachenko`2018b}, \cite{Buryachenko`2018c}, \cite{Buryachenko`2019b}, \cite{Buryachenko`2020},
\cite{Buryachenko`2020b}, \cite{Buryachenko`2020}, \cite{Buryachenko`2022a}, \cite{Buryachenko`2022},
\cite{Buryachenko`2023}, \cite{Buryachenko`2023a}, \cite{Buryachenko`2023b}, \cite{Buryachenko`2023e},
\cite{Buryachenko`2023f}, \cite{Buryachenko`2023g}, \cite{Buryachenko`2023h}, \cite{Buryachenko`2023c},
\cite{Buryachenko`2023d}, \cite{Buryachenko`2023i}, \cite{Buryachenko`2023j}, \cite{Buryachenko`2023k},
\cite{Buryachenko`2024a}, \cite{Buryachenko`2024b}), and, thereby bridging the gap between local and nonlocal theories. By not relying on direct numerical simulations, these equations fall within the category of AMic, offering a more computationally efficient and robust method for analyzing complex composite materials. This work introduces nonlinear GIEs that could lead to novel applications in the material modeling of CMs.

\subsection{Definition of effective elastic moduli}

Peridynamic counterpart of the tensorial decomposition ($^L\!\bfL(\bfx)=^L\!\bfL^{(0)}+^L\!\bfL_1(\bfx)$) for linear local elasticity
\BB
\label{3.14}
\bfsi(\bfx)=^L\!\bfL^{(0)}\bfep(\bfx)+^L\!\bftau(\bfx), \ \ \
^L\!\bftau(\bfx):=\bfsi(\bfx) - ^L\!\bfL^{(0)}(\bfx)\bfep(\bfx)
\EE
can be presented in the next form for
the operator (\ref{2.16}), (\ref{2.17})
\BB
\label{3.15}
\bfcL^{\sigma}(\bfu)(\bfx)=\bfcL^{\sigma (0)}(\bfu)(\bfx)+\bfcL^{\sigma}_1(\bfu) (\bfx),
\EE
were $\bfcL^{\sigma (0)}$ denotes an action of the operator
$\bfcL^{\sigma}$ on the medium with the material properties of the matrix defined by
$\bff^{(0)}$ [for example, for the bond force (\ref{2.21}), $\bfC^{\rm bond}(\bfxi,{\bfx})\equiv \bfC^{{\rm bond}(0)}(\bfxi)$]
and the displacement fields $\bfu({\bfy})$ of the real CM.
The jump operator $\bfcL^{\sigma}_1(\bfu^t) (\bfx)$ defined by Eq. (\ref{3.15}) is called the {\it local stress polarization tensor} [compare with Eq. (\ref{3.14})] and represented
as
\BB
\label{3.16}
\bftau(\bfx)=\bfcL^{\sigma}_1(\bfu) (\bfx)=
{1\over 2} \int_S\int_0^{\infty}\!\!\!\!\int_0^{\infty}\!\! (y+z)^{d-1}\widetilde{\bff}_1(\bfx+y\bfm,\bfx-z\bfm)
\otimes\bfm dzdyd\Omega_{\bf m},
\EE
where
\BB
\label{3.17}
\widetilde{\bff}_1(\bfp,\bfq):=\widetilde{\bff}(\bfp,\bfq)-\widetilde{\bff}^{(0)}(\bfp,\bfq).
\EE
Formal similarity of $^{L}\bftau(\bfx)$ (\ref{3.14}$_2$) and $\bftau(\bfx)$ (\ref{3.16}) is analyzed in \cite{Buryachenko`2022a}.

Subsequent relations are presented for linear properties of the matrix, e.g., (\ref{2.5}) and (\ref{2.21}), and statistically homogeneous media subjected to the homogeneous volumetric boundary conditions (\ref{2.30}) and (\ref{2.31}).
The volume averages are performed for Eqs. (\ref{3.14}) and (\ref{3.15})
\BBEQ
\label{3.18}
\lle \bfsi\rle&=&^L\!\bfL^{(0)}\lle \bfep\rle+\lle ^L\!\bftau\rle,\ \ \
\lle ^L\!\bfL^{(0)}\bfep\rle=^L\!\bfL^{(0)}\lle \bfep\rle,\\
\label{3.19}
\lle\bfcL^{\sigma}(\bfu)\rle(\bfx)&=&\lle\bfcL^{\sigma (0)}(\bfu)\rle(\bfx)+\lle\bfcL^{\sigma}_1(\bfu) \rle(\bfx),
\EEEQ
respectively. For statistically homogeneous media, a peridynamic analog of the locally elastic counterpart (\ref{3.18}$_2$) is presented as
\BBEQ
\label{3.20}
\!\!\lle\bfcL^{\sigma(0)}(\bfu)\rle (\bfx)=\bfL^{(0)}\lle\bfep\rle,\ \ \bfL^{(0)}=\bfcL^{\sigma(0)}(\bfC^{(0)}, \bfu),
\EEEQ
where $\bfL^{(0)}$ is expressed through the equality (\ref{2.17}) ($\bfu=\bfxi,\ \ \bfxi=(y+z)\bfm$):
$
\bfL^{(0)}=\bfcL^{\sigma(0)}(\bfC^{(0)}, \bfu).
$

The representations (\ref{3.18})-(\ref{3.20}) lead to
the representation for the effective elastic modulus similar to their locally elastic counterpart
the simplification of the averaged Eq. (\ref{3.19}) (compare with Eq. (\ref{3.18}$_1$)):
\BBEQ
\label{3.21}
^L\!\bfL^*&=&^L\!\bfL^{(0)}+^L\!\bfR^*,\ \ \ \lle^L\!\bftau\rle^l=^L\!\bfR^*\lle\bfep\rle,\\
\label{3.22}
\bfL^*&=&\bfL^{(0)}+\bfR^*,\ \ \ \lle\bftau\rle^{l\omega}=\bfR^*\lle\bfep\rle.
\EEEQ
The relation $\lle\bftau\rle^{l\omega}_i$ (\ref{3.22}$_2$) can be also expressed through the displacement field
\BBEQ
\label{3.23}
\lle\bftau\rle^{l\omega}_i:= \lle[\bfcL^{\sigma}(\bfu)(\bfs)\otimes\bfn(\bfs)]{\,}{\,}^{\,^S}\!\!\!\!\!\!\otimes \bfs\rle^{l\omega }_i-
\bfL^{(0)}\lle\bfu(\bfs){\,}{\,}^{\,^S}\!\!\!\!\!\!\otimes \bfn(\bfs)\rle^{l\omega }_i,
\EEEQ
which was considered before in \cite{Buryachenko`2020} at the eigenstress $\bfal\equiv {\bf 0}$ for the bond-based case [with $v^l:=v\oplus {\cH}_{\bf 0}$
instead of $v^l:=v\oplus 2{\cH}_{\bf 0}$] and for linear state-based case \cite{Buryachenko`2023b},
\cite{Buryachenko`2023e} (for $\bfal\not\equiv {\bf 0}$ and both the linear matrix and linear inclusions)
as well as for the locally elastic case
(at $v^l=v$) \cite{Buryachenko`2019a}.
The representations (\ref{3.21})-(\ref{3.23}) were obtained for statistically homogeneous media with the averaging
$\lle(\cdot)\rle$. This scheme is also applicable to periodic structure CMs (see for details
\cite{Buryachenko`2022a}) with the averaging $\lle(\cdot)\rle^{\Omega}$ (\ref{2.45}),
and, therefore, the operation $\lle(\cdot)\rle$ in Eqs. (\ref{3.21})-(\ref{3.23}) can be replaced by
$\lle\!\lle(.)\rle\!\rle$, where
\BB
\lle\!\lle(.)\rle\!\rle=\lle(.)\rle\ \ \ {\rm or} \ \ \ \lle\!\lle(.)\rle\!\rle=\lle(.)\rle^{\Omega}
\label{3.24}
\EE
for either the statistically homogeneous or periodic structures, respectively.

The limit of vanishing length scale (scale separation hypothesis)
\BB
\label{3.25}
l_{\delta}/a\to 0
\EE
reduces Eqs. (\ref{3.22}) and (\ref{3.23}) to the locally elastic counterparts
($\bfs\in \partial v_i$)
\BBEQ
\label{3.26}
\!\!\!\!\!\!\!\!\!\!\!\!\!\!\!\!\!\!\!\!\lle\!\lle\bfsi\rle\!\rle&=&^L\!\bfL^{(0)}\lle\!\lle\bfep\rle\!\rle+\lle\!\lle^L\!\bftau\rle\!\rle,\ \ \
\bfL^*=^L\!\bfL^{(0)}+^L\!\bfR^*,\ \ \ \lle\!\lle^L\!\bftau\rle\!\rle=^L\!\bfR^*\lle\!\lle\bfep\rle\!\rle,\\
\label{3.27}
\!\!\!\!\!\!\!\!\!\!\!\!\!\!\!\!\!\!\!\!\!\lle\!\lle^L\!\bftau\rle\!\rle_i&=&\lle\!\lle^L\!\bftau^s\rle\!\rle^{\omega}_i, \ \ ^L\!\bftau^s(
\bfs):= \bft(\bfs){\,}{\,}^{\,^S}\!\!\!\!\!\!\otimes \bfs
-^L\!\bfL^{(0)}\bfu(\bfs){\,}{\,}^{\,^S}\!\!\!\!\!\!\otimes \bfn(\bfs).
\EEEQ

\section {Solution of nonlinear GIEs}

\subsection {Closing assumtions}
\setcounter{equation}{0}
\renewcommand{\theequation}{4.\arabic{equation}}

The so-called effective field hypothesis, the
the main approximate hypothesis of many micromechanical methods, is formulated by Eq. (\ref{3.1}) (see for details \cite{Buryachenko`2007}).
For the closing of Eq. (\ref{3.5}), the following hypothesis is applied:

\noindent {{\bf Hypothesis H2a)}. {\it Each pair of inclusions $v_j$ and $v_j$
is subjected to the inhomogeneous field
$\widetilde{\bfze}_{i,j}(\bfx)$, and statistical average
$\lle\widetilde{\bfze}_{i,j}\rle(\bfx)$ is defined by the formula $(\bfze=\bfu,\bfep$)
\BBEQ
\label{4.1}
\lle\widetilde{\bfze}_{i,j}\rle(\bfx)&=&\lle\overline{\bfze}_{k}\rle(\bfx)
\EEEQ
at $\bfx\in v_k,\ k=i,j$}.

The hypothesis {\bf H2a}, rewritten in terms of the fields
$\bfep(\bfx)$, ($\bfx\in v_i$), is a standard closing assumption
(see for details, e.g., \cite{Buryachenko`2007}, \cite{Kachanov`S`2018}, \cite{Willis`1981})
degenerating to the
``quasicrystalline" approximation by Lax \cite{Lax`1952}, which ignores the
binary interaction of heterogeneities and presumes homogeneity of
the effective fields:

\noindent {\bf Hypothesis H2b, ``quasi-crystalline" approximation}.
{\it It is supposed that the mean value of the effective field at a point
$\bfx\in v_i$ does not depend on the field inside other heterogeneities
$v_j\not = v_i$, $\bfx\in v_k, \ (k=i,j)$}:
\BBEQ
\lle\overline{\bfze}|v_i,\bfx_i;v_j, \rle(\bfx)&=&\lle\overline{\bfze}_k\rle(\bfx)
\nonumber\\
\label{4.2}
\lle\overline{\bfze}_{k}\rle(\bfx)&\equiv&{\rm const.}
\EEEQ

To make further progress, the hypothesis of ``{\it ellipsoidal symmetry}" {\bf H3} for the distribution of inclusions is widely used in LM
(se for references \cite{Buryachenko`2022}):

\noindent {\bf Hypothesis H3, ``ellipsoidal symmetry"}.
{\it The conditional probability density function $\varphi (v_{j},{\bf x}_j|;v_{i},{\bf x}_{i})$ depends on $\bfx_j-\bfx_i$ only through the combination $\rho=|({\bf a}^0_{ij})^{-1} ({\bf x}_{j}-{\bf x}_{i})|$}:
\BB
\label{4.3}
\varphi (v_{j},{\bf x}_j \mid ;v_{i},{\bf x}_{i})
=h(\rho ),
\EE
{\it where the matrix $({\bf a}^0_{ij})^{-1}$ (which is symmetric in the indexes $i$ and
$j$, ${\bf a}^0_{ij}={\bf a}^0_{ji}$)
defines the ellipsoid excluded volume $v^0_{ij}=\{\bfx:\ |({\bf a}^0_{ij})^{-1}\bfx|^2< 1\}$.}

Acceptance of hypothesis {\bf H2a} closes the systems (\ref{3.12}) and (\ref{3.11}) in the following forms,
\BBEQ
\label{4.4}
\!\!\!\!\!\!\!\!\!\!\!\langle \overline{\bfthe}_i\rangle(\bfz)&=&\langle \bfthe\rangle ({\bf z})
+\int \bigl\{[\bfcJ^{I\theta\zeta}_{i,j}(\bfz ,\lle\overline{\bfze}_{i}\rle)+\bfcJ^{J\theta\zeta}_{i,j}(\bfz ,\lle\overline{\bfze}_{j}\rle)]
\nonumber\\
\!\!\!\!\!\!\!\!\!\!\!\!\!\!\!\!\!\!&\times& \varphi (v_j,{\bf x}_j\vert; v_i,{\bf x}_i)-\bfcJ_{i,j}^{\theta\zeta\infty}(\bfz ,\lle\overline{\bfze}_{j}\rle)
\bigl\}d{\bf x}_j,\\
\label{4.5}
\!\!\!\!\!\!\!\!\!\!\!\!\!\langle \overline{\bfthe}_i\rangle(\bfz)&=&\bfthe^{b(0)} ({\bf z})
+\!\!\int\! [\bfcJ^{I\theta\zeta}_{i,j}(\bfz ,\lle\overline{\bfze}_{i}\rle)\!+\!\bfcJ^{J\theta\zeta}_{i,j}(\bfz ,\lle\overline{\bfze}_{j}\rle)]
\varphi (v_j,{\bf x}_j\vert; v_i,{\bf x}_i)\,d{\bf x}_j,
\EEEQ
were a decomposition $\bfcJ^{\theta\zeta}_{i,j}=\bfcJ^{I\theta\zeta}_{i,j}+\bfcJ^{J\theta\zeta}_{i,j}$
was introduced analogously to Eq. (\ref{3.8}).

An integral Eqs. (\ref{4.4}) and (\ref{4.5}) can be solved by the iteration method of the recursion formula
\BBEQ
\label{4.6}
\langle \overline{\bfthe}_i^{[n+1]}\rangle(\bfz)&=&\langle \bfthe\rangle ({\bf z})
+\int \bigl\{
[\bfcJ^{I\theta\zeta}_{i,j}(\lle\overline{\bfze}^{[n]}_{i}\rle)(\bfz)+
\bfcJ^{J\theta\zeta}_{i,j}(\lle\overline{\bfze}^{[n]}_{j}\rle)(\bfz) ]
\\
&\times& \varphi (v_j,{\bf x}_j\vert; v_i,{\bf x}_i)-
\bfcJ^{\theta\zeta\infty}_{i,j}(\lle\overline{\bfze}^{[n]}_{j}\rle)(\bfz)
\bigl\}d{\bf x}_j,\nonumber\\
\label{4.7}
\lle\bfthe^{[n+1]}\rle_i(\bfz)&=&\bfcA^{\theta\zeta}_i(\lle\overline{\bfze}^{[n+1]}_{i}\rle)(\bfz), \\
\label{4.8}
\bar v_i\lle\bftau^{[n+1]}\rle_i({\bf x})
&=&{\bfcR}_i(\lle\overline {\bfze}^{[n+1]}_i\rle) ({\bfx}),
\EEEQ
and
\BBEQ
\label{4.9}
\!\!\!\!\!\!\!\!\!\!\!\!\!\!\!\langle \overline{\bfthe}_i^{[n+1]}\rangle(\bfz)\!&=&\!\bfthe^{b(0)} ({\bf z})
+\int
[\bfcJ^{I\theta\zeta}_{i,j}(\lle\overline{\bfze}^{[n]}_{i}\rle)(\bfz)+
\bfcJ^{J\theta\zeta}_{i,j}(\lle\overline{\bfze}^{[n]}_{j}\rle)(\bfz) ]\nonumber\\
\!\!&\times&\!\!\varphi (v_j,{\bf x}_j\vert; v_i,{\bf x}_i)d{\bf x}_j,\\
\label{4.10}
\!\!\!\!\!\!\!\!\!\!\!\!\!\!\!\!\!\lle\bfthe^{[n+1]}\rle_i(\bfz)\!&=&\!\bfcA^{\theta\zeta}_i(\lle\overline{\bfze}^{[n+1]}_{i}\rle)(\bfz), \\
\label{4.11}
\!\!\!\!\!\!\!\!\!\!\!\!\!\!\!\!\!\bar v_i\lle\bftau^{[n+1]}\rle_i({\bf x})
\!&=&\!{\bfcR}_i(\lle\overline {\bfze}^{[n+1]}_i\rle) ({\bfx}),
\EEEQ
respectively,
with an initial approximation in the form of explicite solution obtained from
Eqs. (\ref{4.3}) and (\ref{4.4}) in the framework of the EFH {\bf H1a)} (\ref{3.1})
(see for details \cite{Buryachenko`2022a}).

Equations (\ref{4.4}) and (\ref{4.5}) can be simplified in the framework of Hypothesis {\bf Hb2}, see Eqs. (\ref{4.2}) which are also used in the LM (see \cite{Buryachenko`2007}).
However, it is possible a straightforward generalization of ``quasicrystalline" approximation by Lax, see (\ref{4.2}), when the assumption (\ref{4.2}) are relaxed:
$\lle\overline{\bfze}_{k}\rle(\bfz)\not\equiv{\rm const}$ at $\bfz\in v_k^l,$ \ $(k=i,j$; a case of the local elasticity was considered in
\cite{Buryachenko`2010a}, \cite{Buryachenko`2010b})
that
is equivalent to the equality
\BBEQ
\label{4.12}
\lle\bfcL^{\theta\zeta}_j(\bfz ,\overline{\bfze}_{j})\vert ; v_i,{\bf x}_i\rle
&=& \lle\bfcL^{\theta\zeta}_j(\bfz ,\overline{\bfze}_{j})\rle.
\EEEQ
It greatly simplifies the problems (\ref{4.5}) and (\ref{4.6}) where in such a case
$\bfcJ^{\theta\zeta}_{i,j}=\bfcJ_{i,j}^{\theta\zeta\infty}$
($\bfx\in v_i$) reducing these equations to
\BBEQ
\label{4.13}
\!\!\!\!\!\!\!\!\!\!\!\!\!\!\!\!\langle \overline{\bfthe}_i^{[n+1]}\rangle(\bfz)\!\!\!\!&=&\!\!\!\!\langle \bfthe\rangle ({\bf z})
\!+\!\!\int\! \!\!\bfcJ^{\theta\zeta\infty}_{i,j}(\lle\overline{\bfze}_{j}^{[n]}\rle)(\bfz)
[ \varphi (v_j,{\bf x}_j\vert; v_i,{\bf x}_i)-n^{(j)}(\bfx_j )]d{\bf x}_j,\\
\label{4.14}
\!\!\!\!\!\!\!\!\!\!\!\!\!\!\!\!\langle \overline{\bfthe}_i^{[n+1]}\rangle(\bfz)\!\!\!\!\!&=&\!\!\!\bfthe^{b(0)} ({\bf z})
+\!\!\int \!\!\bfcJ^{\theta\zeta\infty}_{i,j}(\lle\overline{\bfze}_{j}^{[n]}\rle)(\bfz)
\varphi (v_j,{\bf x}_j\vert; v_i,{\bf x}_i)\, d{\bf x}_j,
\EEEQ
respectively. The second background of LM
proposed in \cite{Buryachenko`2010a}, \cite{Buryachenko`2010b} in the form of Eq. (\ref{4.13}) 
permits to abandonment of the basic
concepts of micromechanics:
the hypothesis of ``ellipsoidal symmetry", and the
effective field hypothesis (EFH {\bf H1a} considered in the next Section 5.2).
Some new effects were discovered that were impossible in the
the framework of the classical (the first) background of micromechanics.

{\color{black}Equations (\ref{4.9}) and (\ref{4.10}) are obtained for general nonlinear cases of either the state-based
(\ref{2.8}) or bond-based (e.g. (\ref{2.15})) PM.
Equations (\ref{4.13}) and (\ref{4.14}) are reduced to the corresponding equations obtained before
for either linear bond-based (\ref{2.10}) PM \cite{Buryachenko`2020},
\cite{Buryachenko`2020b} 
or linear state-based PM (\ref{2.13}) and (\ref{2.14}) \cite{Buryachenko`2023b}.}

\sffamily

\noindent{\bf Comment 4.1.}
Eqs. (\ref{4.4}) and (\ref{4.5}) were obtained using (explicit or implicit) neither the EFH (\ref{3.1}) nor Green functions.
The most intriguing feature of Eqs. (\ref{4.4}) and (\ref{4.5})
is the absence of a constitutive law in these equations. As can be seen, Eq. (\ref{4.4}) completely coincides with the corresponding equations for the locally elastic CMs (see \cite{Buryachenko`2015a}).
It means that the same equation can be exploited for both the peridynamic CMs and locally elastic CMs.
The GIEs (\ref{4.4}) and (\ref{4.5}) proposed are adapted to the straightforward generalizations of corresponding methods of local thermoelastic micromechanics (see, e.g., \cite{Buryachenko`2007}, \cite{Buryachenko`2014b}, \cite{Buryachenko`2015a}).
Equation (\ref{4.4}) was obtained before for the particular linear peridynamic models (bond-based
\cite{Buryachenko`2017} and state-based \cite{Buryachenko`2023b}), which are peridynamic counterparts of the linear GIE
in the LM.
Equation (\ref{4.5}) was also proposed before for the linear bond-based model (see \cite{Buryachenko`2023h}).
However, Eqs.(\ref{4.4}) and (\ref{4.5}) are now proposed for CM (statistically homogeneous and inhomogeneous structures) with the phases described by any peridynamic model (e.g., bond-based and state-based, linear and nonlinear, ordinary and non-ordinary) with nonlinear elastic constitutive law.

\noindent{\bf Comment 4.2.} It should be mentioned that for weakly nonlocal (strain gradient theories, stress gradient theories,
higher-order models) CMs, the GIEs were proposed in \cite{Buryachenko`2022} in a particular form (\ref{4.13}) where the perturbators
(like $\bfcJ^{\theta\zeta\infty}_{i,j}(\lle\overline{\bfze}_{j}^{[n]}\rle)(\bfz)$) were expressed through the corresponding Green functionss and their derivatives. It would be interesting to generalize these GIEs to the case of nonlinear weakly nonlocal media (in the spirit of Eqs.
(\ref{3.13})) with the perturbators estimated numerically. Furthermore, GIE for linear strongly nonlocal media (strain type) was also obtained in
\cite{Buryachenko`2011b} and \cite{Buryachenko`2011c}
in a particular simplified form (\ref{4.13}), where the perturbators were estimated for inclusions of canonical shape and local elastic isotropic matrix. It is desirable to generalize this approach to the general nonlinear strongly nonlocal (strain type) media (in the spirit of Eqs.
(\ref{3.13})) with the perturbators estimated numerically.

\rmfamily

\subsection {Estimation of effective moduli}

{\bf Random structure CMs 4.2.1.} Thus, the GIE (\ref{3.11}) for CM subjected to body force $\bfb(\bfx)$ with compact support was solved for a general representation of the nonlinear perturbator operator $\bfcL^{\theta\zeta}_j(\bfz,\overline{\bfze})$ (\ref{3.5}). Solution (\ref{4.7}) was obtained in the framework of the closing assumption {\bf H2a)} (\ref{4.2}) with estimation of the statistical averages $\lle\bfu\rle_i(\bfx)$ (\ref{4.16}$_1$)
and $\lle\bfu\rle(\bfx)$ (\ref{4.18}$_4$). A set of effective surrogate operators (\ref{6.3}) was estimated for the general nonlinear perturbator operator $\bfcL^{\theta\zeta}_q(\bfz,\overline{\bfze})$ (\ref{3.5}).

However, for the estimation of effective moduli (requering both the medium statistical homogeneity and homogeneous volumetric boundary conditions (\ref{2.30}) ) from the iteration approach (\ref{4.6}), some additional assumptions are needed.
So, for CM with {\it linear elastic} properties of the matrix (i.e. $^L\!\bfL^{(0)}$ and $\bfC^{(0)}$), Eq. (\ref{4.12}) is reduced to
to the representation of the effective elastic modulus similar to their local elastic counterpart
\BB
\label{4.15}
\bfL^*=\bfL^{(0)}+\bfR^*,\ \ \ \lle\bftau\rle^l=\bfR^*\lle\bfep\rle,
\EE
where $\bfL^{(0)}$ is defined by Eq. (\ref{3.18}) with
no assumption about $\bfC^{(0)}$ (e.g., either the spherical shape of ${\cal H}_{\bf x}$ or r isotropy of $\bfC^{(0)}$).
Equation (\ref{4.15}$_2$) implies linearity of the effective properties (although the limiting representation (\ref{3.5}) was obtained for generally nonlinear perturbator operators $\bfcL^{\theta\zeta}_j(\bfz,\overline{\bfze})$). The representation $\lle\bftau\rle^{l}_i=\lle\bftau\rle^{l\omega}_i$ (\ref{4.15}$_2$) can be also expressed through the displacement field
at the external interaction interface $\Gamma^-_i$ of the representative inclusion $v_i$
(see for details \cite{Buryachenko`2022a}, \cite{Buryachenko`2019a})
\BBEQ
\label{4.16}
\lle\bftau\rle^{l\omega}_i&:=&(\overline v_i^l)^{-1}\int_{\Gamma^-_i}\bftau(\bfs)\,d\bfs= \lle[\bfcL^{\sigma}(\bfu)(\bfs)\otimes\bfn(\bfs)]{\,}{\,}^{\,^S}\!\!\!\!\!\!\otimes \bfs\rle^{l\omega }_i
\nonumber\\
&-&
\bfL^{(0)}\lle\bfu(\bfs){\,}{\,}^{\,^S}\!\!\!\!\!\!\otimes \bfn(\bfs)\rle^{l\omega }_i,
\EEEQ
where $\bfu(\bfs)$ ($\bfs\in \Gamma_i^-$) are evaluated by the use of Eqs. (\ref{3.5}$_2$) and (\ref{4.7}).

Subsequent simplifications of Eqs. (\ref{4.6}) and (\ref{4.12}) can be performed in the framework of Hypothesis {\bf H1a} (\ref{3.1}) for the linear operators $\bfcJ^{\theta\zeta}_{i,j}$ (\ref{3.8}) and $\bfcJ^{\theta\zeta\infty}_{i,j}$ (\ref{3.10}) which can be decomposed and reduced to the tensors at the applying to the constant effective fields (\ref{4.1})
\BBEQ
\!\!\!\!\!\!\!\!\!\!\!\!\!\!\!\!\!\!\bfcJ^{I\theta\zeta}_{i,j}(\bfz ,\lle\overline{\bfze}_{i}\rle)&+&\bfcJ^{J\theta\zeta}_{i,j}(\bfz ,\lle\overline{\bfze}_{j}\rle)
=\bfJ^{I\theta\zeta}_{i,j}(\bfz )\lle\overline{\bfze}_{i}\rle(\bfx_i)+\bfJ^{J\theta\zeta}_{i,j}(\bfz )\lle\overline{\bfze}_{j}\rle( \bfx_j ),
\nonumber
\\
\label{4.17}
\!\!\!\!\!\!\!\!\!\!\!\!\!\!\!\!\!\!\bfcJ^{\theta\zeta\infty}_{i,j}(\bfz ,\lle\overline{\bfze}_{j}\rle)
&=&\bfJ^{\theta\zeta\infty}_{i,j}(\bfz )\lle\overline{\bfze}_{j}\rle( \bfx_j ).
\EEEQ
In particular, for the linear operators $\bfcL^{\theta\zeta}_i(\bfz,\overline{\bfze})$ (\ref{3.5})
and homogeneous effective strain $\overline\bfep(\bfx)=\overline\bfep_i$ the corresponding operators are reduced to the
tensor multiplications
\BBEQ
\!\!\!\!\!\!\!\!\!\!\!\!\!\!\!\bfthe(\bfz)-\overline{\bfthe}(\bfz)&=&\bfL^{\theta\varepsilon}_i(\bfz)\overline{\bfep}_i, \ \
\bfthe(\bfz)=\bfA^{\theta\varepsilon}_i(\bfz)\overline{\bfep}_i, \nonumber\\
\label{4.18}
\!\!\!\!\!\!\!\!\!\!\!\!\!\!\!\!\!\!\bfsi(\bfx)&=&\bfL^{\sigma\varepsilon}_i(\bfx),\ \ \
\overline{\bfep}_i , \bftau ({\bf x}) = {\bfR}_i(\bfx)\overline{\bfep}_, \\
\label{4.20}
\!\!\!\!\!\!\!\!\!\!\!\!\!\!\!\!\!\!\!\!\!\!\!\!\bfR_i(\bfx)\lle\overline{\bfep}\rle&=&\bfcL^{\sigma}(\bfC_1\bfeta)(\bfx), \ \ 
\bfeta(\bfz,\bfy)= \big[\bfA^{u\epsilon}_i(\bfz-\bfx_i)-\bfA^{u\epsilon}_i
(\bfy-\bfx_i)\big]\overline{\bfep}.
\EEEQ
Then, substitutions of Eqs. (\ref{4.17}) into Eq. (\ref{4.4}) leads to the linear algebraic equation
\BBEQ
\label{4.20}
\langle \overline{\bfthe}_i\rangle(\bfz)&=&\langle \bfthe\rangle ({\bf z})
+\int \big\{[\bfJ^{I\theta\zeta}_{i,j}(\bfz )\lle\overline{\bfze}_{i}\rle(\bfx_i)+
\bfJ^{J\theta\zeta}_{i,j}(\bfz )\lle\overline{\bfze}_{j}\rle(\bfx_j )]
\nonumber\\
&\times&
\varphi (v_j,{\bf x}_j\vert; v_i,{\bf x}_i)-\bfJ_{i,j}^{\theta\zeta\infty}(\bfz )\lle\overline{\bfze}_{j}\rle(\bfx_j )\bigl\}d{\bf x}_j.
\EEEQ

Equations (\ref{4.20}) and (\ref{4.18}$_4$) can be explicitly solved with the estimation of the effective tensors $\bfR^*$ (\ref{4.15}$_2$)
and the effective moduli $\bfL^*$ (\ref{4.15}). This scheme was realized for the linear bond-based model of constituents of CM \cite{Buryachenko`2020}, and for the linear state-based thermoelastic model \cite{Buryachenko`2023b},
\cite{Buryachenko`2023e}.
In particular, the effective field method (EFM) using the closing assumptions {\bf H2b} (\ref{4.2}) and (\ref{4.3})
leads to
\BBEQ
\label{4.21}
\!\!\!\!\!\!\!\!\!\!\!\!\!\!\!\!\!\!\bfL^{\rm *EFM}&=&\bfL^{(0)}+c^{l(1)}\bfR_i^{l\omega}\bfD^{\rm EFM},\nonumber \\
\!\!\!\!\!\!\!\!\!\!\!\!\!\!\!\!\!\!(\bfD^{\rm EFM})^{-1}&=&\bfI-\int\lle\bfn(\bfs){\,}{\,}^{\,^S}\!\!\!\!\!\!\otimes\bfL_j^{u\epsilon}(\bfs-\bfx_i)\rle^{l\omega}_i
[\varphi (v_j,{\bf x}_j\vert; v_i,{\bf x}_i)-n^{(j)}]d{\bf x}_j,
\EEEQ
where $\bfR_i^{l\omega}=\bfR_i^l$.
For the Mori-Tanaka approach (MTA) using the closing assumption $ \overline{\bfep}(\bfx)=\lle\bfep\rle^{(0)}$
instead of the assumptions (\ref{4.2}) and (\ref{4.3}), the representation $\bfL^{*\rm MTA} $
can be obtained (see for deails \cite{Buryachenko`2022a}).
Numerical results for 1D cases are presented in \cite{Buryachenko`2014b}, \cite{Buryachenko`2014c},
\cite{Buryachenko`2015b}, \cite{Buryachenko`2020}. It was detected a dependence of the
effective moduli estimations
(\ref{4.21}) and (\ref{4.25}) on both the micromodulus profile [e.g. either (\ref{2.26}$_1$) or (\ref{2.26}$_2$)]
and the ratio
$l_{\delta}/a$ of the horizon and inclusion size whereas the estimations obtained by the mixture theory (see \cite{Chen`et`2021}, \cite{Frank`et`2023}, \cite{Mehrmashhadi `et`2019}, \cite{Wu`C`2023}, \cite{Wu`et`2021}) do not depend on ratio $l_{\delta}/a$ (i.e. the scale separation hypothesis ({3.55}) is implicitly used in the mixture theory).

\sffamily
\noindent{\bf Comment 4.3.} Locally elastic counterpart of CAM (\ref{4.6})-(\ref{4.11})
uses the new GIE
proposed in \cite{Buryachenko`2010a}, \cite{Buryachenko`2010b} (see for details \cite{Buryachenko`2022a})
and permits to abandonment of the basic
concepts of micromechanics:
the hypothesis of ``ellipsoidal symmetry" {\bf H3} (\ref{4.3}), and the
effective field hypothesis EFH, {\bf H1a} (\ref{3.1}).
Some new effects (with possible correction of a sign of statistical averages of local field estimations, see \cite{Buryachenko`2022}, p 481) were discovered that were impossible in the framework of the classical (the first) background of micromechanics (see \cite{Buryachenko`2022}).
Furthermore, the LM methods mentioned are linear
whereas the PM methods
(see Eqs. (\ref{3.5})-(\ref{3.13}), (\ref{4.2})-(\ref{4.6}), and (\ref{4.12})) are nonlinear. We used only linearity for the matrix at the obtaining of Eqs. (\ref{4.17}), (\ref{4.17}$_1$), and (\ref{4.18}).

\rmfamily

\noindent {\bf Periodic stucture CMs 4.2.2.}
We now turn our attention to the {\it Computational Homogenization} (CH) in PM for the periodic structure CM. The background of CH is based on the volumetric periodic boundary conditions (VPBC) (\ref{2.35}) introduced in \cite{Buryachenko`2018}, \cite{Buryachenko`2018b},
\cite{Buryachenko`2022a} (see also \cite{Galadima`et`2023}, \cite{Galadima`et`2023b},
\cite{Galadima`et`2023c}, \cite{Hu`et`2022}, \cite{Qi`et`2024}, \cite{Xia`et`2020}, \cite{Xia`et`2019}, \cite{Xia`et`2021a}, \cite{Xia`et`2021b}). Estimation of effective moduli $\bfL^*$ begins from
the evaluation of both the overall macrostress $\{\bfsi\}=\lle\bfsi\rle^{\Omega}$ and the overall macrostrain $\{\bfep\}=\lle\bfep\rle^{\Omega}$ of the UC $\Omega_{00}$ by the use of the Gauss-Ostrogradsky theorem
\BBEQ
\label{4.22}
\lle\bfsi\rle^{\Omega}&:=&|{\Omega_{00}}|^{-1}\int_{\Omega_{00}}\bfsi(\bfx)d\bfx=
|{\Omega_{00}}|^{-1}\int_{\Gamma^0}\bft(\bfs)\,\,\,^{^S}\!\!\!\!\!\! \otimes\bfs d\bfs,\\
\label{4.23}
\lle\bfep\rle^{\Omega}&:=& |{\Omega_{00}}|^{-1}\int_{\Omega_{00}}\bfep(\bfx)d\bfx=
|{\Omega_{00}}|^{-1}\int_{\Gamma^0}\bfu(\bfs)\,\,\,^{^S}\!\!\!\!\!\! \otimes\bfn(\bfs) d\bfs,
\EEEQ
in terms of the traction $\bft(\bfs):=\bfsi(\bfs)\bfn(\bfs)$ and the displacement $\bfu(\bfs)$ on the geometrical boundary of the UC $\bfs\in \Gamma^0$ with the outward normal unit vectors $\bfn(\bfs)$ on $\Gamma^0$.
Both pairs of Eqs. (\ref{4.22}$_1$), (\ref{4.23}$_1$) and
(\ref{4.22}$_2$), (\ref{4.23}$_2$) are equivalent. However, the author \cite{Buryachenko`2018}, \cite{Buryachenko`2018b}, \cite{Buryachenko`2022a} used the averages (\ref{4.22}$_2$) and (\ref{4.23}$_2$) at the UC boundary (these evaluations are most simple) rather than
the volume averages of the stresses (\ref{4.22}$_1$) and strains (\ref{4.23}$_1$) inside the UC
volume in a less general and more cumbersome method of exploiting
\cite{Galadima`et`2023}, \cite{Galadima`et`2023b}, \cite{Galadima`et`2024}, \cite{Hu`et`2022} (there are some difficulties in estimations of local strain fields in UC because the displacement field is not differentiable in UC).

For the periodic peristatic CMs subjected to the remote homogeneous volumetric boundary conditions
either (\ref{2.30}) or (\ref{2.31}), the effective stiffness $\bfL^*$ are estimated as a proportionality factor
between the UC's averages of the stresses $\lle\bfsi\rle^{\Omega}$ and strains $\lle\bfep\rle^{\Omega}$
\BB
\label{4.24}
\lle\bfsi\rle^{\Omega}=\bfL^*\lle\bfep\rle^{\Omega}.
\EE
Equations (\ref{4.22})-(\ref{4.24}) are exactly coincide with the corresponding equations of computational LM (see
e.g., \cite{Kouznetsova`et`2001}, \cite{Miehe`K`2002}, \cite{Matous`et`2017}, \cite{Terada`K`2001}).

Estimating the involved macro variables is performed by the {{ \it micro-to-macro} transition.
Namely, owing to the volumetric PBC (\ref{2.35}), the governing equation (\ref{2.6}) depend only
on the displacement inside the UC $\Omega_{00}$ rather than on the displacement inside the extended UC
$\Omega_{00}\oplus{\cal H}_0$; i.e. the new closed integral equation obtained does not require any additional boundary conditions.
Discretization of this equation in the form either acts as a macro-to-micro transition
of the {\it deformation-driven} type, where the overall deformation $\bfep^{w_{\Gamma}}=\lle\bfep\rle^{\Omega}$ (see remote BC (\ref{2.30})) is controlled. Determination of the microstructural displacements $\bfu(\bfx)$ ($\bfx\in\Omega_{00})$ in an accompany with
the volumetric PBC (\ref{2.35}) allows one to estimate the peristatic traction
$\bft(\bfs):=\bfsi(\bfs)\bfn(\bfs)$ at the geometrical UC's boundary $\bfs\in\Gamma^0$. In so doing, at exploiting
of volumetric periodicity (\ref{2.35}) of displacement $\bfu(\bfx)$ ($\bfx\in w^{\Omega}_{\Gamma}$), the condition of the traction antiperiodicity (\ref{2.36}) is not used and automatically fulfilled.
In its turn, the found traction $\bft(\bfs)$ ($\bfs\in\Gamma^0$) is exploited for estimation of the macroscopic
stresses $\lle\bfsi\rle^{\Omega}$ by the use of Eq. (\ref{4.25}$_1$) with subsequent evaluation of the effective moduli $\bfL^*$ by Eq. (\ref{4.24}). A solution of the current linear problem can be performed in a standard iterative manner
due to the linearity of the fictitious body force $\bfb^{\Omega 0}:=\b{\bfb}^{\Omega 0}\lle\bfep\rle^{\Omega}$ with respect to $\lle\bfep\rle^{\Omega}$.
Indeed, at the initial macroscopic strain $\lle\bfep\rle^{\Omega}_{[0]}$ we found
the microdisplacements $\bfu_{[n]}(\bfx)$ ($n=1$) and the next step of $\lle\bfep\rle^{\Omega}_{[n]}$ (\ref{4.26}$_2$) which is used for estimation of the next step of the of $\bfb^{\Omega 0}_{[n+1]}=\b{\bfb}^{\Omega 0}
\lle\bfep\rle^{\Omega}_{[n]}$. In the considered numerical examples for 1D problems 5 iterations described provide an accuracy of $1.0E-14\%$ for the constraint Eq. (\ref{4.26}$_2$) (see for details \cite{Buryachenko`2018b}).

\sffamily
{\noindent \bf Comment 4.4.}
The constraints (\ref{4.22}$_2$) and (\ref{4.23}$_2$) can be incorporated through Lagrange multipliers in a straightforward manner (see, e.g., \cite{Madenci`et`2016}) which can also be extended to consider the nonlinear problems.
It should be mentioned that in CMic (see Block 2 in (\ref{3.3})) for LM, several high-performance algorithms
with improved convergence rates and solution accuracy have been developed to
study the non-linear response (such as plasticity, viscoplasticity, damage, fracture,
and fatigue, etc.) of microstructures with arbitrary phase contrast subjected
to finite deformations (see for references \cite{Kouznetsova`et`2001}, \cite{Miehe`K`2002}, \cite{Matous`et`2017}).
Generalizations of these methods to the PM are desirable.
\rmfamily

\sffamily
\noindent{\bf Comment 4.5.} It should be mentioned that the AMic method's classification (\ref{3.2}) in LM is also applicable to the PM counterparts.
{\color{black} Model methods Gr1) include the simplified methods such as, e.g. the mixture theory, (see Introduction for details)
\cite{Askari`et`2006}, \cite{Askari`et`2008},
\cite{Askari`et`2015},
\cite{Cheng`et`2024}, \cite{Hu`et`2011}, \cite{Hu`et`2012a}, \cite{Mehrmashhadi `et`2019}, \cite{Wu`C`2023}, \cite{Wu`et`2021},
which have attracted increasing attention for dealing
with complex laminated problems
\cite{Basoglu`et`2022}, \cite{Diyaroglu`et`2016}, 
\cite{Hu`et`2014}, \cite{Madenci`O`2014}, \cite{Madenci`et`2021},
\cite{Madenci`et`2023}, \cite{Ren`et`2022},
\cite{Xu`et`2008}).
The CAM of peridynamic CM belongs to the area of AMic (\ref{3.2}) where the DNS is not used (see Subsection 3.1), although
nothing in Eqs.(\ref{4.4})-(\ref{4.22}) is analytical.
So, Eqs. (\ref{4.4})-(\ref{4.22}) correspond to the LM self-consistent methods Gr4) (\ref{3.2}), i.e. CAM belongs to the grope Gr4).
The term ``computational" in CAM is used to emphasize that background hypotheses of analytical micromechanics are abandoned (see Comments 4.3) which expectedly leads to increasing in the computational complexity of CAM.} The original version of CAM in PM was proposed in \cite{Buryachenko`2020} for linear bond-based properties of phases, it was generalized to any nonlocal elastic properties of phases in \cite{Buryachenko`2023k}.
Perturbation method Gr2) corresponds to the dilute approximation in the PM considered in \cite{Buryachenko`2020b}.
The popular method of the effective medium (MEM) belonging to class Gr1) in the LM has a straightforward generalization to the PM as one peridynamic inclusion in the local effective medium.
Variational methods (grope Gr3)) were considered in \cite{Buryachenko`2020c}. Of special interest in the LM are Hashin-Shtrilman (H-S) bounds, which
were, in particular, obtained from the classical GIE (see \cite{Willis`1981}). However, H-S bounds were also estimated from the new GIE
(LM's counterpart of Eq. (\ref{3.12}),
see pp. 439-442 in \cite{Buryachenko`2022a}) for CMs
with noncanonical inhomogeneous inclusions.
Owing to the formal similarity of GIEs in the LM and PM (\ref{3.12}),
it would be interesting to generalize the H-S bounds obtained in LM to their peridynamic counterpart.

\noindent {\bf Comment 4.6.} Just for completeness, it should be also emphasized that the classification of CMic used DNS (see Eq. (\ref{3.3})) in the LM is also applicable for the PM counterparts (see Introduction for details). Asymptotic homogenization methods (Block 1 in (\ref{3.3}) in PM were considered in
\cite{Allali`L`2012}, \cite{Du`et`2016}, \cite{Du`et`2020}). The papers \cite{Buryachenko`2018}, \cite{Buryachenko`2018b},\cite{Diyaroglu`et`2019a}, \cite{Diyaroglu`et`2019b},
\cite{Galadima`et`2023}, \cite{Galadima`et`2023b}, \cite{Silling`et`2023}
\cite{Qi`et`2024}, \cite{Xia`et`2020}, \cite{Xu`et`2021}
\cite{Xu`F`2020}
\cite{You`et`2022}
\cite{You`et`2021} are dedicated to the generalization of classical computational homogenization approaches (Block 2 in (\ref{3.3})). At last, there are a lot of problems for one or finite numbers of inclusions (or cracks) in a sample (Block 3 in (\ref{3.3})): \cite {Agwai`et`2011}
\cite{Askari`et`2006},
\cite{Askari`et`2008},
\cite{Askari`et`2015},
\cite{Azdoud`et`2013},
\cite{Birner`et`2023} ,
\cite{Bobaru`H`2011},
\cite{Breitenfeld`et`2014},
\cite{Dipasquale`et`2022},
\cite{Ha`B`2010},
\cite{Hu`et`2012b},
\cite{Jenabidehkordi`et`2020},
\cite {Kilic`M`2010},
\cite{Kilic`M`2010b},
\cite{Laurien`et`2023},
\cite{Le`et`2014},
\cite{Macek`S`2007},
\cite{Madenci`G`2015},
\cite{Mousavi`et`2021},
\cite{Nguyen`et`2021},
\cite{Ren`et`2017} ,
\cite {Sarego`et`2016},
\cite {Wang`et`2018},
\cite{Wen `et`2023},
\cite{Yang`et`2023b},
\cite{Zhan`et`2021},
\cite {Zhou`W`2021}.
\rmfamily

\subsection{Body forces with compact support}

For random structure CMs, the scheme (\ref{4.9}) forms the Neumann series for the solution
($\bfx\in v_i^l$)
\BB
\label{4.25}
\langle \overline{\bfu}\rle_i(\bfx) :=\langle \lim_{n\to \infty}\overline{\bfu}^{[n+1]}\rle_i(\bfx) =\widehat{\bfD}_i^{ub}(\bfb,\bfx),
\EE
which yields the representation for the statistical averages of either the displacements (inside the fixed extended inclusion $(\bfx\in v_i^l)$)
or stresses (inside the inclusion $\bfy\in v_i$)
(conditional average)
\BBEQ
\label{4.26}
\lle {\bfu}\rle_i(\bfx) &=& \widehat{\bfD}_i^{ub}(\bfb,\bfx) +
\bfcL_i^{uu}(\bfx-\bfx_i, \widehat{\bfD}_i^{ub}(\bfb,\bfx))\nonumber\\
\lle {\bfsi}\rle_i(\bfy) &=&\bfcL^{\sigma}(\lle {\bfu}\rle_i(\bfx))(\bfy).
\EEEQ
It should be mentioned that the tensor $\widehat{\bfD}_q^{ub}(\bfb,\bfy)$ (\ref{4.25})
is an inhomogeneous function of coordinates of the moving inclusion $\bfy\in v_q^l$.
Moreover, the tensors $\widehat{\bfD}_q^{ub}(\bfb,\bfy)$ (\ref{4.25}) depend on all interecting inclusions (at least at $|\bfx_q|\leq a^{\delta}+l_{\delta}$).

Equation (\ref{4.26}$_1$) allows us to estimate a statistical average of the displacement field in the macropoint $\bfX$
\BB
\label{4.27}
\langle {\bfu}\rle(\bfX) :=c^{l(0)}\lle\bfu\rle^{l(0)}(\bfX)+
c^{l(1)}\lle\bfu\rle^{l(1)}(\bfX)
\EE
expressed through the statistical averages of displacements in the point $\bfX$ in the matrix
$\lle\bfu\rle^{l(0)}(\bfX)$ and inclusions $\lle\bfu\rle^{l(1)}(\bfX)$.
Indeed, at first we built some auxiliary set $v ^1_i (\bfX)$ with the indicator function
$V^1_i(\bfX)$ and
the boundary $\partial v ^1_i (\bfX)$
formed by the centers of translated ellipsoids $v_q({\bf 0})$ around the fixed point $\bfX$.
$v ^1_i (\bfX)$ is constructed as a limit $v^ 0_{ki}\to v_q^1(\bfX) $ if a fixed ellipsoid $v_ k$ is shrinking to the point $\bfX$.
{\color{black} Then $\lle\bfu\rle^{l(1)}(\bfX)$ can be estimated as ($\bfy\in v^l_q)$
\BBEQ
\label{4.28}
\!\!\!\!\!\!\!\!\!\!\!\!c^{l(1)}\lle\bfu\rle^{l(1)}(\bfX)\!&=&\!
c^{l(1)}\bfu^{b(0)}(\bfX)+
\!\!\int_{v^1_i({\bf X})} \!\!\!\!\!\!\!n^{(1)}(\bfx_q)
\bfcL^{uu}_q(\bfX\!-\bfx_q,\bfcD_q^u(\bfb,\bfy))
d{\bf x}_q,
\EEEQ
and, therefore,
the statistical average of displacements (\ref{4.27}) is expressed through the body force density ($\bfy\in v^l_q)$
\BBEQ
\label{4.29}
\lle\bfu\rle(\bfX)\!&=&\!
\bfu^{b(0)}(\bfX)+ c^{(0)}
\int \bfcL^{uu}_q(\bfX-\bfx_q,\bfcD_q^{ub}(\bfb,\bfy))
\varphi (v_q,{\bf x}_q\vert; \bfX) d{\bf x}_q\nonumber\\
\!&+&\! \int_{v^1_i({\bf X})} n^{(1)}(\bfx_q)
\bfcL^{uu}_q(\bfX-\bfx_q,\bfcD_q^{ub}(\bfb,\bfy))
d{\bf x}_q,
\EEEQ
where the first and the second integrals in Eq. (\ref{4.28}) correspond to the first and the second terms in the right-hand side of Eq. (\ref{4.27}), and $\varphi (v_q,{\bf x}_q\vert; \bfX)=0$ if $\bfx_q\in v^1_i(\bfX)$. }

The macroscopic stresses in the effective constitutive law are
\BB
\label{4.30}
\lle\bfsi\rle (\bfx)=\lle\bfcL^{\sigma(0)}(\bfu)\rle(\bfx)+\lle\bftau\rle(\bfx)
\EE
where the first item on the right-hand side is simplified for the linear matrix (see Eq. (\ref{3.20}))
and depends on the statistical averages $\lle\bfu\rle(\bfx)$ ($\bfx\in \mathbb{R}^d$) (\ref{4.29}).
For an estimation of an average $\lle\bftau\rle(\bfx)$, we consider, at first, the fixed inclusion
$v_q$ with the center $\bfx_q$. This inclusion produces value
$\bfcL^{\sigma}_1(\bfx-\bfx_q,\bfu)=\bfcL^{\sigma}(\widetilde{\bftau})$ estimated by Eq. (\ref{2.17}) with replacement $\widetilde\bff^{(0)}\to \widetilde\bff_1$ as in Eq.
(\ref{3.16}). Then
a statistical average of the local polarization tensor $\lle\bftau\rle(\bfX)$ is also obtained by averaging over
the domain $v_i^1$
\BBEQ
\label{4.31}
\!\!\!\!\!\!\!\!\!\!\!\!\!\!\!\!\lle\bftau\rle(\bfX)&=&\int_{v^1_i({\bf X})} n^{(1)}
\bfcL^{\sigma}(\bfC_1\bfeta^D)(\bfy-\bfX)d\bfy,%
\\
\label{4.32}
\!\!\!\!\!\!\!\!\!\!\!\!\!\!\!\!\! \bfeta^D(\bfx,\bfy)&=&\Big[\bfcD_i^{ub}(\bfb,\bfx)+\bfcL_i^{uu}(\bfx-\bfx_i,
\bfcD_i^{ub}(\bfb,\bfx) \nonumber\\
&-& \bfcD_i^{ub}(\bfb,\bfy)-\bfcL_i^{uu}(\bfy-\bfx_i,
\bfcD_i^{ub}(\bfb,\bfy))\Big],
\EEEQ
where $v_q^1(\bfx)$ is defined by (\ref{2.41}) and one used the representation for the statistical average of the displacement field in
the inclusion $\bfx\in v_q$: $\lle\bfu\rle_q(\bfx)$ (\ref{4.28}). {\color{black} Estimation of effective micromodulus
\cite{Yang`et`2024} (nonlocal parameter) through averaging of the strain energy of peridynamic CM (corresponding to the homogeneous boundary conditions) is questionable.}

The new nonlocal effective constitutive low (\ref{4.32}) looks very similar to the nonlocal effective constitutive law
\BB
\label{4.33}
\lle\bfsi\rle(\bfx) ={^L}\!\bfL^{(0)}\lle\bfep\rle(\bfx) +\lle ^L\!\bftau\rle(\bfy), \ \ \ ^L\!\bftau(\bfy):={^L}\!\bfL_1(\bfy)\bfep(\bfy),
\EE
corresponding to locally elastic constitutive law (\ref{2.2}) at the microlevel. But,
the first term ${^L}\!\bfL^{(0)}\lle\bfep\rle(\bfx) $ in the right-hand side of Eq. (\ref{4.33})
corresponds to the arbitrary inhomogeneous $\lle\bfep\rle(\bfx)\not\equiv$const. whereas $\lle\bfcL^{\sigma(0)}(\bfu)\rle(\bfx)$
(\ref{4.30})
is reduced to $\!\bfL^{(0)}\lle\bfep\rle(\bfx) $ only for homogeneous $\lle\bfep\rle(\bfx)\equiv$cons.

Numerical analysis was performed for 1D statistically homogeneous random structure bar \cite{Buryachenko`2023}, \cite{Buryachenko`2023a} for self-equilibrated body force (\ref{2.4}): $\bfb(\bfx) = 0$ for $|\bfx| > B^b$ and $\bfx(\bfx)=-\bfb(-\bfx)$ for $|\bfx| \leq v^b$. The engineering approach using the scale separation hypothesis
(\ref{2.28}$_2$) can be described in the following scheme: estimation of $\bfL^*$ (\ref{4.28}) at the homogeneous VBC (\ref{2.30}) , and evaluation of $\lle\bfu^{\rm EA}\rle(\bfx)$ by Eq. (\ref{2.1})-(\ref{2.3}) with replacement $\bfL\to \bfL^*$. It leads to monotonical $\lle\bfu^{\rm EA}\rle(\bfx)$ (see the solid curve 4 in Fig. 3). In Fig. 3 the parameters $l_{\delta}/a=1$ and $c^{(1)}=0.5$ are fixed whereas $ B^b/a$ takes the values  $B^b/a=0.25,\ 0.375,\ 0.5$ (see the curves 1, 2, 3, respectively).
As can be seen in Fig. 3, simultaneous consideration of peridynamic constitutive equation (\ref{2.5}) and inhomogeneity of $\bfb(\bfx)$)
leads to strongly nonmonotonical distributions $\lle\bfu\rle(\bfx)$ dirrering from $\lle\bfu^{\rm EA}\rle(\bfx)$ by a factor 9 for curve 1.
Owing to nonlocal effects, the domain of the long-range action is limited not only by $|\bfx|\leq  B^b$ but by a domain $|\bfx|\leq a^{\rm l-r}$ $(a^{\rm l-r}\approx 3 B^b$) i.e. the problem domain $\bfx\in R^1$ is reduced to a finite size domain without inconsistencies mentioned before \cite{Buryachenko`2022b}.
It means that the domain size $|\bfx|\leq a^{\rm l-r}\approx 3 B^b$ (at
the considered ratios $a/a^{b}/l_{\delta}$)
required stabilization of displacement fields $\lle\bfu\rle(\bfx)\approx$const. (at $|\bfx|>a^{\rm l-r}$)
significantly exceeds the sum of the body force scale $ B^b$ and the horizon $l_{\delta}$ (synergism effect);
the parameter $a^{\rm l-r}$ is to be learned. Although the term RVE (see Comment 4.12) was not used, but, in fact, the domain
$|\bfx|\leq a^{\rm l-r}$ is the RVE for $\lle\bfu\rle(\bfx)$ and this RVE depend on the scale ratios $a/a^{b}/l_{\delta}$. Generalizing the found phenomena for a general body force $\bfb(\bfx)$ (\ref{2.4}), we can define RVE as the area for which $\exists B^{\rm RVE}$ (called the size of RVE) such as
\BB
\label{4.34}
\lle\bfu\rle(\bfx)=\bfu^{\infty}\equiv {\rm const}., \ \ {\rm for}\ \ \forall \bfx\geq B^{\rm RVE},
\EE
where equality is provided with prescribed tolerance.

For the linear bond-based model of constituents, Eqs. (\ref{4.32})-(\ref{4.32}) are reduced to the corresponding equations in \cite{Buryachenko`2023}, \cite{Buryachenko`2023a}. The representations (\ref{4.33}) and (\ref{4.32}) can be generalized to the nonlinear
inclusion properties analogous to Eqs. (\ref{4.6}). A nonlocal effective operator similar to (\ref{4.32}) was also obtained
from Eq. (\ref{4.26}$_1$) for the linear bond-based model in \cite{Buryachenko`2023}; generalization to the linear state-based model is obvious. For random structure CMs with locally elastic properties of phases, nonlocal effective operator \cite{Luciano`W`2001} is expressed in terms of microstructural information on the constituents and the applied body forces
through the Green function of the matrix (\ref{2.5}); generalization of this approach to PM is not obvious.
Possible practical applications of micromechanical modeling of CMs subjected to body force with compact support (e.g.
laser heating, see \cite{Yang`et`2019}, \cite{Isakari`et`2017}, \cite{Yilbas`2013}, and \cite{Yang`et`2019}) were considered in
\cite{Buryachenko`2024b}.

\vspace{-0.mm} \noindent \hspace{-5mm} \parbox{6.2cm}{
\centering \epsfig{figure=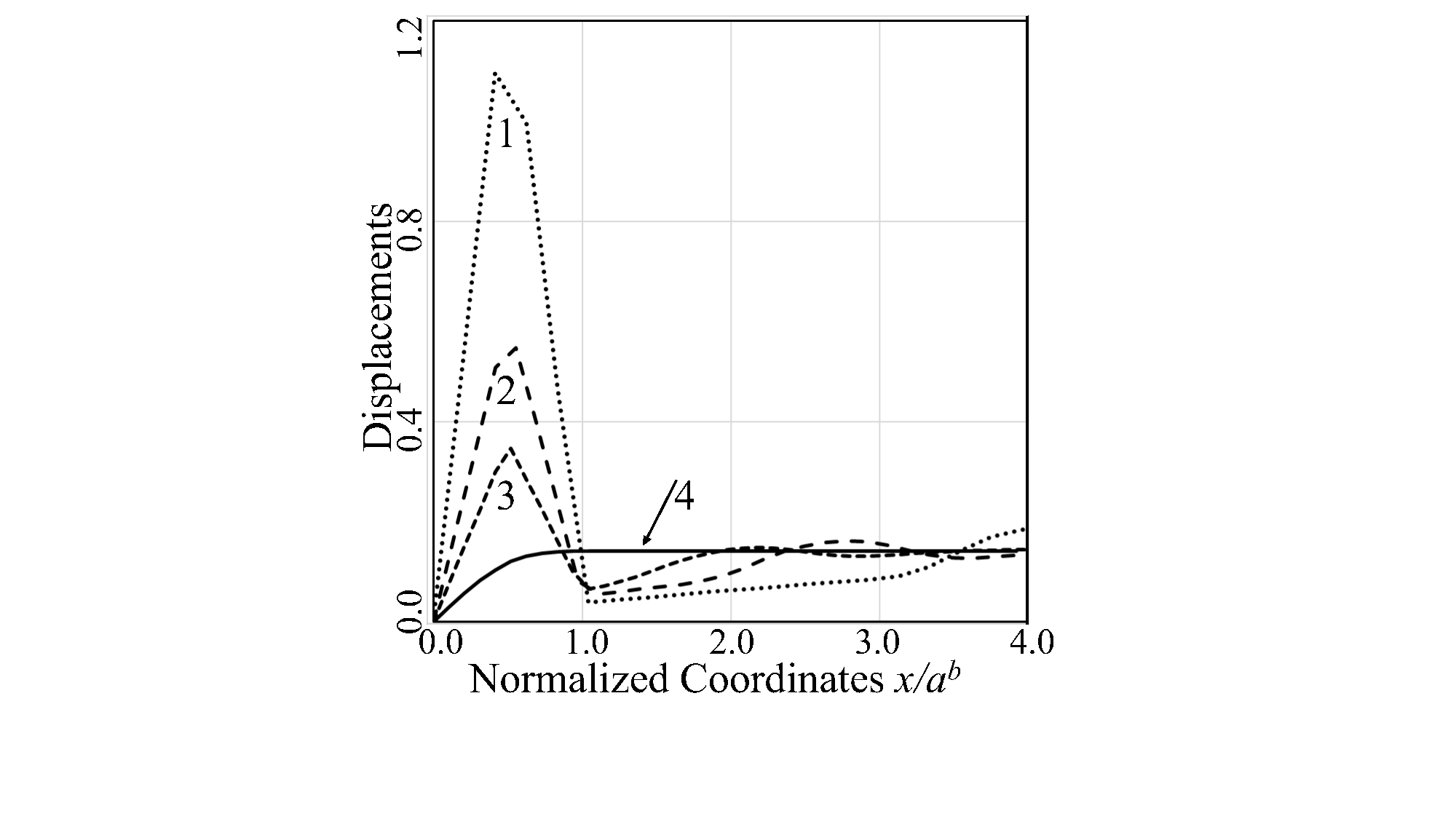, width=13.2cm}\\ \vspace{-22.mm}
\vspace{22.mm}
\vspace{-6.mm} \tenrm \baselineskip=8pt
\centerline{{\sc Fig. 3:} Displacements $\lle u\rle( x)$ vs $ x/ B^b$ }}

\subsection{Datasets for body force with compact support}

\rmfamily
For arbitrary body force with compact support $\bfb(\bfx)$ we obtained representations for effetive parameters (fields)
$\lle\bfu\rle(\bfx),\ \lle\bfsi\rle(\bfx),\ \lle\bfu\rle_i(\bfz,\bfx),\ \lle\bfsi\rle_i(\bfz,\bfx)$. For subsequent applications, we can consider
a dataset for effective parameters of random structure CM
\BBEQ
\label{4.35}
\!\!\!\!\!\!\!\!\!\!\!{\cal D}^{\rm r}\!=\!\{\lle{\bfu}_k\rle(\bfb_k,\bfx),\!\lle\bfsi_k\rle(\bfb_k,\bfx),\!\lle{\bfu}_{ik}\rle(\bfb_k,\bfz,\bfx),
\!\lle{\bfsi}_{ik}\rle(\bfb_k,\bfz,\bfx), \bfb_k(\bfx)
\}_{k=1}^N,
\EEEQ
where each effective parameters $\lle{\bfu}_k\rle(\bfx)$, $\lle{\bfsi}_k\rle(\bfx)$, $\lle\bfu_{ik}\rle(\bfz,\bfx):=\lle\bfu_{k}\rle_i(\bfz)$,
and $\lle{\bfsi}_{ik}\rle(\bfb_k,\bfz,\bfx)=\lle\bfsi_{k}\rle_i(\bfz)$
(\ref{4.35})
are estimated for each $\bfb_k$ in the offline stage in $k$-th realization; the coordinates $\bfx\in R^d$ are a macrocoordinate whereas
$\bfz\in v_i$ are local coordinates in the representative inclusion $v_i$.

Let us turn to the analysis of periodic structure CMs. We assume that the body force is fixed whereas the CM defined by the the grid $\bfx_{\bf m}$ is moving. Just for concreteness, we
consider
a peridynamic model (\ref{2.12}) and
we designate some configuration $\bfx_{\bf m}$ as an initial and mark them by the lower index $_0$ with the corresponding properties
$\bfC_0(\bfx,\bfq)$ (\ref{2.13}) and solution at the microlevel $\bfu_0(\bfx)$ (\ref{2.12}). At the parallel transition of the grid
$\bfx_{\bf m}$ over the vector $\bfchi$ ($\bfLa_0\to \bfLa_{\bf \chi}$) the corresponding stiffness $\bfC(\bfx,\bfq,\bfchi)$ (\ref{2.13}) , indicator function $V_i(\bfx,\bfchi)$,
and the solution $\bfu(\bfx,\bfchi)$ (\ref{2.12}) will also change
\BBEQ
\label{4.36}
\bfC(\bfx,\bfq,\bfchi)&=&\bfC_0(\bfx-\bfchi, \bfq-\bfchi), \ \ V_i(\bfx,\bfchi)=V_{i0}(\bfx-\bfchi),\\
\bfb(\bfx,\bfchi)&=&\bfb_0(\bfx),\ \ \bfu(\bfx,\bfchi)\not\equiv \bfu_0(\bfx-\bfchi),
\label{4.37}
\EEEQ
where the inequality (\ref{4.37}$_2$) holds because $\bfb(\bfx)$ is fixed. Of course, the same solution $\bfu(\bfx,\bfchi)$ can be obtained
in an alternate case when the field $\bfL(\bfx)$ is fixed whereas the body-force $\bfb(\bfx)$ is mowing
\BB
\bfC(\bfx,\bfq,\bfchi)=\bfC_0(\bfx,\bfq), \ \ \bfb(\bfx,\bfchi)=\bfb_0(\bfx-\bfchi),\ \ \bfu(\bfx,\bfchi)\not\equiv \bfu_0(\bfx-\bfchi).
\label{4.38}
\EE
However, we consider the case (\ref{4.36}) and (\ref{4.37}) as more convenient.

Thus, for each $\bfchi\in \cV_{\rm \bf x}$ we get the solution $\bfu(\bfx,\bfchi)$ obtained in the framework of (\ref{4.36}) and (\ref{4.37})
that makes it possible to define an effective (or macroscopic) displacement ($\bfx\in w$)
\BBEQ
\label{4.39}
\!\!\!\!\!\!\!\!\!\!\!\!\!\!\!\!\!\!\!\!\!\!\lle\bfu\rle(\bfx)\!&=&\!{1\over\overline
{\cV}_{\bf x}}\int_{{\cal V}_{\rm \bf x}}\bfu(\bfx,\bfchi)~d\bfchi, \
\lle\bfu\rle^{l(1)}(\bfx)= {1\over\overline{\cV}_{\bf x}}\int_{{\cal V}_{\rm \bf x}}\bfu(\bfx,\bfchi)V_i(\bfx,\bfchi)~d\bfchi, \\
\label{4.40}
\!\!\!\!\!\!\!\!\!\!\!\!\!\!\!\!\!\!\!\!\!\!\!\!\lle\bfsi\rle(\bfx)\!&=&\! {1\over\overline
{\cV}_{\bf x}}\int_{{\cal V}_{\rm \bf x}}\bfsi(\bfx,\bfchi)~d\bfchi,\
\lle\bfsi\rle^{(1)}(\bfx)= {1\over\overline
{{\cal V}}_{\bf x}}\int_{\cV_{\rm \bf x}}\bfsi(\bfx,\bfchi)V_i(\bfx,\bfchi)~d\bfchi,
\EEEQ
It is interesting that Eqs. (\ref{4.39}$_1$) and (\ref{4.40}$_1$) formally looks as averaging over the moving-window ${\cV}_{\bf x}$ (\ref{2.46}) although the operations (\ref{4.39}$_1$) and (\ref{4.40}$_1$) are conceptually different and obtained by averaging over the number of the displacement realizations $\bfu(\bfx,\bfchi)$ produced by a parallel transform of $\bfC(\bfx,\bfq,\bfchi)$ (\ref{4.38}) rather than by averaging of one realization
$\bfu_0(\bfx)$ over the moving-window ${\cV}_{\bf x}$ (\ref{2.46}) (see Eq. (\ref{4.38})).
The everages (\ref{4.39}) and (\ref{4.40}) are ``ensemble averaging'' with respect to the statistics of all ``translated'' realisations of the periodic microstructure which are constructed from a given periodic microstructure by all possible translations $\bfchi$
(with uniform random distribution on $\cV_{\bf x}$)
within the periodicity cell. Translated averaging (\ref{4.39}) and (\ref{4.40}) are applicable for periodic structure CMs with any constitutive laws of phases and any inhomogenious loading.
  The partiqular case of averaging (like (\ref{4.40}$_1$)) was proposed in assymptotic homogenization approach by
\cite{Smyshlyaev`C`2000} and \cite{Ameen`et`2018} (the authors of \cite{Smyshlyaev`C`2000} referred a personal communication by J.R. Willis)
for periodic media subjected to periodic loading. It is interesting that the statistical averages in Eqs. (\ref{2.44}), (\ref{4.28}), (\ref{4.39}$_2$), and (\ref{4.40}$_2$) are only reformulation of the student probability problem about ``random dropping of coin (inclusion $v_i$) on the fixed pont $\bfx\in R^d$".
But in a cases of (\ref{4.39}$_2$) and (\ref{4.40}$_2$), the dropping coin $v_i$ belongs to the randomly translating grid $\bfLa_{\bf \chi}$.

A scheme of averaging (\ref{4.40}) for 1D case is presented in Fig. 4. So, the fixed macropoint $\bfX\in b({\bf0}, B^b)$
corresponds to the local points $\bfx_j=\bfX-\bfchi$ in the translated grid $\bfLa_{\bf \chi}=\bfLa_j$ ($j=1,\ldots,6)$.
Estimation $\lle\bfsi\rle(\bfX)$ is performed by summation of $\bfsi(\bfX,\bfchi)=\bfsi(\bfx_j)$ over all realizations of
$\bfLa_{\bf \chi}=\bfLa_j$ ($j=1,\ldots,6)$. However, the average stresses inside inclusions $\lle\bfsi\rle^{(1)}(\bfX)$ is evaluated
by summations of $\bfsi(\bfX,\bfchi)=\bfsi(\bfx_j)$ in the realizations of $\bfLa_{\bf \chi}=\bfLa_j$ ($j=1,2,3,6)$.
It is interesting that owing to nonlocal effects, for $\bfY\not\in b({\bf0}, B^b)$, we can get $\lle\bfsi\rle(\bfY)\not= {\bf 0}$ althogh
$\bfb(\bfY)={\bf 0}$. $\lle\bfsi\rle(\bfY)$ is estimated by summation of $\bfsi(\bfX,\bfchi)=\bfsi(\bfx_j)$ for all realizations
$\bfLa_{\bf \chi}=\bfLa_j$ ($j=1,\ldots,6)$, whereas $\lle\bfsi\rle^{(1)}(\bfY)$ is obtained by summation of
$\bfsi(\bfX,\bfchi)=\bfsi(\bfx_j)$ ($j=1,2,6)$.

\vspace{-0.mm} \noindent \hspace{10mm} \parbox{6.2cm}{
\centering \epsfig{figure=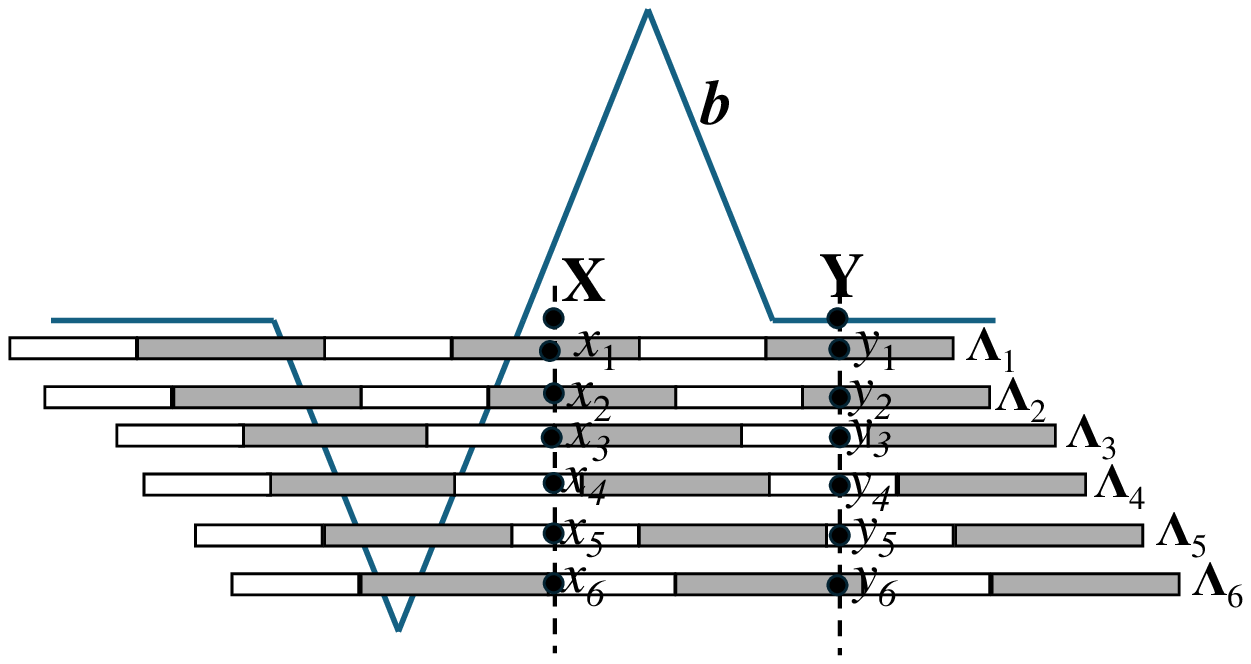, width=10.2cm}\\ \vspace{-22.mm}
\vspace{20.mm}
\vspace{-72.mm} \tenrm \baselineskip=8pt
{{\sc Fig. 4:} Scheme of translated averaging}}
\vspace{2.mm}

It should be also mentioned that the authors \cite{Xu`et`2021} (see also \cite{Xu`F`2020}) and
\cite{You`et`2021} used the averaged displacement field $
\lle \widetilde {\bf u}\rle^{\Omega}_i(\bfx) =|\Omega_i|^{-1}\int_{\Omega_i} \widetilde \bfu(\bfchi)~d\bfchi
$ for the fixed grid $\bfx_i$; i.e. the found averaged fields $\lle \widetilde {\bf u}\rle^{\Omega}_i(\bfx)$ are assugned to the specific grid
${\bf \Lambda}$ (in contrast with the translation averages (\ref{4.39}) and (\ref{4.40})).
It leads to obtaining a discrete kernel $\bfK(|\bfx|)$ as opposite to the continuous ones obtained by averaging over the ``moving averaging" cell (\ref{4.39}) and (\ref{4.40}) (see for details p. 895 in \cite{Buryachenko`2022a}).

Analogously to the dataset $\bfcD^{\rm r}$ (\ref{4.35}) for random structure CMs, we can define a dataset $\bfcD^{\rm p}$ for periodic structure CMs
\BBEQ
\label{4.41}
\!\!\!\!\!\!\!\!\!\!\!\!\!{\cal D}^{\rm p}\!=\!\{\lle{\bfu}_k\rle(\bfb_k,\bfx),\!\lle\bfsi_k\rle(\bfb_k,\bfx),\!\lle{\bfu}_{k}\rle^{l(1)}(\bfb_k,\bfx),
\!\lle{\bfsi}_{k}\rle^{(1)}(\bfb_k,\bfx), \bfb_k(\bfx)
\}_{k=1}^N,
\EEEQ
where each effective parameters are estimated for each $\bfb_k$ by Eqs. (\ref{4.39}) and (\ref{4.40}) in the offline stage in $k$-th realization. The coordinates $\bfx\in R^d$ are a macro coordinates, i.e. the dataset $\bfcD^{\rm p}$ is less detailed than
datased $\bfcD^{\rm r}$ depending on both the macro coordinates $\bfx\in R^d$ and local coordinats $\bfz\in v_i$ inside representative inclusion $v_i$.

To the best of the author’s knowledge, the method of reduction of DNS to smoothed (effective) parameters (\ref{4.41}) is new. Moreover, a critical issue in analyses of periodic structure CMs is the choice of
PBC (\ref{2.36}) and VPBC (\ref{2.35}) at the interface of UCs provides a connection to field distributions, which is the neighboring UCs. The known PBC (\ref{2.36}) and VPBC (\ref{2.35})  are definetly correspond to both the homogeneous remote loading ((\ref{2.30}) or (\ref{2.31})) and
zero body force $\bfb({\bfx})\equiv {\bf 0}$ (\ref{2.4}). For a general case of body force (\ref{2.4}) being considered, both
PBC (\ref{2.36}) and VPBC (\ref{2.35}) are incorrect (see Subsection 2.4). However, prescription of any PBC (\ref{2.36}) or VPBC ((\ref{2.35}) are not required if DNS is exploited in Eqs. (\ref{4.39}) and (\ref{4.40}) are estimated on RVE (containing, perhaps, a few UCs) rather than on one UC;
the size of RVE (\ref{2.38}) (see Comment 4.13) is a controlled parameter that should be learned with prescribed tolerance. Although the integrants
in Eqs. (\ref{4.39}) and (\ref{4.40}) are estimated by DNS in CMic, combining of Eqs. (\ref{4.39}) and (\ref{4.40}) with the RVE concept (see Eq. (\ref{2.38}) and Comment 4.13) allow the dataset $\bfcD^{\rm p}$ , and, because of this, the method (\ref{4.39}) and (\ref{4.40}) are called
CAM's version for periodic structure CMs.

Both datasets of effective macroscopic parameters $\bfcD^{\rm r}$ and $\bfcD^{\rm r}$ (with the scale $\Lambda$ (\ref{2.28}) of body force $\bfb(\bfx)$ (\ref{2.4})) remember neither the microstructure of CMs (either random, see Subsection 2.2, or periodic, see Subsection 2.4)
nor the method how the initial fields (see either the left-hand sides of Eqs. (\ref{4.26}), (\ref{4.29}), (\ref{4.30}) or the integrands in
Eqs. (\ref{4.39}) and (\ref{4.40})) were estimated.

It should be mentioned that Eq. (\ref{4.14}) is reduced by centering to more general Eq. (\ref{4.13}). However, Eq.
(\ref{4.14}) has a fundamental advantage to Eq. (\ref{4.13}) because precisely Eq. (\ref{4.14}) is used for the generation of the dataset $\bfcD^{\rm r}$ or $\bfcD^{\rm p}$ exploited for the construction of machine learning (ML) technique such as the surrogate operators (see Section 6). ML technique, in turn, allows us to obtain a surrogate dataset for any macroscopic loading (either $\lle\bfu\rle(\bfx)$ or $\bfb(\bfx)$) without solution of the micro-problem as in Eq. (\ref{4.13}).

\section {Representative volume element (RVE) }

\setcounter{equation}{0}
\renewcommand{\theequation}{5.\arabic{equation}}

\subsection {RVE for CMs subjected to remote homogeneous loading}

The concept of Representative Volume Element (RVE, introduced by Hill \cite{Hill`1963}) has a very long and dramatic history (see for references \cite{Ostoja`et`2016}). To properly reproduce Hill's definition and its significance, here's a citation from Hill's \cite{Hill`1963} providing a rigorous foundation for the RVE concept.

\noindent{\bf Definition 5.1.} {\it Representative volume element (RVE)
(a) is structurally entirely typical of the whole mixture on average, and
(b) contains a sufficient number of inclusions for the apparent overall moduli ({\scshape and statistically averaged field distributions in the phases} \rmfamily) to
be effectively independent of the surface values of traction and displacement, so
long as these values are ‘macroscopically uniform.... The contribution of this surface layer to any average can be negligible by taking the sample large enough.}

In the origin text definition \cite{Hill`1963}, one added a {\scshape small caps} font fragment explained later. We emphasize the background concepts of RVE reflecting the essence of this definition:

1. RVE is defined for either statistically homogeneous or periodic structure CMs subjected to remote homogeneous BC (\ref{2.32})
(or (\ref{2.33})), although Hill's original work \cite{Hill`1963} did not explicitly use terms like "statistically homogeneous" or "periodic structure CMs"
(and, moreover, the methods of Blocks 1 and 2 were unknown at the Hill's time). Farthemore, the media of ``functionally graded stucture" (this term was also unknown at the Hill's time) don't fall within the application of RVE concept.

2. The destiny of RVE is estimation of effective moduli $\bfL^*$ based, in turns, on the evaluation of average fields in the phases.
For periodic structure CMs, the RVE is equivalent to UC (RVE$\equiv$UC), and remote homogeneous BC (\ref{2.32}) (or (\ref{2.33})) are reduced to PBC (\ref{2.36}). Furthermore, in Definition 5.1 for statistically homogeneous CMs,
any ``statistically representative" geometrical descriptions (obtained by either computational simulation or image-based methods) of so-called RVE are not required. However, in engineering practice, the RVE is sampled using computational or image-based methods;
quantitative proof of the correctness of replacing remote homogeneous BCs (\ref{2.32}) (or (\ref{2.33})) with finite-size counterparts is essential, as finite-size effects could distort the estimation of effective properties.

3. Though Hill \cite{Hill`1963} applied his definition to linear locally elastic CMs, the term ``constitutive low" is absent in Definition 5.1.
Definition 5.1 is generalized to any peridynamic CMs by only replacement of one word ``surface" (see (\ref{2.32}) and (\ref{2.33}))
by ``volumetric" (see (\ref{2.30}) and (\ref{2.31})).

The RVE concept is so popular and natural that, over time, the RVE concept has become more of a heuristic or rule-of-thumb in practice, rather than a strict mathematical definition. Engineers and practitioners often use "intuitive" or simplified versions of RVEs based on their experience and common sense, such as choosing a volume element that simply "looks representative" to them. This is particularly true in cases where the microstructure of the material is complex but doesn't lend itself easily to analytical or computational homogenization methods. The key takeaway is that while intuitive RVEs might provide quick, rough estimates, they can often be misleading, especially when material behavior is strongly dependent on the details of its microstructure.
So, the term RVE in an intuitive (or common) sense was implemented to finite samples (subjected to either BC or VBC) generated by two popular methods. In the first case, one or a few random inclusion fields (which is used as a synonym of the phrase ``an arbitrary simulated field") are generated (see, e.g., \cite{Ongaro`et`2022}, \cite{Wildman`et`2017}, \cite{Yang`et`2024}, \cite{Yang`et`2023}).
In the second one, image-based methods (for example, microcomputer tomography, micro-CT) are used for the
modeling of ``real" RVE (see \cite{Ahmadi`et`2022}, \cite{Anbarlooie`H`2024}, \cite{Talamadupula`et`2020}).

The indicated inaccuracies of the intuitive RVE in PM are overcome in LM a long time ago (see Introduction for references). So,
the mentioned sample response is estimated from DNS of microstructural volume elements (MVEs) simulated or extracted, e.g. from micro-CT. The homogenized (apparent) properties $\bfL^{\rm A}_{\rm KUBC}$ and $\bfM^{\rm A}_{\rm SUBC}$ determined for the KUBC (\ref{2.32}) or SUBC (\ref{2.32}), respectively, are different.
The convergence trend of $\bfL^{\rm A}_{\rm KUBC}-(\bfM^{\rm A}_{\rm SUBC})^{-1}$, as the MVE
increases, allows one to approximate the RVE size and estimate
the effective moduli $\bfL^*$; i.e. RVE concept requires
the adoption of very large (theoretically infinite) material domains (see for references and details \cite{Ostoja`et`2016} and \cite{Buryachenko`2022a}, p. 226).
In so doing, the convergence of estimation of $\bfL^*$ (which is, in fact, an average response of a sample) with grooving of realization number of this intuitive EVE (see, e.g. \cite{Ongaro`et`2022}) does not guarantee the elimination of edge scale effect.
The mentioned approaches for RVE investigations in LM are based on the methods of Block 3 (\ref{3.3}). Pridynamic counterparts of these methods are well developed (see Comment 4.6.), and, therefore, their realization for RVE in PM is straightforward.

Another sort of inconsistency related to the replacement of PBC by the homogeneous boundary BC is also well known in LM (see, e.g., p. 508 in \cite{Buryachenko`2007}); the same difficulties are expected in PM at the modeling of infinite (e.g. periodic) media by a finite sample (see, e.g.,
\cite{Ahmadi`et`2022},
\cite{Askari`et`2006},
\cite{Askari`et`2008},
\cite{Askari`et`2015},
\cite{Cheng`et`2024},
\cite{Decklever`S`2016}
\cite{Jenabidehkordi`et`2020},
\cite{Yang`et`2023b},
\cite{Zhan`et`2021},
\cite{Zhang`Q`2021}).
The use of body force with compact support (like (\ref{3.11})) as a training parameter enables the sample size and edge effects to be eliminated.

We considered RVE interpretation for random structure CMs in the framework of CMic. Implementation of AMic (see Comment 4.5) to the random structure CMs has no potential difficulties and ambiguities related to eithe sample size or edge effects as well as with the prescription of ``surface values" of the sample. The situation is drastically simplified because the domain of interest is a full space $R^d$ and the homogeneous BCs
(see (\ref{2.32}) and (\ref{2.33})) or PBC (see (\ref{2.30}) and (\ref{2.31})) are remote. ``Taking the sample large enough" has no sense anymore and $\bfL^*$ is completely defined by a micromechanical model used.
So,
in the framework of the hypothesis (\ref{2.28}$_2$) in LM, the estimations of effective moduli $\bfL^*$ (\ref{3.21}) (``upscale") and the field concentration factors $\bfA^*(\bfz)$ ($\bfz\in v_i$) (``downscale")
\BB
\label{5.1}
^L\!\bfL^*=^L\!\bfL^{(0)}+^L\!\bfR^*, \ \ \ \lle\bfep\rle_i(\bfz)=^L\!\bfA^*(\bfz)\lle \bfep\rle
\EE
are identical for either random or periodic structure CMs; these are mutual-coupling problems (as two sides of one
coin). The mentioned identity of the problems (\ref{5.1}$_1$) and (\ref{5.2}$_1$) is a reason for addition part (see the {\scshape small caps} font) in Definition 5.1.
Then for random structure CMs, the size of RVE is completely defined by the estimation method of micromechanics.
So, in one particle methods (such as, e.g. effective field method (EFM), Mori-Tanaka method (MTM), and effective medium method (EMM), see, e.g., \cite{Buryachenko`2007})
we can accept RVE=$v_i$ and $a^{\rm int}=a$ (\ref{2.28}). In the multiparticle EFM (MEFM) taking a binary interaction into account leads to increasing of $a^{\rm int}\approx 6a$ (\ref{2.28}) with a corresponding size increase of RVE.
PM's counterparts of all mentioned one particle methods (see Comment 4.5) lead to different estimations of $\bfL^*$ with the same RVE$=v_i^l$.
Thus, a
computational bulkiness in Definition 5.1 is eliminated: we don't need to control that the sample is large enough by the CMic methods with a subsequent definition of RVE as a limit of this sample size. The case of AMic's implementation is significantly simpler.
For the periodic structure CMs, the RVE coincides with UC (therefore, the size of the RVE coincides with the size of UC). Thus, in the framework of hypothesis (\ref{2.28}$_2$) in LM, the equality $\rm RVE^{\bf L^*}\equiv RVE^{\lle\bf \epsilon\rle_i({\bf z})}$ holds for both random and periodic structure CMs.

For the periodic structure CMs with scale separation hypothesis (\ref{2.29}$_2$) accepted, the homogeneous BC (see (\ref{2.32}) and (\ref{2.33})) are definitely reduced to the PBC (\ref{2.36}), and RVE is uniquely determined by the equality RVE=UC.
Then Definition 5.1 implies the implementation of CMic methods (peridynamic counterparts of Block 1 or Block 2, see (\ref{3.3}) and Comment 4.6).

\subsection {RVE for CMs subjected to inhomogeneous loading}

The RVE concept
serving as a bridge between different scales is rigorously justified if a field scale
(field inhomogeneity) infinitely exceeds a material scale (material inhomogeneity, see (\ref{2.28}$_2$) and (\ref{2.29}$_2$) ).
Coupling of scales—when the material scale is comparable to the scale of internal inhomogeneities or stress gradients—naturally leads to nonlocal material behavior.
Nonlocal constitutive laws describe how the material's response is influenced by points far from the location of interest, which is especially relevant for materials with internal heterogeneities or discontinuities.

The nonlocal character of material behavior becomes evident when the assumption of statistical homogeneity is removed, leading to a more accurate representation of the material’s true behavior in the presence of microstructural heterogeneity or internal stresses.
In these models, the material response is described by statistical averages of stress and strain tensors, which are weighted by a tensorial kernel, reflecting the nonlocal interactions between different material points.
This approach requires the use of an effective elastic operator that incorporates the nonlocal coupling between stress and strain, and it leads to an integral formulation of the constitutive law.
There are known strongly nonlocal (strain type and displacement type, peridynamics) and weakly nonlocal
(strain-gradient, stress-gradient, and higher-order models) forms of these nonlocal operators.
Perhaps the most
challenging issue is how micromechanics can contribute to the understanding of the
bridging mechanism between the coupled scales, which is described by the nonlocal
constitutive equations involving the parameters of a relevant effective nonlocal operator.

\noindent{\bf Definition 5.2.} {\it RVE (a) is structurally entirely typical of the whole mixture in $R^d$ on statistical average sense, and
(b) contains a sufficient number of inclusions for the apparent effective nonlocal operator (and statistically averaged field distributions in the phases) to be effectively independent of the applied inhomogeneous field.}

The RVE concept in Definition 5.2 is foundational in Amic, as it allows the upscaling of heterogeneous material behavior without needing full-scale numerical simulations for every material point.
We emphasize the fundamental difference between Definition 5.1 and Definition 5.2 reflecting the essence of these definitions.
So, the notion of ``sample large enough" (implying the loading of the sample boundary, see Definition 5.1) is absent in Definition 5.2. Instead, a heterogeneous medium (of either statistically homogeneous or periodic structure) is considered in a full space $R^d$.

At first, we consider statistically homogeneous media. If the field $ \langle \bfep \rangle ({\bf y})$ is varying sufficiently
slowly in the neighborhood of an arbitrary point ${\bf y}$, then it permits both the use of
the Taylor expansion for the statistical average of this field
and the application of a Fourier transform method. The ``quasi-crystalline" approximation by Lax \cite{Lax`1952}
(\ref{4.2}) is often used for truncation of the hierarchy of integral
equations involved and allows one to derive explicit relations for the nonlocal
overall differential operator by the different methods:
via the effective field method \cite{Kanaun`L`1994},
or via the method of conditional moments \cite{Khoroshun`1996}.
An advantage of the rigorous approach \cite{Drugan`2000}, \cite{Drugan`2003}, \cite{Drugan`W`1996}
is that it is based on variational principles,
providing the bounds of approximations; Drugan and Willis \cite{Drugan`W`1996} obtained an elegant explicit
representation of the nonlocal effective differential operators of the second order involving a term proportional to the second gradient of the ensemble average of strain
\BB
\label{5.2}
\lle\bfsi\rle(\bfx)=\bfL^*\lle\bfep\rle(\bfx)+\bfcL^*\nabla\otimes\nabla\lle\bfep\rle(\bfx).
\EE
It was shown that the minimum RVE size is at most twice the reinforcement diameter for any reinforcement concentration level, for several sets of matrix and reinforcement moduli characterizing large classes of important structural materials. This RVE size $B^{\rm RVE{\cal L}*}=2a$ is defined by a size of excluded volume $v_i^0$ where $\nabla\otimes\nabla\lle\bfep\rle(\bfx)$ is not negligible
whereas $B^{\rm RVE{\bf L}*}=a$ for EFM, see Subsection 5.1. Combining the MEFM with the standard scheme of
the Fourier transform method \cite{Drugan`W`1996} allowed
\cite{Buryachenko`1998} and \cite{Buryachenko`R`1998a}
to obtain the explicit representation of a nonlocal overall
operator in the form of the second-order differential operator (\ref{5.2}).
However, $B^{\rm RVE{\bf L}*}=6a$ (see Subsection 5.1) for the MEFM, and the RVE size for the corresponding nonlocal operator is $B^{\rm RVE{\cal L}*}=18a$ (see \cite{Buryachenko`1998} and \cite{Buryachenko`1999b}). The mentioned approach was proposed in the framework of the first background of micromechanics in LM (see \cite{Buryachenko`2007}), which was generalized to the second background of LM in \cite{Buryachenko`2022}.

Moreover, this author's approach can be easily generalized to PM. Indeed, simplifications of Eqs. (\ref{4.4}) can be performed in the framework of Hypothesis {\bf H1a} (\ref{3.1}) for the linear operators $\bfcJ^{\theta\zeta}_{i,j}$ and $\bfcJ^{\theta\zeta\infty}_{i,j}$ (\ref{4.4}) which can be decomposed and reduced to the tensors at the applying to the constant effective fields (\ref{3.1}) ($\bfz\in v_i$)
\BBEQ
\label{5.3}
\!\!\!\!\!\!\!\!\!\!\!\langle \overline{\bfthe}_i\rangle(\bfz)&=&\langle \bfthe\rangle ({\bf z})
+\int \bigl\{[\bfJ^{I\theta\zeta}_{i,j}(\bfz )\lle\overline{\bfze}_{i}\rle(\bfx_i)+\bfJ^{J\theta\zeta}_{i,j}(\bfz)\lle\overline{\bfze}_{j}\rle
(\bfx_j)] \nonumber\\
\!\!\!\!\!\!\!\!\!\!\!\!\!\!\!\!\!\!&\times& \varphi (v_j,{\bf x}_j\vert; v_i,{\bf x}_i)-\bfJ_{i,j}^{\theta\zeta\infty}(\bfz) \lle\overline{\bfze}_{j}\rle(\bfx_j))
\bigl\}d{\bf x}_j.
\EEEQ
Reformulating of Eq. (\ref{5.3}) by the use of notation $\bfF^{J\theta\zeta}_{i,j}(\bfz)=\bfJ^{J\theta\zeta}_{i,j}(\bfz)
\varphi (v_j,{\bf x}_j\vert;$ $ v_i,{\bf x}_i)-\bfJ_{i,j}^{\theta\zeta\infty}(\bfz)n^{(j)}$ leads to the equation
\BBEQ
\label{5.4}
\!\!\!\!\!\!\!\!\!\!\!\langle \overline{\bfthe}_i\rangle(\bfz)&=&\langle \bfthe\rangle ({\bf z})
+\int [\bfJ^{I\theta\zeta}_{i,j}(\bfz )\varphi (v_j,{\bf x}_j\vert; v_i,{\bf x}_i)+\bfF^{J\theta\zeta}_{i,j}(\bfz)]~d\bfx_j\lle\overline{\bfze}_{i}\rle
(\bfx_i)\nonumber\\
&+&
\int\bfF^{J\theta\zeta}_{i,j}(\bfz) [\lle\overline{\bfze}_{j}\rle
(\bfx_j)-\lle\overline{\bfze}_{i}\rle
(\bfx_i)]d\bfx_j,
\EEEQ
which can be solved by different methods. The most general method is an iteration scheme (see LM counterpart in
\cite{Buryachenko`2022}) allowing a general integral effective nonlocal operator without any {\it a priori} prescribed form. A fundamental deficiency of this method (at least in light of the goal of the current manuscript) is that at the different external loading
$\langle \bfthe\rangle ({\bf z})$, we need completly repeet a solution of micromechanical problem (\ref{5.4}). If the field
$\lle\overline{\bfze}_{j}\rle
(\bfx_j)$ is smooth enough, it can be approximated
by the first terms of its Taylor expansion (see for details \cite{Eskin`1981})
\BB
\langle \overline{\bfze}_j\rangle (\bfx_j)\approx\sum_{|\alpha|=0}^m
{1\over \alpha !}[\otimes(\bfx_j-\bfx_i)]^{\alpha}
\nabla^{\alpha} \langle \overline{\bfze}_j\rangle (\bfx_i),
\label{5.5}
\EE
where the notation for the multi-indices of non-negative integers
$\alpha=(\alpha_1, \ldots,\alpha_d)\in Z^d_+$ is used. Substitution of Eq. (\ref{5.5}) into (\ref{5.4}) yields Eq. (\ref{5.4}) to differential equation, which is reduced to the algebraic equation by Fourier transform. Solution of this algebraic equation in a ``long-wave" approximation with subsequent inverse Fourier transform leads to the differential effective operator (like(\ref{5.2})). RVE for this operator is defined by the ball $b(\bfx_i,B^{\rm RVE})$ of radius $B^{\rm RVE}$
centered at $\bfx_i$ at which the first integral in Eq. (\ref{5.4}) (defining the local effective moduli) is stabilized. RVE of the second order differential part of this operator is define by convergence of the integral with the integrand $\bfF^{J\theta\zeta}_{i,j}(\bfz)\otimes (\bfx_j-\bfx_i)^{\bf \alpha}$ ($|{\bf \alpha}|=2$); this integral vanishes slower at infinity than the integrand $\bfF^{J\theta\zeta}_{i,j}(\bfz)$ that define increasing of $B^{\rm RVE{\cal L}*}$ with respect to $B^{\rm RVE{\bf L}*}$.

In the framework of the EFH (\ref{3.1}) for locally elastic CM
(i.e. LM is considered) with simple cubic packing of rigid spherical inclusions, it was demonstrated (see p. 358 in \cite{Buryachenko`2007}) that the size of RVE required for estimation of the effective second order differential operator (\ref{5.2}) is triple as large as that RVE for the effective moduli estimations (\ref{5.1}). This method can be easily generalized to PM with any prescribed accuracy of the operator (\ref{5.2})
and moduli (\ref{5.1}). For this purpose, we need to substitute the probability densities (\ref{2.37}) to the exact Eq. (\ref{4.6}).
After that, we can use the same scheme exploited for solutions of Eqs. (\ref{5.4}) and (\ref{5.5}).

We now turn to an analysis of periodic structure CMs in LM for CMic methods.
For an infinitely extended periodic elastic medium with the periodicity cell of small size $\epsilon$, a higher order (so-called strain gradient) homogenized equations are rigorously derived in the presence of a fixed periodic body force $\bfb$, via a combination of variational and asymptotic expansions techniques (see Block 1 in (\ref{3.3})). The approach (see \cite{Smyshlyaev`C`2000}) uses
the asymptotic techniques allowing us to construct in a rigorous way ``higher-order''
homogenized equations which are elliptic, and to show that their solutions are ``closest'' to the actual solution in a certain variational
sense. Truncation of asymptotic representation of solution on its ``zero" is the
``zeroth-order" homogenized solution of the homogenized equation (see Subsection 4.1). Despite the generality advantage, the numerical examples and RVE definition were not considered in \cite{Smyshlyaev`C`2000}. Quantitative assessment of this approach
is performed in \cite{Ameen`et`2018} on a two-dimensional elastic two-phase composite consisting of stiff circular particles in a
softer matrix material and subjected to periodic anti-plane shear.

Alternate approach is considered in a second-order computational homogenization \cite{Kouznetsova `et`2004a},
\cite{Kouznetsova `et`2004b} based on a proper incorporation
of the macroscopic gradient of the deformation tensor and the associated higher-order stress measure into the multiscale framework.
The macroscopic homogenized continuum obtained through this scheme is the full second gradient continuum.
The analytical second-order homogenization of linearly elastic CMsl leads to the second gradient elastic Mindlin’s
continuum (see a short description in, e.g., \cite{Buryachenko`2022a}) on the macroscale, where the resulting macroscopic length scale parameter is proportional to the UC size.
It demonstrated the significance of the contribution of the physical and geometrical
nonlinearities in the relation between the RVE size and the calculated macroscopic response.

Thus, for the construction of the effective strain gradient operator of random structure infinite CMs with local properties of phases, AMick
methods have a fundamental advantage. These AMic methods are easily generalized for CMs with nonlocal (strain type and displacement type, peridynamic) properties of phases when the effective strain gradient operator is estimated in the framework of Eqs.
(\ref{5.2})-(\ref{5.4}). AMic methods can be also applied to periodic structure CMs by substitution of the probability densities (\ref{2.37}) into the exact Eq. (\ref{4.6}) with subsequent implementation of the scheme (\ref{5.2})-(\ref{5.4}). In so doing, the CMic methods are well developed for periodic structure CMs with local properties of phases. The effective second-order gradient operator was constructed by both asymptotic homogenization \cite{Ameen`et`2018}, \cite{Smyshlyaev`C`2000} and computational homogenization
\cite{Kouznetsova `et`2004a}, \cite{Kouznetsova `et`2004b} approaches. Any generalization of these approaches to CMs with peridynamic properties of phases is unknown to the author.

However, the methods considered in Subsection 5.2 have two joint shortcomings. The first deficiently is that all methods are adopted
to an estimation of the prescribed effective operator of precisely second order gradient model; applicability for construction of general (e.g. integral) effective operator is questionable. The second drawback is that each method is obtained by special transformations of the initial corresponding method used in Subsection 5.1 which leads to the creation of many specific methods for CMs of either the random or periodic structure. Thus, RVE depends on the problem being considered (see, e.g. p. 358 in \cite{Buryachenko`2007}; analyses of other nonlocal effects in LM can be found in Sections 11 and 12 in \cite{Buryachenko`2007}) and the method used for the solution of this problem.
In the next subsection, we will consider only one universal method applicable for the estimation of any (nonprescribed) form
of effective (called surrogate) nonlocal operator for CMs of either the random or periodic structure; the computational complexity of this method is the same as the computational complexity of the initial version of the corresponding approach adopted in Definition 5.1.
Therefore, from the point of view of generality and universality (see Subsection 5.3 and Section 6), the methods of the current Subsection 5.2 are
dead end.

\subsection{RVE for CMs subjected to body force with compact support}

Reformulation and generalization of the classical definition \cite{Hill`1963}
enables one to formulate a flexible definition sufficient for our current interests in PM with self-equilibrated body force $\bfb(\bfx)$ (\ref{2.4}):

\noindent {\bf Definition 5.3.} {\it RVE is structurally entirely typical of the whole CM area which is sufficient for all apparent effective parameters $\bfcD^{\rm r}$ (\ref{4.35}) or $\bfcD^{\rm p}$
(\ref{4.41}) to be effectively stabilized outside RVE (as, e.g. in Eq. (\ref{4.24})) in the infinite (random or periodic structure) CMs
 with any constitutive laws of phases.}

\noindent This definition formalized by Eq. (\ref{4.24}) conceptually differs from Definitions 5.1 and 5.2.
So, the notion of “sample large enough” (see Definition 5.1) is absent in
Definition 5.3. Instead, a heterogeneous medium (of either statistically homogeneous
or periodic structure) is considered in a full space $R^d$. Any reference to a sought ``nonlocal operator" (see Definition 5.2) is also absent. Instead, we consider the areas $\overline {\rm RVE}=R^d\setminus {\rm RVE}$ where the datasets either
$\bfcD^{\rm r}$ (found by AMic) or $\bfcD^{\rm p}$ (found by CMic) are stabilized.
Stabilization of $\bfcD^{\rm r}$
(or $\bfcD^{\rm p}$) and corerct choice of RVE mean that e.g.
all effective parameters between the surfaces $|\bfx|=B^{\rm RVE}$ and $|\bfx|=B^{\rm RVE}+B^{b}/2$ are the same
(with prescribed tolerance). In such a case, the area $|\bfx|>B^{\rm RVE}+B^{b}/2$ can be removed and the infinite medium can be modeled by a finite size sample; i.e. a correct choice of $RVE$ eliminate edge effect, wich is manifested in LM in the boundary layer with the thickness $5|\Omega|$ if $\lle\bfep\rle^{\Omega} (\bfx)$ is not vanished in the vicinity of the edge (this is not our case), see e.g. p. 129 in
in \cite{Buryachenko`2007}.
If ${\rm RVE}$ is incorrectly chosen (i.e. $B^{\rm RVE}$ is not large enough) then a subsequent implementation of $\bfcD^{\rm r}$ (or $\bfcD^{\rm p}$) (see Section 6) will lead to a numerical error produced by both the sample size and edge effects.

It should be mentioned a very promising generalized Maxwell approach (the second born of the Maxwell \cite{Maxwell`1873}
approach, see comprehensive review \cite{Sevostianov`et`2019})
reducing the modeling of infinite statistically inhomogeneous CM to the inclusion cloud inside the infinite matrix (i.e. the domain $\overline {\rm RVE}$ is replaced by the matrix). However, Buryachenko \cite{Buryachenko`2022b} proved that
23 most fundamental papers in this direction are fundamentally wrong for the case of homogeneous remote BC (\ref{2.33})
being considered. Nevertheless, we can replace BC (\ref{2.33}) by freeloading at infinity with additional body force
(\ref{2.4}). Then the mentioned inclusion cloud (which size is to be learned) in the infinite matrix can be used for the estimation of RVE,
following definition 5.3. After the Monte-Carlo simulation of ``random" realizations of inclusion sets in the cloud, we can estimate
$\bfcD^{\rm p}$ in the same form as (\ref{4.41}); in so doing, the uniformly distributed $\bfchi$ in the scheme (\ref{4.39}) and (\ref{4.40}) (see Fig. 4 ) is replaced by ``random" realizations of inclusion sets.
Now we can forget about both the microstructure of the inclusion cloud and the method (e.g. multipole expansion method, see for details \cite{Sevostianov`et`2019}) how this $\bfcD^{\rm r}$ was estimated.
The next step is the evaluation of a surrogate nonlocal operator by any available ML and NN technique (see Section 6), which can be used for the modeling of infinite statistically homogeneous media. Thus, the third born of the Maxwell \cite{Maxwell`1873} approach appears to be quite successful.

In so doing, both datasets $\bfcD^r$ and $\bfcD^p$ are defined by specific micromechanical methods (e.g. CAM) used for their estimations.
However, in the framework of the scale separation hypothesis (\ref{2.28}$_2$) in LM, Definition 5.3 is reduced to the classical one \cite{Hill`1963} for the apparent effective moduli $^L\!\bfL^*$.

The significance of RVE concept is drastically increased in PM where there are three sorts of nonlocal effects generated by both
inhomogeneous applied fields (e.g. inhomogeneous body force with the scale $B^b$ (\ref{2.4})), material nonlocality $l_{\delta}$, and interactions between inclusions, $a^{int}$ (\ref{2.4})); the effect of interactions of these phenomena increases owing to synergism effect whereas in the LN \cite{Hill`1963} (in the framework of the scale separation hypothesis (\ref{2.28}$_2$)), the nonlocal effect of binary interaction of inclusions degenerates to the constant.
For sell-equilibrated body force (\ref{2.4}),
the statistical average displacements $\lle\bfu\rle(\bfx)$ ($\bfx\in R^1$) are estimated in \cite{Buryachenko`2023} for random heterogeneous bar (with different scale ratios $a/a^{b}/l_{\delta}$) in the framework of AMic (without DNS, see Subsections 3.1 and 4.3). It was detected that at $|\bfx|\geq a^{\rm l-r}$ the solution $\lle\bfu\rle(\bfx)\approx$const.,
i.e. the domain $|\bfx|\leq a^{\rm l-r}$ is RVE$^{\lle\bf u\rle(\bf x)}$ for $\lle\bfu\rle(\bfx)$ (depended on the scale ratios $a/a^{b}/l_{\delta}$) and
the papers \cite{Buryachenko`2023} and \cite{Buryachenko`2023a} can be recognized as initiators for the investigation of RVE estimations in PM of random structure CMs (although the term RVE was not used in these papers). Moreover, it is also necessary to estimate the RVEs for
other effective parameters of $\bfcD^{\rm r}$ (\ref{4.35}) and $\bfcD^{\rm p}$ (\ref{4.41}) for different scale ratios $a/B^{b}/l^{(1)}_{\delta}/
l^{(0)}_{\delta}$; there is no warranty that all these RVEs for each scale ratio $a/B^{b}/l^{(1)}_{\delta}/
l^{(0)}_{\delta}$ are the same (as in LM (\ref{2.28}$_2$): $\rm RVE^{\bf L^*}\equiv RVE^{\lle\bf \epsilon\rle_i({\bf x})}$).

It should be mentioned that in peridynamic CMic, the term RVE is prevalent in both groups of methods Block 3 (for the samples with finite numbers of inclusions) and Block 2 (for periodic structure CMs, see (\ref{2.36})), see (\ref{3.3}).
General deficiencies of Block 3 were described in Subsection 5.1.
For infinite periodic media, to the best of the author's knowledge, the term RVE was only used in Block 2 as a synonym (or shortcut) of the cumbersome phrase ``UC of CM subjected to remote homogeneous loading (\ref{2.30}) corresponding to the scale separations
$\Lambda/L=\infty$, $\Lambda/|\Omega_{00}|=\infty$, and $L/|\Omega_{00}|=\infty$", see \cite{Madenci`et`2017}, \cite{Madenci`et`2018}, \cite{Diyaroglu`et`2019a}, \cite{Diyaroglu`et`2019b},
\cite{Galadima`et`2023}, \cite{Galadima`et`2023b},
\cite{Galadima`et`2023c}, \cite{Galadima`et`2024}, \cite{Hu`et`2022},
\cite{Li`et`2022b}, \cite{Qi`et`2024}, \cite{Xia`et`2020}, \cite{Xia`et`2019}, \cite{Xia`et`2021a}, \cite{Xia`et`2021b}, and \cite{WangQ`et`2024}.
This shortcut is rigorous
in LM (in the case of acceptance of the hypothesis (\ref{2.28}$_2$)) and
has been well-known in LM for at least 50 years. However, automatic exploitation of this shortcut in PM of periodic structure CMs
(even at the mentioned scale separations) should be used more carefully. So, the use of PBC (\ref{2.36}) (which is correct in LM) is incorrect in PM (see Subsection 2.4). Moreover, the VPBC (\ref{2.36}) originally proposed in \cite{Buryachenko`2023} should be generalized for arbitrary scale ratios $|\Omega_{00}|/a/l^{(1)}_{\delta}/l^{(0)}_{\delta}$ and geometry of UC (see also Subsection 2.4).

Of course, for estimation of any effective nonlocal operator by any method (e.g. surrogate operator in any ML and NN technique), analyses of field distributions (and evaluation of the RVE's size) for the scale ratio $\Lambda/|\Omega_{00}|\not =\infty$ (or $B^b/|\Omega_{00}|\not =\infty$) is a necessary ingredient. To the best of the author's knowledge, the term RVE was never used in any sense (even in the common sense) in neither LM nor PM for the body force (\ref{2.4}). On the other hand, the new rigorous Definition
5.3 applies to both the periodic and random structure CMs. In the case of statistically homogeneous media, RVE is invariant to parallel translation: if $\bfx\to \bfy$ ($\bfy=\bfx+\bfz$) then RVE($\bfx)\to$RVE$(\bfy)$ for any $\bfz$. For statistically inhomogeneous (functionally graded) media, this invariantness is lost RVE($\bfx)\not\to$RVE$(\bfy)$.
Although this problem was not considered before, it can be easily solved by CAM (for both random and periodic structure CMs) in a straightforward manner.

It should be mentioned some restrictions of the RVE concept in Definition 5.3 in comparison with Definitions 5.1 and 5.2. So, in Definition 5.1, RVE does not depend on the surface loading. In Definition 5.2, the parameter $\bfcL^*$ of the effective operator (\ref{5.2}) does not depend on the everage field $\lle\bfep\rle(\bfx)$. However, in Definition 5.3, RVE depending on the ratios
$a/B^{\rm RVE}/l^{(1)}_{\delta}/l^{(0)}_{\delta}$ will also depend on the gradient $\nabla \bfb(\bfx)$; i.e. for the different
$\bfb(\bfx)$ and $\bfb'(\bfx)$ with the same $a/B^{\rm RVE}/l^{(1)}_{\delta}/l^{(0)}_{\delta}$ , we can obtain the different RVE.
However, even this restrictive RVE concept is beneficial because it can be used as a necessary critical value: if the domain $w$ of the solution
of Eq. ({4.25})-({4.30}) covers RVE (RVE$\subset w$), then the found lelement $\bfcD^{\rm r}_k=\{\lle{\bfu}_k\rle(\bfb_k,\bfx),\!\lle\bfsi_k\rle(\bfb_k,\bfx),\!\lle{\bfu}_{ik}\rle(\bfb_k,\bfz,\bfx),
\!\lle{\bfsi}_{ik}\rle$ $(\bfb_k,\bfz,\bfx), \bfb_k(\bfx)\}$ (or $\bfcD^{\rm p}_k$) can be exploited in subsequent evaluation of surrogate operator in Section 6, otherwise, the domain $w$ should be increased (or the gradient $|\nabla\bfb(\bfx)|$ should be decreased). 
This means that the RVE for the $k$-th implementation acts as a governing parameter (or necessary ingredient) for constructing the dataset for the surrogate operator. (see Section 6): if RVE$\subset w$ then
$\bfcD^{\rm r}_k$ (or $\bfcD^{\rm p}_k$) is included into $\bfcD^{\rm r}$ (or $\bfcD^{\rm p}$), otherwise the mentioned elements
$\bfcD^{\rm r}_k$ (or $\bfcD^{\rm p}_k$) will not be used for subsequent investigation of the surrogate operator.

The original version of CAM in nonlocal surrogate modeling (see Section 6)
was proposed in \cite{Buryachenko`2023h} for linear bond-based properties of phases used for the estimation of effective parameters in
$\bfcD^{\rm r}$ (\ref{4.35}) and $\bfcD^{\rm p}$ (\ref{4.41}). However, the mentioned parameters in (\ref{4.35}) and (\ref{4.41}) can be estimated for any nonlocal elastic properties of phases, see \cite{Buryachenko`2023k}. Therefore, CAM can be looked upon as being proposed for any nonlocal elastic properties of phases
used in Eqs. ({4.25})-({4.30}).
Thus, we obtain the formally similar datasets for both
the random structure CMs $\bfcD^{\rm r}$ (\ref{4.35}) and periodic structure CMs $\bfcD^{\rm p}$ (\ref{4.41}) estimated by conceptually different tools of AMic and
CMic, respectively (see Subsection 3.1).
Moreover, due to compact support of the body forces $\{\bfb_k(\bfx)\}_{k=1}^N$ being considered, the problem on $R^d$ (for both the random structure CMs and periodic structure CMs) is reduced, in fact, to the analysis on the finite size domain RVE that significantly simplifies the problem. Of course,
any difficulties related to either sample size or edge effects are lacking.


\rmfamily


\rmfamily

\section {Estimation of a set of surrogate operators}
\setcounter{equation}{0}
\renewcommand{\theequation}{6.\arabic{equation}}

The perspective direction of the data-driven Machine Learning (ML) technique in the modeling of CMs was initiated by Silling \cite{Silling`2020}, \cite{You`et`2020} (see also \cite{You`et`2024}). Construction of a surrogate nonlocal operator for the infinite medium (based on DNS) was performed at the example of a finite size 1D heterogeneous bar (of periodic and random structure) with two sorts of inhomogeneous loading by either the wave loading at the boundary or the oscillating body force.

In the following model, the Slilling's approach \cite{Silling`2020}, \cite{You`et`2020}, and \cite{You`et`2024} is modified
by replacement of
input DNS dataset by the dataset $\bfcD^{\rm r}$ (\ref{4.35}) (or $\bfcD^{\rm p}$ (\ref{4.41}) ).
The loading by inhomogeneous body force $\bfb(\bfx)$ can also be exploited as a prospective tool for data-driven learning of the surrogate nonlocal constitutive laws of CMs rather than just a particular loading parameter.

So, two datasets are defined
\BBEQ
\label{6.1}
\!\!\!\!\!\!\!\!\!\!\!\!\!\!\!\!{\cal D}^{\rm r}\!&=& \!\{\lle{\bfu}_k\rle(\bfb_k,\bfx),\!\lle\bfsi_k\rle(\bfb_k,\bfx),\!\lle{\bfu}_{ik}\rle(\bfb_k,\bfz,\bfx),
\!\lle{\bfsi}_{ik}\rle(\bfb_k,\bfz,\bfx), \bfb_k(\bfx)\}_{k=1}^N, \\
\label{6.2}
\!\!\!\!\!\!\!\!\!\!\!\!\!\!\!\widetilde{\cal D}^{\rm r}\!&=&\!\{\widetilde\bfu_k(\bfb_k,\bfx),\!\widetilde\bfsi_k(\bfb_k,\bfx),\!\widetilde{\bfu}_{ik}(\bfb_k,\bfz,\bfx),
\!\widetilde{\bfsi}_{ik}(\bfb_k,\bfz,\bfx), \bfb_k(\bfx)
\}_{k=1}^N,
\EEEQ
where each effective parameters $\lle{\bfu}_k\rle(\bfx)$, $\lle{\bfsi}_k\rle(\bfx)$, $\lle\bfu_{ik}\rle(\bfz,\bfx):=\lle\bfu_{k}\rle_i(\bfz,\bfx)$,
and $\lle{\bfsi}_{ik}\rle(\bfb_k,\bfz,\bfx)=\lle\bfsi_{k}\rle_i(\bfz,\bfx)$
(\ref{4.35})
are estimated for each $\bfb_k$ in the offline stage in $k$-th realization; the coordinates $\bfx\in R^d$ are a macrocoordinate whereas
$\bfz\in v_i$ are local coordinats in the representative inclusion $v_i$.
The surrogate dataset $\widetilde{\cal D}^{\rm r}$ is close to the
dataset ${\cal D}^{\rm r}$ of the effective field parameters presenting, in fact, a combination of both the generation and compression stages;
dataset ${\cal D}^{\rm r}$ is called field PM dataset.
Dataset $\cal D^{\rm r}$ defined by statistical averages in the micromechanical problem is approximated by a surrogate model
\BBEQ
\!\!\!\!\!\!\!\!\!\bfcL_{\rm \gamma}[\lle{\bfu}_k\rle](\bfx)&=&\bfGa(\bfx), \nonumber\\
\label{6.3}
\!\!\!\!\!\!\!\!\! \bfcL_{\rm \gamma}[\lle{\bfu}_k\rle](\bfx)&=&
\int \bfK_{\gamma}(|\bfx-\bfy|) (\lle{\bfu}_k\rle(\bfy)-\lle{\bfu}_k\rle(\bfx))~d\bfy,
\EEEQ
where $\gamma:=b,{\rm \sigma},{\rm u_i},{\rm \sigma}_i$ and ${ \bfGa}(\bfx):=-\bfb(\bfx), \lle\bfsi\rle(\bfx), \lle\bfu\rle_i(\bfz,\bfx),\lle\bfsi\rle_i(\bfz,\bfx)$, respectively, correspond to four surrogate operators $\bfcL_{\gamma}$.
Obtaining
an optimal {\it surrogate} model for the kernel functions $\bfK^*_{\gamma}$ corresponding to the surrogate dataset $\widetilde {\cal D}$
is achieved by
solving four optimization problems
\BBEQ
\label{6.4}
\!\!\!\bfK_{\gamma}^*={\rm arg}\!\min_{\!\!\!\!\!\!\!\!\!\! {{\bf K}_{\gamma}}}\!\sum_{k=1}^N\!|| \bfcL_{\rm {\gamma}}[\lle{\bfu_k}(\bfb_k)\rle](\bfx)- \bfGa_k(\bfx)||^2_{l_2} +{\cal R}(\bfK_{\gamma}),
\EEEQ
by the use, e.g., Adam optimizer \cite{Kingma`B`2014} after each step of gradient descent,
see for details \cite{You`et`2020} (see also \cite{Fan`et`2023}, \cite{You`et`2021}, \cite{You`et`2022}, and \cite{You`et`2024}) where the kernel $\bfK_{\gamma}$ is presented as
a linear combination of Bernstein-based polynomials.
Here, the $l_2$ norm is taken over $\bfx\in R^d$, and ${\cal R}(\bfK_{\gamma})$ is a regularization term on the coefficients
that improves the conditioning of the optimization problem (e.g., Tikhonov regularization).

It should be mentioned that Eqs. (\ref{4.16}$_1$) and (\ref{4.20}) obtained for general nonlinear cases of either the state-based
(\ref{2.8}) or bond-based (e.g. (\ref{2.15})) PM are reduced to their linear counterparts of bond-based PM
\cite{Buryachenko`2023}, \cite{Buryachenko`2023a}. However, construction of the surrogate model (\ref{6.3}) 
was considered in \cite{Buryachenko`2023} only for the linear bond-based model. Moreover, Eqs. (\ref{6.1}) and (\ref{6.2}) for random structure CMs can be recast for the dataset $\bfcD^{\rm p}$ (\ref{4.41}) for periodic structure CMs.
The rest Eqs. (\ref{6.3}) and (\ref{6.4}) are not altered.
So, the datasets $\bfcD^{\rm r}$ (\ref{6.1}) (or $\bfcD^{\rm p}$ (\ref{4.41})) are formed offline from the statistical averages
$\lle{\bfu}\rle(\bfx),\lle\bfsi\rle(\bfx), \lle{\bfu}\rle_i(\bfz,\bfx),
\!\lle{\bfsi}\rle_i(\bfz,\bfx)$ (\ref{6.1}) (or $\lle{\bfu}\rle(\bfx),\lle\bfsi\rle(\bfx), \lle{\bfu}\rle^{l(1)}(\bfx),
\!\lle{\bfsi}\rle^{(1)}(\bfx)$ (4.41))
presenting, in fact, a combination of both the generation and compression stages.
Loading
by dynamic body force $\bfb(\bfx, t)$ is necessary for a surrogate model with nonlocal
dynamics properties (see, e.g., \cite{Askes`A`2011} and \cite{Buryachenko`2024b}).

In so doing
the approach \cite{You`et`2020} and \cite{You`et`2024} is
based on the datasets
\BBEQ
\label{6.5}
{\cal D}^{\rm DNS}=\{\bfu_k(\bfb_k,\bfx),\bfb_k(\bfx)\}^N_{k=1}
\EEEQ
generated without compression that is the dissimilarity from the approach (\ref{6.1})-(\ref{6.4}). ${\cal D}^{\rm DNS}$ contains all information about $\bfu_k(\bfb_k,\bfx)$ at each micropoint $\bfx\in w$ for each $\bfb_k(\bfx)$; i.e. even for the same $\bfb_k(\bfx)$ ($k=1,\ldots, N)$ in (\ref{6.1}) and (\ref{6.5}), the dataset ${\cal D}^{\rm DNS}$ (\ref{6.5}) is much bigger than the dataset
${\cal D}^{\rm r}$ (\ref{6.1}) (or ${\cal D}^{\rm p}$ (\ref{4.41})) for effective parameters.
However, the term compressed applied to the field PM dataset ${\cal D}^{\rm r}$ (\ref{6.1}) does not mean that the full-field dataset
${\cal D}^{\rm DNS}$ for a big sample should be preliminary estimated. In general, the destiny of micromechanics (in both the LM and PM) is providing the required estimations by cheaper, faster, more robust, and more flexible methods (although it takes additional intellectual complexity to the implementations) than DNS.
On the other side, the micromechanics scheme (\ref{4.25})-(\ref{4.32}) is not a single way for estimation of the field PM dataset ${\cal D}^{\rm r}$ (\ref{6.1}) (or ${\cal D}^{\rm p}$ (\ref{4.41})).

In the original method proposed in \cite{Silling`2020}, \cite{You`et`2020},
the dataset
${\cal D}^{\rm DNS}$ makes it possible to estimate the kernel $\bfK^*_b$ by the solution of optimization problem (\ref{6.4}).
However, it does not provide reason enough to assume that the kernel $\bfK^*_b$ can be used for the estimation of
statistical average stresses $\lle\bfsi\rle(\bfx)$ (as a micromodulus was exploited for the estimation of stresses $\bfsi(\bfx)$).
Then, any information about the statistical averages of displacements inside inclusions
$\lle\bfu\rle_i(\bfx)$ ($\bfx\in v_i$) is lost for the online stage in the case of dataset ${\cal D}^{\rm DNS}$, although
it is precisely the operators $\bfK^*_{\rm u_i}$ and $\bfK^*_{\rm \sigma_i}$
(rather than $\bfK^*_{b}$) is potentially used for any nonlinear problem (e.g., fracture and plasticity).
We described the exploiting of the forcing term with compact support for the construction of surrogate operators. Estimation of the effective moduli by the use of limiting
passing $ B^b\to \infty$ is not necessary; this problem can be solved more effectively in a straightforward manner (see Subsection 4.2).

Some sort of compression stage is proposed in a coarse-grained,
homogenized continuum model by \cite{You`et`2022}, \cite{You`et`2023} and \cite{Silling`et`2023}
presenting a learning framework to extract, from MD data, an optimal Linear Peridynamic Solid (LPS) model as a surrogate for MD displacements. The coarse-graining method allows the dependence of bond force on bond length to be determined, including
the horizon and allows substantial reductions in computational cost compared with MD.
One (see \cite{You`et`2022} and \cite{You`et`2023}) considers the periodic boundary conditions with the self-equilibrated and periodic loads.
{\color{black} A fundamental advantage of the approach
\cite{You`et`2022}, \cite{You`et`2023} and \cite{Silling`et`2023} is a direct coupling of MD simulation and PD without any additional assumptions. In comparison, the authors \cite{Izadi`et`2024} assume that withstanding force for a single
fiber simulated in PD matches the value obtained from MD simulation, i.e. a scale separation hypothesis is implicitly used for the mentioned coupling.

In the framework of computational micromechanics for {\color{black} 2D particular realizations of random structures,
one \cite{Silling`et`2024} proposed a coarse-graining model of the upscaled mechanical properties of CMs with particular Monte-Carlo simulated structures in a finite-size square box $w$ loaded by self-equilibrated body force with compact support $\bfb(\bfx)$ (\ref{2.4}) (corresponding to the domain of the long-range action \cite{Buryachenko`2023}). In so doing, the ratios of scales ${\rm dist}(\partial {\rm RVE}, \partial w)\approx 10a=100\lambda$ ($\lambda$ is a lattice spacing) satisfy the conditions of RVE's Definition 5.3 that is attractively demonstrated by vanishing of effective strains $\lle\bfep\rle(\bfx)$ in the vicinity of $\partial w$ (see colored Fig. 12 in \cite{Silling`et`2024} which is compatible with Fig. 3); the term RVE was not used in \cite{Silling`et`2024}.
The upscaled peridynamic model proposed can have a much coarser discretization than the original small-scale model (exploiting DNS, see Subsection 3.1), allowing larger-scale simulations to be performed efficiently [estimations of $\lle\bfu\rle_i(\bfz,\bfx),\
\lle\bfsi\rle_i(\bfz,\bfx)$, $\bfz\in v_i, \ \bfx\in R^di$ are beyond the scope of the study].

Deficiently of the above work of construction of the surrogate operators (\ref{6.3}) and (\ref{6.4}) is that these operators are predefined
and can only be applied to linear response prediction. Recently, the nonlocal neural operator was proposed as a way to learn a surrogate mapping between function spaces, see \cite{Lanthaler`et`2024} and \cite{Li`et`2003}.
The neuron layers of an ordinary
neural network (NN) actually gives rise to a nonlinear local operator.
So, let us consider an $L$-layer fully connected neural network (FCNN) $\Psi(\bfx)$: $\bfR^{\rm d_{\bf x}}\to \bfR^{\rm d_{\rm \bf u}}$, which maps from input vector $\bfx$ to output vector $\bfu$ through a few layers. Each layer consisting of a few neurons computes the output to the next layer $l$ based on the input from the previous $(l-1)$th layer. The neurons in the same layer are not connected whereas every pair of neurons in neighboring layers are assigned by network parameters such as a weight matrix ${\bf w}^l$ and a bias vector $\bfb^l$ in the $l$ layer $(1\leq l\leq L-1)$:
\BBEQ
\label{6.6}
\bfz^l(\bfx)=\bfcA(\bfw^l \bfz^{l-1}(\bfx)+\bfb^l), \ \ \ \bfu(\bfx)=\bfw^L \bfz^{L-1}(\bfx)+\bfb^L,
\EEEQ
where $\bfcA$ denotes a nonlinear activation function (e.g. tanh). The tractable parameters $\bfthe:= \{\bfw^l,\bfb^i\}_{1\leq l\leq L}$ used for calculation for each neuron are optimized to minimize the loss function.
This function $\Psi$ is local, in the sense that the value $\Psi(\bfx)$ at a given evaluation point $\bfx$ depends only on the value of the input function $\bfz(\bfx)$ at that
same point $\bfx$. }
Instead, in the nonlocal neural operator, the increment between neuron layers is modeled as a nonlocal
(integral) operator to capture the long-range dependencies in the feature space
\BBEQ
\label{6.7}
\bfz^l(\bfx)= \bfcA(\bfw^l \bfz^{l-1}(\bfx)+\bfb^l+(\bfcK^l(\bfz^{l-1})(\bfx)),
\EEEQ
where the nonlocal operator $\bfcK^l$ is given by integration against a matrix-valued
integral kernel $\bfK^l$: $ (\bfcK^l(\bfz)(\bfx)=\int\bfK^l(\bfx,\bfy)\bfz^{l-1}(\bfy)d\bfy)$.
Several neural operator architectures have been proposed:
deep operator networks (DeepONet), methods that combine ideas from principal component analysis with neural
networks (PCA-Net), graph neural operators, Fourier neural operator (FNO), Laplace neural operator (LNO), and others; the recent
reviews \cite{Gosmani`et`2022}, \cite{Hu`et`2024}, \cite{Kumara`Y`2023}, \cite{Lanthaler`et`2024} extensively compared these neural operators.

\rmfamily

A new integral neural operator architecture called
the Peridynamic Neural Operator (PNO) \cite{Jafarzadeh`et`2024} provides a surrogate operator $\bfcG$ ($\bfcG(\lle\bfu\rle)(\bfx)\approx -\bfb(\bfx)$) for physically consistent predictions of the overall behavior of highly nonlinear, anisotropic, and heterogeneous materials and achieves improved accuracy
and efficiency compared to baseline models that use prior expert-constructed knowledge
of predefined constitutive laws (as, e.g., in Eq. (\ref{6.3})).
The heterogeneous PNO (HeteroPNO) approach is proposed in \cite{Jafarzadeh`et`2024b} for data-driven constitutive modeling of heterogeneous fiber orientation field in anisotropic materials. It was considered two cases of loading for finite square sample:
At first, volumetric Dirichlet BC with no body force are considered. In the second case the body force $\bfb(\bfx)$ is produced
by FFT from Gaussian white noise random field (rather than $\bfb(\bfx)$ with compact support (\ref{2.4})). Eliminating the sample size and edge effects without the RVE concept (see Definition 5.3 and Fig. 12 in \cite{Silling`et`2024}) with subsequent implementation of a surrogate operator to infinite medium is not obvious.

Physics-informed neural networks (PINN, see \cite{Cuomo`et`et`2022}, \cite{Haghighata`et`2021}, \cite{Harandi`et`2024},\cite{Hu`et`2024}, \cite{Karniadakis`et`2021}, \cite{Kim`L`2024}, \cite{Raissi`et`2019}, and \cite{Ren`L`2024}) have attracted considerable attention
because this framework embeds physics equations (like (\ref{2.5}) or (\ref{2.6})) into the NN as
constraints, ensuring the training results correspond to physical laws. In this way, the adding
difference between the iterations and the physics equations to the loss function of the established
NN is also involved in the training process. A
neural operator can be combined with the PiNN methods (see \cite{Faroughi `et`2024}, \cite{Gosmani`et`2022},
\cite{Wang`Y`2024}) to train a model that can learn complex nonlinearity, multi-material heterogeneity, and nonlocality in
physical systems with extremely high generalization accuracy. Traditionally, finite-size samples are considered without attempting to generalize the approaches proposed to infinite media.

In particular, the PD differential operators are
applied in PINNs to solve the problems with sharp gradients in \cite{Haghighat`et`2021}, but the constraint equation is based on
the solid mechanics PDEs.
PD governing equation is enforced in the PINN’s residual-based loss function
for analyses
of the displacement fields of homogeneous and especially heterogeneous elastic plates \cite{Ning`et`2023}.
PD theory with PINN is presented in \cite{Eghbalpoor`S`2024} to predict quasi-static damage and crack propagation in brittle materials.
The proposed PD-PINN can learn and capture intricate displacement patterns associated with different geometrical parameters, such as pre-crack position and length. The total loss function $\cL_{\rm tot}=\omega_{\rm g}\cL_{\rm gov}+\omega_{\rm u}\cL^u_{\rm BC}+\omega_{\rm f}\cL_{\rm BC}^f+\omega_{\rm d}\cL_{\rm data}$
(see \cite{Gosmani`et`2022}, \cite{Karniadakis`et`2021}, \cite{Raissi`et`2019}) contains the loss corresponding to the forces of internal material points and free
boundaries (governing equation loss $\cL_{\rm gov}$, e.g. $\cL_{\rm gov}=||\bfcG(\bfu)(\bfx) +\bfb(\bfx)||$ \cite{Jafarzadeh`et`2024}), the loss due to the applied boundary conditions like Dirichlet BCs
$\cL^u_{\rm BC}$ (\ref{2.30}), a local balance between the internal and external forces $\cL_{\rm BC}^f$ (\ref{2.31}) at the volumetric boundary, and a data loss that aims to match the neural network output with the given data $\cL_{\rm data}$, each of them adequatly weighted. The
network parameters $\bfthe^*$ are determined through minimizing
the total loss function $\cL_{\rm tot}$ below certain tolerance until a certain target
accuracy or a prescribed maximum number of iterations is
reached (see, e.g., \cite{Kingma`B`2014}, \cite{Paszke`et`2019})
\BBEQ
\label{6.8}
\bfthe^*={\rm arg} \!\min_{\!\!\!\!\!\!\!\!\!\! {\bf \theta}}~\cL_{\rm tot}(\bfthe).
\EEEQ

Instead, in a nonlocal energy-informed neural
network (EINN, \cite{Yu`Z`2024}, \cite{Yu`Z`2024b}, \cite{Zhou`Y`2024}), the total loss function consists of the total potential
of the deformed body (representing the loss in the internal
strain energy and the external work), and the loss means the
residuals of boundary material points to enforce a boundary condition of a finite-size sample.
One \cite{Difonzo`et`2024} proposed to apply radial basis functions (RBFs) as activation functions in suitably designed PINNs
to solve the inverse problem of computing the peridynamic kernel. It was shown that selecting an RBF is necessary to achieve meaningful solutions, that agree with the physical expectations of the data.
}

Of course, nonlinear micromechanical model of CAM (\ref{4.15})-(\ref{4.22}) can be easily incorporated into PNO (and PINN) by replacement of ${\cal D}^{\rm DNS}$ (one dataset of full-field, spatial measurements of displacement and loading) used in PNO by ${\cal D}^{\rm r}$ (\ref{6.1}) (a few field PM datasets of compressed sets of statistical average fields) exploited in (\ref{4.15})-(\ref{4.22}); we will call it CAMNN. As before, it leads to the elimination of both the edge and size effects. The incorrectness of using a finite size sample for estimation of effective behavior is well known in local micromechanics (see, e.g. \cite{Buryachenko`2007}, p. 593); for a finite sample of soft biological tissue (modeling porcine tricuspid valve anterior leaflet), NN learning model was used for construction of a surrogate operator in \cite{You`et`2022b}, that can accurately predict the overall displacement field at unseen loading scenarios (the question of a sample size influence is opened).
The known difficulties for generalizability to different domain shapes for neural operators
(see for references \cite{Jafarzadeh`et`2024}) are also avoided simply because the domain of our interests is the entire space $R^d$.
The last feature of ${\cal D}^{\rm r}$ (\ref{6.1}) (or ${\cal D}^{\rm p}$ (\ref{4.41})) leads to additional
qualitative benefits following immediately from the absence of the loss meaning the residuals of boundary material points to enforce boundary conditions (as in \cite{Eghbalpoor`S`2024}, \cite{Ning`et`2023}, \cite{Yu`Z`2024}, \cite{Yu`Z`2024b}, and \cite{Zhou`Y`2024}).
However, the mentioned edge effect
(appearing in \cite{Jafarzadeh`et`2024}, \cite{Jafarzadeh`et`2024b}, and \cite{You`et`2022b}) can also be eliminated by a naive
way proposed in the minus-sampling method (see Eq. (5.27) in \cite{Buryachenko`2022a}) which, of course, leads to a large loss of information because only the inner points outside of the boundary layer can be used for estimation of ${\cal D}^{\rm DNS}$.
Exploiting of ${\cal D}$ (\ref{6.1}) instead of ${\cal D}^{\rm DNS}$ will lead to four surrogate operators $\bfcG_{\gamma}(\lle\bfu\rle)(\bfx)$ ($\gamma=b,\sigma,u_i,\sigma_i$ as in Eq. (\ref{6.1})) insted of one
$\bfcG(\lle\bfu\rle)(\bfx)$ \cite{Jafarzadeh`et`2024}.
This approach is similar to the mixed formulation for PINNs \cite{Haghighata`et`2021}, \cite{Harandi`et`2024} in the LM, where the PDE is reformulated as a system of equations
where the primary unknowns are the fluxes or gradients of the solution and the
secondary unknowns are the solution itself.
Moreover, what is even more important is that the use of ${\cal D}^{\rm r}$ (\ref{6.1}) (or ${\cal D}^{\rm p}$ (\ref{4.41})) allows one to get a nonlocal operator counterpart
of a basic concept of the LM such as effective concentration factor (\ref{5.1}$_2$);
precisely these (``downscale") operators (rather than ``upscale" tensor $\bfL^*$ (\ref{5.1}$_1$) or operator $\bfcG$) are potentially used for any nonlinear problem (e.g., fracture and plasticity).

However, the main feature of the apparent technical replacement $\bfcD^{\rm DNS}\to \bfcD^{\rm r}$ (\ref{6.1}) (or ${\cal D}^{\rm p}$ (\ref{4.41})) is that $\bfcD^{\rm r}$ (\ref{6.1}) (or ${\cal D}^{\rm p}$ (\ref{4.41})) is obtained for body force with compact support (\ref{2.4}) and construction of this dataset $\bfcD^{\rm r}$ (\ref{6.1}) (or ${\cal D}^{\rm p}$ (\ref{4.41})) has a fundamentally new ingredient called RVE (see Subsection 5.3). Only because of this, any potential incorrectness related to either the size scale or edge effects are initially eliminated. Ignoring of RVE concept at the preparation of dataset $\bfcD^{\rm r}$ (or $\bfcD^{\rm p}$) with necessity leads to both the size scale and edge effects for subsequent evaluation of four surrogate operators $\bfcG_{\gamma}(\lle\bfu\rle)(\bfx)$ ($\gamma=b,\sigma,u_i,\sigma_i$ as in Eq. (\ref{6.1})). But, after estimation of these surrogate operators
$\bfcG_{\gamma}(\lle\bfu\rle)(\bfx)$, any eliminations of the mentioned size scale and edge effects are already questionable.
For comparison of the significance of RVE concepts in Definitione 5.2 and 5.3 we can remind that RVE in Subsection 5.2 acts as ``the icing on the cake" reflecting the micromechanical nature of the effective operator (\ref{5.2}) which was before estimated without RVE (see Section 5.2 and \cite{Ameen`et`2018}, \cite{Drugan`W`1996},
\cite{Kouznetsova `et`2004a}, \cite{Kouznetsova `et`2004b}, \cite{Smyshlyaev`C`2000}). Thus, RVE in Definition 5.3 is a real working tool for correct implementation of both ML and NN techniques.

The scheme for obtaining the set of surrogate models is presented in Fig. 5.
Block 1 DNS (see $\bfcD^{\rm DNS}$ (\ref{6.5})) is used in \cite{Fan`et`2023}, \cite{Jafarzadeh`et`2024}, \cite{You`et`2020}, \cite{You`et`2021}, and \cite{You`et`2022}.
In the proposed approach CAMNN, we consider either the random structure CMs
(see Eqs. (\ref{4.9})-(\ref{4.11}), and (\ref{4.30})) or periodic structure CMs (see Eqs. (\ref{4.39}) and
(\ref{4.40})). In these cases, we use the tools of AMic and CMic, respectively, but in both cases, we estimate the similar datasets
(either $\bfcD^{\rm r}$ (\ref{6.1}) or $\bfcD^{\rm p}$ (\ref{4.41})) placed in
the block 2 Field PM dataset (Compression) used in Block 3 as an input. Block 3 Optimization contains any well-developed
method of ML and NN technique considered in Section 6. We only need to replace the input Bloc 1 DNS (see $\bfcD^{\rm DNS}$ (\ref{6.5})) by the input Bloc 2 Field PM Data. The dataset $\bfcD^{\rm DNS}$ (\ref{6.5}) is much bigger than the compressed datasets ${\cal D}^{\rm r}$ (\ref{6.1}) (or ${\cal D}^{\rm p}$ (\ref{4.41})) for effective parameters, and, because of this, the mentioned replacement is very beneficiary. Only adjustment of Bloc 2 with Block 3 (analogous to adjustment of Bloc 1 with Block 3) is required. Any corrections of Block 3 are not assumed. Solution of optimization problems (Block 3) leads to
obtaining either one surrogate operator (e.g. $\bfK^*_b$ or $\bfcG$, Block 4) or a set of surrogate models (e.g. Eq. (\ref{6.4}), Block 5).

\vspace{-0.mm} \noindent \hspace{5mm} \parbox{11.2cm}{
\centering \epsfig{figure=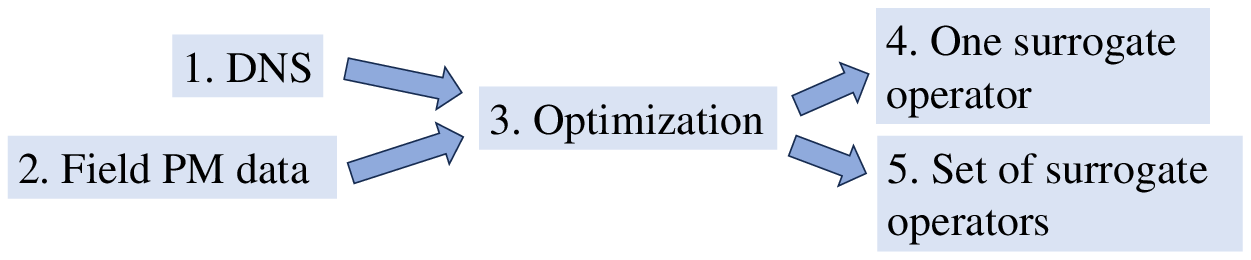, width=11.2cm}\\ \vspace{-119.mm}
\vspace{119.mm}
\vspace{-120.mm} \tenrm \baselineskip=8pt
{{\sc Fig. 5:} The scheme of obtaining of surrogate model set}}
\vspace{0.mm}

NN techniques in PM can be subdivided, loosely speaking, in two directions. The first direction D1 is totally dominated in PM and associated with PM's counterparts of CMic presented by Blocks 2 and 3 in Eq. (\ref{3.3}) (the most popular methods are NO \cite{Gosmani`et`2022}, \cite{Hu`et`2024}, \cite{Lanthaler`et`2024}; PNO \cite{Jafarzadeh`et`2024},\cite{Jafarzadeh`et`2024b}; PINN \cite{Eghbalpoor`S`2024}, \cite{Faroughi `et`2024}, \cite{Gosmani`et`2022}, \cite{Haghighat`et`2021}, \cite{Karniadakis`et`2021}, \cite{Kingma`B`2014}, \cite{Ning`et`2023}, \cite{Paszke`et`2019},
\cite{Raissi`et`2019}; and EINN \cite{Yu`Z`2024}, \cite{Yu`Z`2024b}, \cite{Zhou`Y`2024}). The second direction D2 (CAMNN) of NN technique in PM is based on the PM's counterparts of AMic presented in Gr2) and Gr4) in Eq. (\ref{3.2}) (see \cite{Buryachenko`2022a}, \cite{Buryachenko`2023}). The difference between D1 and D2 is defined by the difference in their backgrounds (see Subsection 3.1).
There is a known series of advantages and disadvantages between AMic and CMic in one another particular problem, and
it is crucial for the analyst to be aware of their range of applications. In particular, for the periodic structure of CMs with periodic (or VPBC
(\ref{2.35})), exploitation of CMic is very profitable for the estimation of effective moduli.
However, estimation of the nonlocal surrogate operator, in any way, requires evaluation of the system response subjected to the inhomogeneous field (either applied stress or body force). In its turn,
it requires the investigation of rations either $a/B^b/l_{\delta}$ (in D2, see (\ref{2.28})) or $L ({\rm or}\,\Omega_{00})/B^b /l_{\delta}$ (in D1). To the best of the author’s knowledge, the rations $L ({\rm or}\,\Omega_{00})/B^b /l_{\delta}$ were not investigated in D1 for any known NN technique (i.e. the background concept of micromechanics RVE was ignored and
the papers \cite{Buryachenko`2023} and \cite{Buryachenko`2023a} can not compensate the absence of research in RVE's investigation
by NN technique).
However, analyses of the rations $a/B^b /l_{\delta}$ in D2 by different NN techniques are affordable solvable problems, especially for inhomogeneous body force (because the solution of Eqs. (\ref{4.9})-(\ref{4.11}) is simpler than the solution of Eqs. (\ref{4.6})-(\ref{4.8})). The fundamental advantage of D2 to D1 is the total elimination in D2 of both the sample size effects $L ({\rm or}\,\Omega_{00})/B^b$ and the enge layer
effect of BC. Thus, CAMNN of both versions (for random structure CMs and periodic structure CMs) is based on the offline dataset (formally similar) ${\cal D}^{\rm r}$ (\ref{6.1}) (or ${\cal D}^{\rm p}$ (\ref{4.41})); in doing so, the version of CAMNN for periodic structure CMs uses CMic in Eqs. (\ref{4.39}) and (\ref{4.39})).
Moreover, the use of ${\cal D}^{\rm r}$ (\ref{6.1}) (or ${\cal D}^{\rm p}$ (\ref{4.41})) in D2 (estimated by the unified approach either for the random or periodic structure CMs) instead of ${\cal D}^{\rm DNS}$ (\ref{6.5}) in D1 allows one to get the multifield surrogate operators (see Block 5 in Fig. 5) instead of one surrogate operator $\bfcG$ in D1 \cite{Jafarzadeh`et`2024} (see Block 4 in Fig. 5), that is critical for any nonlinear phenomena (e.g. strength, plasticity, and fracture). Block 3 Optimization is well developed in D1 (see, e.g., NO, PNO, PINN, and EINN mentioned above). A single modification
for implementation of Block 3 to CAMNN is a simplification of the offline dataset from ${\cal D}^{\rm DNS}$ (\ref{6.5}) to
${\cal D}^{\rm r}$ (\ref{6.1}) (or ${\cal D}^{\rm p}$ (\ref{4.41})).

\sffamily
\noindent{\bf Comment 6.1.} Let us summarize the main peculiarities, restrictions, and difficulties in implementing ML and NN techniques to PM exploiting CMic. At first, we consider a generation of dataset $\bfcD^{\rm DNS}$ (\ref{6.5}) by one of the methods of Block 3 (of PM's counterpart of Eq. (\ref{3.3})) for a sample with a finite number of inclusions loaded at the sample boundary. For any method, it leads to both the sample size and edge effects (the exception is the case of body force mentioned after Definition 5.3). There are known attempts in LM (see, e.g. p. 240 in \cite{Buryachenko`2007}) to estimate effective elastic moduli $\bfL^*$ (even for homogeneous loading (\ref{2.30})) with a reduction of the mentioned effects; all of them lead to significant numerical errors. Attempts for estimation of similar errors at the evaluation of effective nonlocal operator for the general case of body force (\ref{2.4}) in LM (and, especially, in PM)
are unknown to the author. For the periodic structure CMs, the main difficulty for the implementation of ML and NN in PM exploiting CMic
is using PBC (\ref{2.36}) or VPBC (e.g. (\ref{2.35})) which are incorrect for the general case of body force (\ref{2.4})
(doubtfully that this incorrectness will be eliminated in the nearest years, see Subsection 2.4).

However, all the mentioned difficulties and incorrectness are initially absent in CAMNN (even though there is no sense in overcoming these difficulties). For both random and periodic structure CMs, the medium is assumed to be infinite and, therefore, any potential sample size and edge effects do not exist in CAMNN. Furthermore, for the periodic structure CMs, CAM implementation is based on the solution of problems (\ref{2.39}) and (\ref{4.40}) on RVE (containing, perhaps, a few UCs) where any prescribed PBC (or VPBC) at the UC interface (or interaction interface) are not required (see the text after Eq. (\ref{4.41})). Thus, the new concept of RVE (see Definition 5.3) for the general case of body force (\ref{2.4}) is a necessary ingredient for CAMNN (and its subsequent incorporation in ML and NN techniques considered in this Section 6) implementation for CMs of both random and periodic structures.


\rmfamily

\rmfamily


\section{Conclusion}


The proposed universal tool (called CAM) is sufficiently flexible and based on physically clear
hypotheses that can be modified and improved if necessary (up to abandonment
of these hypotheses that are forced by challenging achievement rather than
only by pursuing an abstruse theoretic exercise) in the framework of a unique scheme for analyses of a wide class of
micromechanical problems, e.g., statistically homogeneous and inhomogeneous media, inhomogeneous loading (inhomogeneous body force is included), nonlinear and nonlocal constitutive laws of phases, and coupled
physical phenomena.
While the PM was derived as an extension of methods in the LM, CAM highlights the bidirectional enrichment of these methodologies. Some innovations originating in PM (notably, e.g., the nonlinear GIEs (\ref{3.11}) and (\ref{3.12});
body force with compact support, see Subsection 4.3; Definition 5.3; creation of special
datasets $\bfcD^{\rm r}$ (\ref{6.1}) and $\bfcD^{\rm p}$ (\ref{4.41}) for implementation into ML and NN technique, see Sction 6) have inspired new methods in LM.
The author concentrates on his own vision of PM as the area of micromechanics (and its generalization and universal representation).
One of the key points of the approach proposed is that the presented PM is reduced to the classical LM
(see, e.g., \cite{Buryachenko`2007}, \cite{Dvorak`2013}, \cite{Ghosh`2011}, \cite{Kachanov`S`2018}, \cite{Kanaun`L`2008}, \cite{Shermergor`1977}, \cite{Torquato`2002}, \cite{Willis`1981}) at the acceptance of both the scale separation hypothesis $L\gg\Lambda\gg a\gg l_{\delta}$ (see (\ref{2.28}$_2$) and (\ref{2.29}$_2$)) and some simplified assumptions (see Subsections 4.1). Loosely speaking, we consider in more detail the methods of PM reducing (at $l_{\delta}/a\to 0)$ to the corresponding approaches of LM in the framework of the hypotheses
(\ref{2.28}$_2$ and (\ref{2.29}$_2$)) and some simplified assumptions (see Subsection 4.1). In so doing, LM performs the role of the red flag: if some background concepts of LM are violated (or ignored) then the correctness of the corresponding PM model is questionable. On the other side, the correctness of the local counterpart of PM (at $l_{\delta}/a\to 0)$ is not a warranty of the correctness of the corresponding PM model (see, e.g. Subsection 2.4).

Furthermore, Definitions 5.1 and 5.2 of the RVE can be considered as a straightforward generalization of the correcponding notions for CMs with locally elastic properties of phases to their nonlocal counterparts (strongly nonlocal and weakly nonlocal ones).
However, powerful tools such as ML and NN techniques often ignore the background concepts of micromechanics (of both LM and PM) such as size scale and edge effects, and RVE. To eliminate this disadvantage, the proposed CAM methods generate fundamentally new compressed datasets for either random or periodic structure CMs.
The new RVE concept (see Definition 5.3) is revolutionary because: it does not rely on the constitutive laws of the phases; it is independent of the form of the predicted surrogate operator. Instead, it focuses on field concentration factors within the composite material (CM) phases. This independence makes it more flexible and universally applicable.
The mentioned datasets intrinsically incorporated the new RVE concept as a necessary
 ingredient which can be implemented into any known ML and NN technique used for the prediction of nonlocal surrogate operators.
This method could significantly expand the scope and accuracy of ML and NN in fields requiring micromechanical analysis. Exploitation of the new RVE concept eliminates size, boundary layer, and edge effects that ensure more reliable predictions in complex systems.

\medskip
\noindent {\bf Acknowledgements:}

The author acknowledges Dr. Stewart A. Silling for the fruitful personal discussions, encouragements, helpful
comments, and suggestions. 

\medskip
\noindent{\bf Declarations}

\noindent{\bf Ethical Approval}:
(applicable for both human and/ or animal studies. Ethical committees, Internal Review Boards, and guidelines followed must be named. When applicable, additional headings with statements on consent to participate and consent to publish are also required) 
“not applicable.”\\
\noindent{\bf Competing interests}:
(always applicable and includes interests of a financial or personal nature)
No competing interests.\\
\noindent{\bf Authors' contributions}:
(applicable for submissions with multiple authors). 
“not applicable”. There is only one author.\\
\noindent{\bf Funding}:
The author received no financial support for the research, authorship, and/or publication of this article.




\begin{thebibliography}{99}


{\baselineskip=9pt
\parskip=0.1pt
\tenrm
{


\bibitem{Aguiar`F`2014}
Aguiar, A.R., Fosdick, R., 2014.
A constitutive model for a linearly elastic peridynamic body
{\tenit Mathematics and Mechanics of Solids},
{\tenbf 19}: 502--523

\bibitem{Agwai`et`2011}
Agwai, A., Guven, I., Madenci, E. (2011)
Predicting crack propagation with peridynamics: a comparative study
{\tenit Int. J. Fracture},{\tenbf 171}: 65-78

\bibitem{Ahmadi`et`2022}
Ahmadi M, Sadighi M, Hosseini-Toudeshky H. (2022) Microstructure-based deformation and fracture modeling of particulate reinforced composites with ordinary state-based peridynamic
theory. {\tenit Compos Struct}, {\tenbf 279}:114734.

\bibitem{Aksoylu`P`2011}
{\color{black} Aksoylu B, Parks ML (2011) Variational theory and domain decomposition for nonlocal
problems. {\tenit Applied Mathematics and Computation}, {\tenbf 217}: 6498–6515}

\bibitem{Alali`A`2020}
Alali, B., Albin, N. (2020) Fourier spectral methods for nonlocal models. {\tenit J. Peridynamics Nonlocal
Modeling}, {\tenbf 2}, 317-335.

\bibitem{Allali`G`2015}
Alali B, Gunzburger M (2015)
Peridynamics and material interfaces
{\tenit J. Elast.}, { 120}: 225-–248

\bibitem{Allali`L`2012}
Alali B, Lipton R, (2012) Multiscale dynamics of heterogeneous media in the peridynamic
formulation. {\tenit J. Elast.}, 106: 71--103

\bibitem{Ameen`et`2018}
 Ameen MM,  Peerlings RHJ,  Geers MGD. (2018)
quantitative assessment of the scale separation limits of classical and
higher-order asymptotic homogenization
{\tenit European J. Mech. A. Solids}, {\tenbf 71}:  89–100


\bibitem{Anbarlooie`H`2024}
Anbarlooie B, Hosseini-Toudeshky H. (2024) 
Damage mechanisms analyses in DP steels using SEM images, FEM, and nonlocal peridynamics methods
{\tenit Mech. Adv. Materials and Structures}, https://doi.org/10.1080/15376494.2024.2367010


\bibitem{Artur`V`2007}
Arthur, D., Vassilvitskii, S. $k$-means++: the advantages of careful seeding.
{\tenit Proc. of the 18th Annual
ACM-SIAM Symposium on Discrete Algorithms}, 2007; 1027--1035.


\bibitem{Askari`et`2006}
Askari, E., Xu, J., Silling S.A., 2006.
Peridynamic analysis of damage and failure in composites.
{\tenit 44th AIAA Aerospace Sciences Meeting and Exhibition,} {\tenbf AIAA 2006--88}, Reno, NV, 1--12.

\bibitem{Askari`et`2008}
Askari E, Bobaru F, Lehoucq RB, Parks ML, Silling SA, Weckner O (2009) Peridynamics
for multiscale materials modeling. {\tenit Journal of Physics: Conference Series}, 125:012078

\bibitem{Askari`et`2015}
Askari, A., Azdoud, Y., Han, F., Lubineau, G., Silling, S. 2015.
Peridynamics for analysis of failure in advanced composite materials
{\tenit Numerical Modelling of Failure in Advanced Composite Materials},
Woodhead Publishing Series in Composites Science and Engineering,
331--350

\bibitem{Askes`A`2011}
{\color{black} Askes H, Aifantis EC (2011) Gradient elasticity in statics and dynamics: An overview of
formulations, length scale identification procedures, finite element implementations and new
results. {\tenit Int. J. Solids Structures} {\tenbf 48}: 1962–1990 }

\bibitem{Azdoud`et`2013}
Azdoud Y, Han F, Lubineau G. (2013)
A Morphing framework to couple non-local and local anisotropic continua
{\tenit Int. J. Solids Struct.}, {\tenbf 50}, 1332--1341


\bibitem{Babuska`1976}
Babuska I (1976) Homogenization and its application. Mathematical and computational problems.
In: Lions, J.-L., Glowinski, R. (Eds.)
{\tenit Numerical Solution of Partial Differential Equations}. III. Academic Press, New York, pp. 89--116


\bibitem{Bakhvalov`P`1984}
Bakhvalov NS, Panasenko GP (1984) {\tenit Homogenisation: Averaging Processes in Periodic Media}. Nauka, Moscow (in Russian; English translation: Kluwer, 1989)

\bibitem{Bargmann`et`2018} 
Bargmann~S, Klusemann~B, Markmann~J, Schnabel~JE, Schneider~K, Soyarslan~C, Wilmers~J
(2018) Generation of 3D representative volume elements for heterogeneous materials: A review. 
{\it Progress in Materials Science}, {\bf 96}:322--384

\bibitem{Barrett`et`1994}
Barrett R, Berry M, Chan TF, Demmel J, Donato J, Dongarra J, Eijkhout V., Pozo R, Romine C, der
Vorst HV. (1994)
{\tenit Templates for the Solution of Linear Systems: Building Blocks for Iterative Methods}, 2nd Edition, SIAM.

\bibitem{Basoglu`et`2022}
Basoglu MF, Kefal A, Zerin Z, Oterkus E. (2022). Peridynamic modeling of toughening enhancement in unidirectional fiber-reinforced
composites with micro-cracks. {\tenit Composite Structures}, {\tenbf 297}: 115950

\bibitem{Bazant`et`2022}
Ba$\check{\tenrm z}$ant, Z, Nguyen N.H., D$\ddot{\tenrm o}$nmez, A.A. (2022)
Critical Comparison of Phase-Field, Peridynamics, and Crack Band Model M7 in Light of Gap Test and Classical Fracture Tests
{\tenit J. Appl. Mech.}, {\tenbf 89}: 061008 (26 pages)

\bibitem{Beckmann`et`2013}
Beckmann R, Mella R, Wenman MR (2013) Mesh and timestep sensitivity of fracture from
thermal strains using peridynamics implemented in Abaqus. {\tenit Comput. Methods Appl. Mech.
Engrg.}, { 263}: 71--80

\bibitem{Benaimeche`et`2022}
Benaimeche MA, Yvonnet J, Bary B, He Q-C (2022) A k-means clustering machine learning-based multiscale
method for inelastic heterogeneous structures with
internal variables.
{\tenit Int J Numer Methods Eng.} { 123}: 2012–2041

\bibitem{Benner`et`2020}
{\color{black}
Benner P, Grivet-Talocia S., Quarteroni A, Rozza G, Schilders W, Silveira LM. (Eds) (2020)
{\tenit Model Order Reduction.
V.2: Snapshot-Based Methods and Algorithms}.
Walter de Gruyter GmbH, Milano}


\bibitem{Benveniste`1987}
{Benveniste Y} (1987) A new approach to application of
Mori-Tanaka's theory in composite materials. {\tenit Mech. Mater.}
{ 6}: 147—157

\bibitem{Bessa`et`2014}.
Bessa MA, Foster JT, Belytschko T, Liu WK. (2014) A meshfree unification: reproducing kernel peridynamics.
{\tenit Comput. Mech.}, {\tenbf 53}, 1251–1264.

\bibitem{Białecki et`2005}
{\color{black} Białecki RA, Kassab AJ, Fic A. (2005)
Proper orthogonal decomposition and modal analysis for
acceleration of transient FEM thermal analysis
{\tenit Int. J. Numer. Meth. Engng}, {\tenbf 62}:774–797}

\bibitem{Bie`et`2024}
Bie Y, Ren H, Rabczuk T, Bui TQ, Wei Y. (2024)
The fully coupled thermo-mechanical dual-horizon peridynamic correspondence damage model for homogeneous and heterogeneous materials
{\tenit Comput. Meth. Appl. Mech. Engng}, {\tenbf 420}: 116730

{\color{black} \bibitem{Birner`et`2023}
Birner M, Diehl P, Lipton R, Schweitzer MA. (2023)
A fracture multiscale model for peridynamic
enrichment within the partition of unity method. {\tenit Adv. Engineering Software}, {\tenbf 176}
103360.}



\bibitem{Bobaru`et`2016}
Bobaru, F., Foster, J., Geubelle, P., Silling, S.,
(Editors) 2016. {\tenit Handbook of Peridynamic Modeling},
CRC Press, Boca Raton, FL.

\bibitem{Bobaru`H`2011}
Bobaru, F., Ha, Y.D., 2011. Adaptive refinement and multiscale modeling in 2D peridynamics.
{\tenit Int. J. Multiscale Comput. Eng.}, {\tenbf 9}, 635–-659

\bibitem{Bobaru`et`2009}
Bobaru, F., Yang, M., Alves, L.F., Silling, S.A., Askari, A., Xu, J., 2009.
Convergence, adaptive refinement, and scaling in 1d peridynamics. {\tenit Int.
J. Numerical Methods Engng}, {\tenbf 77}, 852-–877

{\color{black} \bibitem{Bode`et`2022}
Bode T, Weißenfels C, Wriggers P. (2022)
Peridynamic Galerkin method: an attractive alternative
to finite elements. {\tenit Computational Mechanics}. {\tenbf 70}: 723–743.}

{\color{black} \bibitem {Bonnet`et`1996}
Bornert M, Stolz C, Zaoui A (1996) Morphologically representative pattern-based
bounding in elasticity. {\tenit J Mech Phys Solids}, {\tenbf 44}: 307–331}

\bibitem{Boyaval`2008}
{\color{black}Boyaval}, S. Reduced-basis approach for homogenization beyond the periodic setting, {\tenit Multiscale
Modeling \& Simulation} 2008; {\tenbf 7}: 466-494.

\bibitem{Breitenfeld`et`2014}
Breitenfeld MS, Geubelle PH, Weckner O, Silling SA. (2014)
Non-ordinary state-based peridynamic analysis of stationary crack problems
{\tenit Comput. Meth. Appl. Mech. Engng}, {\tenbf 272}, 233--250


\bibitem{Brunton`K`2022}
Brunton SL, Kutz JN (2022)
{\tenit Data-driven science and engineering: Machine learning, dynamical systems, and control.}
Cambridge University Press, Cambridge, UK


\bibitem{Buryachenko`1998} 
Buryachenko~VA (1998) Some nonlocal effects in graded random structure matrix
composites. {\it Mech Res Commun,} {\bf 25}:117--122 


\bibitem{Buryachenko`1999b} 
Buryachenko~VA (1999b) Effective thermoelastic 
properties of graded 
doubly periodic particulate composites in varying external stress fields.
{\it Int J Solids Struct,} {\bf 36}:3861--3885 

\bibitem{Buryachenko`2007}
Buryachenko VA (2007) {\tenit Micromechanics of Heterogeneous Materials}. Springer, NY.


\bibitem{Buryachenko`2010a}
Buryachenko VA (2010) On the thermo-elastostatics of heterogeneous materials. I. General integral equation.
{\tenit Acta Mechanica}, 213: 359-374

\bibitem{Buryachenko`2010b}
Buryachenko VA (2010) On the
thermo-elastostatics of heterogeneous materials. II. Analyze and
generalization of some basic hypotheses and propositions. {\tenit
Acta Mech}., { 213}: 375-398.

\bibitem{Buryachenko`2011a} 
Buryachenko~VA (2011a) Inhomogeneity of the first and second statistical moments of stresses inside the heterogeneities of random structure matrix composites. {\it Int. J. Solids and Structures}, {\bf 48}:1665--1687.

\bibitem{Buryachenko`2011b} 
Buryachenko~VA (2011b) On thermoelastostatics of composites with nonlocal properties of constituents. I. General representations for effective material and field parameters. {\it Int. J. Solids and Structures}, {\bf 48}:1818--1828.

\bibitem{Buryachenko`2011c} 
Buryachenko~VA (2011c) On thermoelastostatics of composites with nonlocal properties of constituents. II. Estimation of effective material and field parameters. {\it Int. J. Solids and Structures}, {\bf 48}:1829--1845.

\bibitem{Buryachenko`2013} 
Buryachenko~VA (2013) General integral equations of micromechanics of composite materials with imperfectly
bonded interfaces. {\it Int. J. Solids and Structures}, {\bf 50}:3190--3206.

\bibitem{Buryachenko`2014a}
Buryachenko V A (2014) Some general representations in thermoperidynamics of random structure composites. {\tenit Int. J. Multiscale Comput. Enging}, { 12}: 331–-350.

\bibitem{Buryachenko`2014b}
Buryachenko V (2014) Solution of general integral equations of micromechanics of heterogeneous materials. {\tenit Int. J. Solids and Structures}, { 51}: 3823–-3843

\bibitem{Buryachenko`2014c}
Buryachenko V (2014) Effective elastic modulus of heterogeneous peristatic bar of random
structure. {\tenit J. Solids and Structures}, {\tenbf 51}: 2940-2948


\bibitem{Buryachenko`2015a}
Buryachenko V (2015) General integral equations of micromechanics of heterogeneous materials. {\tenit J. Multiscale Comput. Enging.}, { 13}: 11--53

\bibitem{Buryachenko`2015b} 
Buryachenko VA (2015) Effective thermoelastic properties of heterogeneous thermoperi-
static bar of random structure. {\tenit Int. J. Multiscale Comput. Enging.}, {\tenbf 13}: 55-71

\bibitem{Buryachenko`2017}
Buryachenko VA (2017) Effective properties of thermoperidynamic random structure composites: some background principles.
{\tenit Math. Mech. of Solids}., {\tenbf 22}: 366-–1386

\bibitem{Buryachenko`2018}
Buryachenko VA (2018)
Computational homogenization in linear elasticity of peridynamic periodic
structure composites. {\tenit Math. Mech. of Solids}, { 23}:
2497--2525

\bibitem{Buryachenko`2018b}
Buryachenko V (2018) Effective elastic modulus of heterogeneous peristatic bar of periodic
structure. {\tenit Computers \& Structures}, {\tenbf 202}:129-139

\bibitem{Buryachenko`2018c}
Buryachenko V (2018) Effective elastic modulus of damaged peristatic bar of periodic
structure. {\tenit J. Multiscale Comput. Enging}, {\tenbf 16}:101–118


\bibitem{Buryachenko`2019a}
Buryachenko VA (2019) Interface integral technique for the
thermoelasticity of random structure
matrix composites. {\tenit Math. Mech. of Solids}, { 24}: 2785–2813


\bibitem{Buryachenko`2019b}
Buryachenko VA (2019) Modeling of one inclusion in the infinite peridynamic matrix subjected to homogeneous remote loading
{\tenit J. Peridynamics and Nonlocal Modeling},{\color{black} { 1}: 75–87}


\bibitem{Buryachenko`2020}
Buryachenko V (2020a) Generalized effective field method in peridynamic micromechanics
of random structure composites. {\tenit Int. J. Solid Structure}, { 202}: 765-786

\bibitem{Buryachenko`2020b}
Buryachenko V (2020b) Generalized Mori-Tanaka approach in micromechanics of peristatic random structure composites. {\tenit J. Peridynamics and Nonlocal Modeling}, { 2}: 26–49

\bibitem{Buryachenko`2020c}
Buryachenko V. (2020c) Variational principals and generalized Hill’s bounds in micromechanics of peristatic random structure composites.
{\tenit Math. Mech. of Solids}, {\tenbf 25}: 682–704


\bibitem{Buryachenko`2022a}
Buryachenko VA (2022) {\tenit Local and Nonlocal Micromechanics of Heterogeneous Materials.} Springer, NY.

\bibitem{Buryachenko`2022}
Buryachenko V. A. (2022) Multiscale and multiphysics modelling of advanced heterogeneous materials. 
{\tenit International Association of Advanced Materials Award Lecture}, see 20min Fellow of IAAM video presentation 
https://lnkd.in/gYp3SJmy); Vid. Proc. Adv. Mater., Volume 3, DOI: 10.5185/vpoam.2022.08325)


\bibitem{Buryachenko`2022b}
Buryachenko, V. (2022) Critical analysis of generalized Maxwell homogenization schemes and related prospective problems. {\tenit Mechanics of Materials}, {\tenbf 165}, 104181


\bibitem{Buryachenko`2023}
Buryachenko VA (2023a)
Effective nonlocal behavior of peridynamic random structure
composites subjected to body forces with compact support and related prospective problems.
{\tenit Math. Mech. of Solids}, {\tenbf 28}: 1401-1436


\bibitem{Buryachenko`2023a}
Buryachenko V (2023b) Effective displacements of peridynamic heterogeneous bar loaded by body force with compact support.
{\tenit J. Multiscale Comput. Enging}, {\tenbf 21}: 27--42

\bibitem{Buryachenko`2023b}
Buryachenko V. A. (2023c). Linearized ordinary state-based peridynamic micromechanics of composites.
{\tenit J. Materials and Structures}, {\tenbf 18}, 445--477



\bibitem{Buryachenko`2023e}
Buryachenko V. A. (2023d). Second moment of displacement state and effective energy-based criteria in peridynamic micromechanics of random structure composites {\tenit J. Peridynamics and Nonlocal Modeling} {\tenbf 5}, Published: 23 September 2023

\bibitem{Buryachenko`2023f}
Buryachenko V. A. (2024e) Generalized Mori-Tanaka approach in peridynamic micromechanics of multilayered composites of random structure. {\tenit J. Peridynamics and Nonlocal Modeling} {\tenbf 6}, Published: 12 February 2024

\bibitem{Buryachenko`2023g}
Buryachenko V. A. (2023f)
Transformation field analysis as a background of clustering discretization methods in micromechanics of composites. {\tenit Math. Mech. of Solids}, {\tenbf 28} 2677–2703

\bibitem{Buryachenko`2023h}
Buryachenko V. (2023g) Effective nonlocal behavior of peridynamic random structure composites subjected to body forces with compact support. {\tenit Math. Mech. of Solids}, {\tenbf 28}: 1401--1436

\bibitem{Buryachenko`2023c}
Buryachenko VA (2024)
Estimations of energy-based criteria in nonlinear
phenomena in peridynamic micromechanics of
random structure composites.
{\tenit J. Peridynamics and Nonlocal Modeling}, {\tenbf 6}: 250–269 

\bibitem{Buryachenko`2023d}
Buryachenko V. A. (2024) Transformation field analysis and clustering discretization method in pyridynamic micromechanics of composites.
{\tenit J. Peridynamics and Nonlocal Modeling} {\tenbf 6}: Published: 22 February 2024

\bibitem{Buryachenko`2023i}
Buryachenko V. A. (2024) Fast Fourier transform method for peridynamic bar of periodic structure. {\tenit J. Multiscale Comput. Enging}, {\tenbf 22}: 1-17

\bibitem{Buryachenko`2023j}
Buryachenko V. A. (2024) Fast Fourier transform in peridynamic micromechanics of composites {\tenit Math. Mech. of Solids}, {\tenbf 29}, https://doi.org/10.1177/10812865241236878

\bibitem{Buryachenko`2023k}
Buryachenko V., (2024) Nonlinear general integral equations in micromechanics of random
structure composites. {\tenit Math. Mech. of Solids}, {\tenbf 29}

\bibitem{Buryachenko`2024a}
 Buryachenko V. A. (2024)  Peridynamic micromechanics of composites: a review. J. Peridynamics and Nonlocal Modeling 6, 531–601  https://arxiv.org/abs/2402.13908v3 (89pp., 400 refs., extended version). 

\bibitem{Buryachenko`2024b}
 Buryachenko V. A. (2024)  Peridynamic micromechanics of composites: a critical review.   https://arxiv.org/abs/2402.13908v3 (109pp., 466 refs., extended version of JPER, 2024, 6, 531–601)

\bibitem{Buryachenko`R`1998a} 
Buryachenko~VA, Rammerstorfer~FG 
(1998a) Micromechanics and nonlocal effects
in graded random structure matrix composites. In: Bahei-El-Din~YA, Dvorak~GJ (eds)
{\it IUTAM Symp. on Transformation Problems in Composite and Active Materials.} 
Kluwer, Dordrecht, 197--206

\bibitem{Chen`G`2011}
Chen, X., Gunzburger, M. (2011) Continuous and discontinuous finite element methods for a peridynamics model of mechanics.
{\tenit Computer Methods in Applied Mechanics and Engineering}, {\tenbf 200}, 1237-1250.


{\color{black} \bibitem{Chen`et`2021}
Chen Z, Jafarzadeh S, Zhao J, Bobaru F. (2021) A coupled mechano-chemical peridynamic model
for pit-to-crack transition in stress-corrosion cracking, {\tenit J. Mechanics Physics Solids} {\tenbf 146}: 104203.}


\bibitem{Cheng`et`2024}
Cheng Z-Q, Liu H, Tan W. (2024)
Advanced computational modelling of composite materials.
{\tenit Engng Fract. Mechanics}, {\tenbf 305}: 110120


\bibitem{Chiu`et`2013}
Chiu, S.N., Stoyan, D., Kendall, W.S., Mecke, J. (2013)
{\tenit Stochastic Geometry and its Applications.} Third Edition.
J. Wiley \& Sons, Chichester, NY.


\bibitem{Coclite`et`2022a}
Coclite GM, Dipierro S, Maddalena F, Valdinoci E (2022a)
Wellposedness of a nonlinear peridynamic model.
{\tenit Nonlinearity}, {\tenbf 32}(1), {\tenit arXiv:1804.00273v1}

\bibitem{Coclite`et`2022b}
Coclite GM, Dipierro S, Fanizza G, Maddalena F, Valdinoci E (2022b)
Dispersive effects in a scalar nonlocal wave equation inspired by peridynamics
{\tenit Nonlinearity}, {\tenbf 35}(11), {\tenit arXiv:2105.01558v2}



\bibitem{Cuomo`et`et`2022}
Cuomo S, Schiano V, Cola D, Giampaolo F, Rozza G, Raissi M, Piccialli F. (2022) Machine learning through physics–informed
neural networks: where we are and what’s next.
{\tenit J. Scientific Computing}, {\tenbf 92}:88 (62pp)


\bibitem{Dahal`et`2023}
Dahal B, Seleson P, Trageser J. (2023)
The Evolution of the Peridynamics Co Authorship Network
{\tenit J.Peridynamics Nonlocal Modeling}, {\tenbf 5}: 311–355


\bibitem{Decklever`S`2016}
Decklever J, Spanos P. Nanocomposite material properties estimation and fracture analysis via peridynamics and Monte Carlo simulation.
(2016) {\tenit Probab Eng Mech}, {tenbf 44(SI)}:77--88

\bibitem{DElia`et`2020}
D’Elia M, Du Q, Glusa C, Gunzburger M, Tian X, Zhou Z (2020) Numerical
methods for nonlocal and fractional models. {\tenit Acta Numerica},
{\tenbf 29}, 1 - 124


\bibitem{DElia`et`2017}
D'Elia M, Du Q, Gunzburger M. (2017)
Recent progress in mathematical and computational aspects of peridynamics
{\tenit Handbook of Nonlocal Continuum Mechanics for Materials and Structures}.
Springer International Publishing.

\bibitem{DElia`et`2022}
D’Elia, M., Xingjie Li, X., Seleson, P., Tian, X., Yu, Y. (2022)
A review of Local-to-Nonlocal coupling methods in nonlocal
diffusion and nonlocal mechanics. {\tenit Journal of Peridynamics and Nonlocal Modeling}, {\tenbf 4}, 1--50

\bibitem{Desai`2024}
Desai S. (2024)
A novel equation of motion to predict elastoplastic deformation of 1 D stochastic bars
{\tenit J. Peridyn. Nonlocal Modeling}, {\tenbf 6}: 468–504

\bibitem{Diana`2023}
{\color{black} Diana V. (2023)
Anisotropic continuum molecular models: a unified framework based
on pair potentials for elasticity, fracture and diffusion type problems.
{\tenit Archives Comput. Methods in Engng}, {\tenbf 30}:1305–1344}

\bibitem{Diana`et`2022}
Diana V, Bacigalupo A, Lepidi M, Gambarotta L. (2022). Anisotropic peridynamics for homogenized microstructured materials. {\tenit Comput. Meth. Applied Mech. Engng}, {\tenbf 392}: 114704.


\bibitem{Diehl`et`2019}
Diehl P, Prudhomme S, Levesque M. (2019) A review of benchmark experiments for the validation
of peridynamics models. {\tenit J. Perid. Nonlocal Modeling}, {\tenbf 1}: 14–35.

\bibitem{Difonzo`et`2024}
Difonzo FV, Lopez L, Pellegrino SF. (2024)
Physics informed neural networks for an inverse problem
in peridynamic models. {\tenit Engineering with Computers}
https://doi.org/10.1007/s00366-024-01957-5

\bibitem{Dimola`et`2022}
Dimola N, Coclite A, Fanizza G, Politi T (2022)
Bond-based peridynamics, a survey
prospecting nonlocal theories of
fluid-dynamics.
{\tenit Advances in Continuous and Discrete Models}, 2022:60 (27pp.)

\bibitem{Ding`S`2024}
Ding W, Semperlotti F. (2024)
Two-dimensional nonlocal Eshelby’s inclusion theory:
eigenstress-driven formulation and applications.
{\tenit Proc. R. Soc.}, {\tenbf A 480}: 20230842

\bibitem{Dipasquale`et`2022}
Dipasquale D, Sarego G, Prapamonthon P, Yooyen S, Shojaei A.
(2022) A stress tensor-based failure criterion for ordinary
state-based peridynamic models.
{\tenit J. Appl. Computat. Mechanics}, {\tenbf 8}, 617--628

\bibitem{Diyaroglu`et`2016}
Diyaroglu C, Oterkus E, Madenci E, Rabczuk T, Silliq A (2016) Peridynamic
modeling of composite laminates under explosive loading. {\tenit Composite Structures},
{ 144}: 14-23


\bibitem{Diyaroglu`et`2019a}
Diyaroglu C, Madenci E, Phan N (2019a) Peridynamic homogenization of microstructures with orthotropic constituents in a finite element framework. {\tenit Composite Structures},
227:111334

\bibitem{Diyaroglu`et`2019b}
Diyaroglu C, Madenci E, Stewart RJ, Zobi SS (2019b) Combined peridynamic and finite
element analyses for failure prediction in periodic and partially periodic perforated structures
{\tenit Composite Structures}, 227:111481


\bibitem{Dorduncu`et`2024}
Dorduncu M, Ren H, Zhuang X, Silling S, Madenci E, Rabczuk T. (2024)
A review of peridynamic theory and nonlocal operators along with their
computer implementations.
{\tenit Computers and Structures}, {\tenbf 299}: 107395

\bibitem{Drugan`2000} 
Drugan~WJ (2000) Micromechanics-based variational estimations 
for a higher-order nonlocal constitutive equation and optimal choice 
of effective moduli for elastic composites.
{\it J Mech Phys Solids,} {\bf 48}:1359--1387

\bibitem{Drugan`2003} 
Drugan~WJ (2003) Two exact micromechanics-based nonlocal constitutive equations for random linear elastic composite materials.
{\it J Mech Phys Solids}, {\bf 51}:1745--1772 

\bibitem{Drugan`W`1996} 
Drugan~WJ, Willis~JR (1996) A micromechanics-based
nonlocal constitutive equation and estimates of representative volume elements
for elastic composites. {\it J Mech Phys Solids,} {\bf 44}:497--524


\bibitem{Dvorak`2013}
Dvorak, GJ. {\tenit Micromechanics of Composite Materials.}
Dordrecht: Springer, 2013.

\bibitem{Du`et`2020}
{\color{black} Du Q, Engquist B, Tian X. (2020) Multiscale modeling, homogenization and nonlocal effects: Mathematical
and computational issues. {\tenit Contemporary mathematics}, {\tenbf 754}, 115–140.}

\bibitem{Du`et`2013}
Du Q, Gunzburger M, Lehoucq RB, Zhou, K (2013)
Analysis of the volume-constrained peridynamic Navier equation of linear elasticity.
{\tenit J. Elast.}, {113}: 193--217.

\bibitem{Du`et`2016}
Du Q, Lipton R, Mengesha T. (2016) Multiscale analysis of linear evolution equations with applications to nonlocal models for heterogeneous media. {\tenit ESAIM: Mathematical
Modelling and Numerical Analysis}, {\tenbf 50}, 1425--1455.


\bibitem{Duan`et`2022}
Duan H, Wang J, Huang Z. (2022)
Micromechanics of composites with interface effects
{\tenit Acta Mech. Sin.}, {\tenbf 38}: 

\bibitem{Emmrich`W`2006}
Emmrich, E., Weckner, O., 2006.
The peridynamic equation of motion in non-local elasticity theory.
In: C. A. Mota Soares {\tenit et al.} (eds.), {\tenit III European Conference on Computational Mechanics. Solids, Structures, and Coupled Problems in Engineering}. Springer, Dordrecht

\bibitem{Emmrich`W`2007a}
Emmrich, E., Weckner, O., 2007a. Analysis and numerical approximation of an
integro-differential equation modeling non-local effects in linear elasticity.
{\tenit Math. Mech. Solids,} {\tenbf 12}, 363–-384.

\bibitem{Emmrich`W`2007b}
Emmrich, E., Weckner, O., 2007b. On the well-posedness of the linear peridynamic model
and its convergence towards the Navier equation of linear elasticity.
{\tenit Commun. Math. Sci.} {\tenbf 5}, 851--864.

\bibitem{Eghbalpoor`S`2024}
Eghbalpoor R, Sheidaei A. (2024)
A peridynamic-informed deep learning model for brittle damage prediction
{\tenit Theoret. Appl. Fracture Mech.}, {\tenbf 131}: 104457

\bibitem{Eriksson`S`2021}
Eriksson, K, and Stenström, C.
Homogenization of the 1D peri-static/dynamic bar with triangular micromodulus.
{\tenit Journal of Peridynamics and Nonlocal Modeling}, 2021; {\tenbf 3}: 85--112.

\bibitem{Eshelby`1957}
Eshelby, J.D., 1957. The determination of the elastic field of an ellipsoidal inclusion,
and related problems. {\tenit Proc. Roy. Soc. Lond.}, {\tenbf A 241}, 376-–396.

\bibitem{Eskin`1981} 
Eskin GI (1981) {\it Boundary Value Problems for Elliptic Pseudodifferential Equations.} American Mathematical Society, Providence, RI 

\bibitem{Fan`et`2023}
{\color{black} Fan Y, D’Elia M, Yu Y, Najm HN, Silling S. (2023) Bayesian nonlocal operator regression: A datadriven
learning framework of nonlocal models with uncertainty quantification. {\tenit J. Engig
Mech.}, {\tenbf 149}, 04023049.}

\bibitem{Fan`et`2022a}
Fan Y, Tian X, Yang X, Li C, Webster C, Yu Y. (2022)
An asymptotically compatible probabilistic
collocation method for randomly heterogeneous nonlocal problems. {J. Comput. Physics},
{\tenbf 465}: 111376

\bibitem{Fan`et`2022b}
Fan Y, You H, Tian X, Yang X, Li C, Prakash N. (2022)
A meshfree peridynamic model for brittle fracture in randomly heterogeneous materials
{\tenit Comput. Meth. Appl. Mech. and Engng}, {\tenit 399}: 115340

\bibitem{Fan`et`2024}
Fan Y, You H, Yu Y. (2024)
OBMeshfree: An optimization-based meshfree solver for nonlocal diffusion and peridynamics models
{\tenit J. Peridyn. Nonlocal Modeling}, {\tenbf 6}: 4–32

\bibitem{Faroughi `et`2024}
Faroughi SA, Pawar NM, Fernandes C, Raissi M, Das S, Kalantari NK, Kourosh Mahjour S. (2024)
Physics-guided, physics-informed, and physics-encoded neural networks and operators in scientific computing: Fluid and solid mechanics
{J. Computing and Information Science}, {\tenbf 24}: 040802

\bibitem{Ferreira`et`2021}
Ferreira BP, Pires FM, Bessa MA (2021)
Adaptive clustering-based reduced-order modeling framework: fast and accurate modeling of localized history-dependent phenomena
{\tenit arXiv preprint arXiv}, 2109.11897


\bibitem{Fish`2014}
Fish J (2014) {\tenit Practical Multiscaling}. Chichester: John Wiley \& Sons.

\bibitem{Francqueville`et`2019} 
Francqueville~F, Gilormini~P, Diani~J (2019)
Representative volume
elements for the simulation of isotropic composites highly lled with monosized spheres -
{\it Int. J. Solids and Structures}, {\bf 158}:277--286

\bibitem{Frank`et`2023}
Frank X, Lampoh K, Delenne J-Y. (2023)
From stress concentrations between inclusions to probability of breakage:
A two-dimensional peridynamic study of particle-embedded materials
{\tenit Physics Review}, {\tenbf E108}: 034903


\bibitem{Galadima`et`2019}
Galadima Y, Oterkus E, Oterkus S (2019) Two-dimensional
implementation of the coarsening method for linear peridynamics.
{\tenit AIMS Mater Sci} { 6}: 252–275

\bibitem{Galadima`et`2023}
Galadima YK, Xia W, Oterkus E, Oterkus S (2023)
A computational homogenization framework for non-ordinary
state-based peridynamics { \tenit Engineering with Computers}
{\tenbf 39}: 461–487

\bibitem{Galadima`et`2023b}
Galadima YK, Xia W, Oterkus E, Oterkus S (2023)
Peridynamic computational homogenization theory for materials
with evolving microstructure and damage. { \tenit Engineering with Computers}
{\tenbf 39}:{\tenbf 39}: 2945–2957

\bibitem{Galadima`et`2023c}
Galadima Y K, Oterkus S, Oterkus E, Amin I, El-Aassar AH, Shawky H.
(2023) A nonlocal method to compute effective properties of viscoelastic composite materials based on peridynamic computational homogenization theory
{\tenit Composite Structures}, {\tenbf 319}, 117147.


\bibitem{Galadima`et`2024}
Galadima YK, Oterkus S, Oterkus E, Amin I, El-Aassar A-H, Shawky H. (2024)
Effect of phase contrast and inclusion shape on the effective response of
viscoelastic composites using peridynamic computational homogenization
theory.
{\tenit Mwch. Advanced Mater. Structures.} {tenbd 31}, 155--163


\bibitem{Geers`et`2010}
Geers, MGD, Kouznetsova, VG, Brekelmans, WAM. (2010) Multi-scale computational
homogenization: Trends and challenges. {\tenit J. Comput. Applied Mathematics};
{\tenbf 234}: 2175-2182.


\bibitem{Ghosh`2011}
Ghosh, S. {\tenit Micromechanical Analysis and Multi-Scale Modeling Using the Voronoi Cell Finite
Element Method (Computational Mechanics and Applied Analysis)}. Boca Raton: CRC Press, 2011.

\bibitem{Goodfellow`et`2016}
Goodfellow I, Bengio Y, Courville A (2016) {\tenit Deep learning}. MIT Press, Cambridge, MA

\bibitem{Gosmani`et`2022}
Goswami S, Bora A, Yu Y, Karniadakis GE. (2022) Physics-Informed
Neural Operators, arXiv preprint arXiv:2207.05748.

\bibitem{GrahamBrady`et`2003}
Graham-Brady~LL, Siragy~EF, Baxter~SC (2003)
Analysis of heterogeneous composites based
on moving-window techniques.
{\it J Engng Mech}, {\bf 129}:1054--1064

\bibitem{Ha`B`2010}
Ha, Y.D., Bobaru, F. (2010) Studies of dynamic crack propagation and crack branching with peridynamics, {\tenit Int. J. Fract.} {\tenbf 162}, 229–244

\bibitem{Haghighat`et`2021}
Haghighat E, Bekar AC, Madenci E, Juanes R. (2021) A nonlocal physics-informed deep learning framework using the peridynamic
differential operator. {\tenit Comput. Methods Appl. Mech. Engrg.} {\tenbf 385}: 114012

\bibitem{Haghighata`et`2021}
Haghighata E, Raissi M, Moure A, Gomez H, Juanes R. (2021)
A physics-informed deep learning framework for inversion and
surrogate modeling in solid mechanics
{\tenit Comput. Methods Appl. Mech. Engrg.}, {\tenbf 379}: 113741

\bibitem{Han`et`2016}
Han, F., Lubineau, G. and Azdoud, Y., (2016) Adaptive coupling between damage mechanics and peridynamics: a route for objective simulation of material degradation up to complete failure. {\tenit J. Mechanics Physics of Solids}, {\tenbf 94}, 453--472


\bibitem{Harandi`et`2024}
Harandi A, Moeineddin A, Kaliske M, Reese S. (2024)
Mixed formulation of physics‐informed neural networks for thermo‐mechanically coupled systems and heterogeneous domains.
{\tenit Int J Numer Methods Eng}, {\tenbf 125}: e7388

\bibitem{Harper`et`2012} 
Harper~ LT, Qian~C, Turner~TA, Li~S, Warrior~NA (2012)
Representative volume elements for discontinuous carbon fibre
composites—Part 2 Determining the critical size. {\it Compos. Sci. Technol.},
{\bf 72}:204--210


\bibitem{Hill`1963}
Hill R (1963) Elastic properties of reinforced solids: some theoretical principles. {tenit J Mech
Phys Solids}, 11:357–372

\bibitem{Hill`1965}
Hill, R. (1965) A self-consistent mechanics of composite materials. {\tenit J. Mech. Phys.
Solids} {\tenbf 13}, 212--222

\bibitem{Hobbs`et`2024}
Hobbs M, Rappel H, Dodwell T. (2024)
A probabilistic peridynamic framework with an application to the
study of the statistical size effect.
{\tenit Applied Math. Modelling}, {\tenbf 128}, 137–153

\bibitem{Hu`et`2024}
Hu H, Qi L, Chao X. (2024)
Physics-informed Neural Networks (PINN) for computational solid mechanics: Numerical frameworks and applications
Author links open overlay panel. {tenit Thin-Walled Structures}: 112495

\bibitem{Hu`et`2010}
Hu, W., Ha, Y.D., Bobaru, F., (2010) {\tenit Numerical integration in peridynamics},
Tech. rep., University of Nebraska-Lincoln

\bibitem{Hu`et`2011}
Hu W, Ha YD, Bobaru F (2011) Modeling dynamic fracture and damage in a fiber-
reinforced composite lamina with peridynamics. {\tenit Int J Multiscale Comput Eng}.
{ 9}: 707–726

\bibitem{Hu`et`2012a}
Hu, W., Ha, Y.D., Bobaru, F. (2012a)
Peridynamic model for dynamic fracture in unidirectional
fiber-reinforced composites.
{\tenit Comput. Methods Appl. Mech. Engrg.}, {\tenbf 217–-220}, 247–-261.

\bibitem{Hu`et`2012b}
Hu, W., Ha, Y.D., Bobaru, F., Silling, S.A., (2012b)
The formulation and computation of the nonlocal J-integral
in bond-based peridynamics. {\tenit Int. J. Fract.}, {\tenbf 176}, 195-–206.

\bibitem{Hu`et`2014}
Hu Y-L, Yu Y, Wang H (2014) Peridynamic analytical method for progressive damage
in notched composite laminates. {\tenit Composite Structures},
{ 108}: 801—810

\bibitem{Hu`et`2022}
Hu YL, . Wang JY, Madenci E, Mu Z, Yu Y. (2022)
Peridynamic micromechanical model for damage mechanisms
in composites. {\tenit Composite Structures}, {\tenbf 301},
116182

\bibitem{Hu`et`2024}
Hu Z, Daryakenari NA, Shen Q, Kawaguchi K.
Karniadakis GE. (2024)
State-space models are accurate and efficient neural operators for
dynamical systems. arXiv:2409.03231


\bibitem{Isakari`et`2017}
Isakari S, Asakura T, Haraguchi Y, Yano Y, Kakami A. (2017)
Performance evaluation and thermography of solid-propellant microthrusters with laser-based throttling.
{\tenit Aerospace Science Technology}, {\tenbf 71}: 99--108.

\bibitem{Isiet`et`2021}
{\color{black} Isiet M, Mi\^skovi\'c I, Mi\^skovi\'c S. (2021)
Review of peridynamic modelling of material failure and damage due to
impact. {\tenit Int. J. Impact Engng}, {\tenbf 147}, 103740}

\bibitem{Izadi`et`2024}
Izadi R, Das R, Fantuzzi N, Trovalusci P. (2024) 
Fracture properties of green nano fibrous network with random and aligned fibre distribution: a hierarchical molecular dynamics and peridynamics approach.
-{\tenit Available at SSRN}, 4790967

\bibitem{Jafarzadeh`et`2022}
Jafarzadeh S, Mousavi M, Larios A, Bobaru F (2022)
A general and fast convolution-based method for peridynamics: Applications to elasticity and brittle fracture.
{\tenit Comp. Meth. Appl. Mech. Enging}, {\tenbf 392}: 114666

\bibitem{Jafarzadeh`et`2024a}
Jafarzadeh S, Mousavi M, Wang L, Bobaru F. (2024)
PeriFast/Dynamics: A MATLAB code for explicit fast
convolution based peridynamic analysis of deformation
and fracture.
{\tenit Journal of Peridynamics and Nonlocal Modeling}, {\tenbf 6}, 33--61


{\color{black} \bibitem{Jafarzadeh`et`2024}
Jafarzadeh S, Silling S, Liu N, Zhang Z, Yu Y. (2024)
Peridynamic neural operators: a data-driven nonlocal constitutive model for complex material responses.
{\tenit arXiv preprint arXiv:2401.06070}}

\bibitem{Jafarzadeh`et`2024b}
Jafarzadeh S, Silling S, Zhang L, Ross C, Lee CH, Rahman SM, Wang S, Yu Y. (2024)
Heterogeneous peridynamic neural operators: discover biotissue constitutive law and microstructure from digital image correlation measurements. {\tenit ArXiv preprint arXiv:2403.18597}.

\bibitem{Javili`et`2019}
Javili A, Morasata R, Oterkus E (2019) Peridynamics review. {\tenit Mathematics Mechanics
of Solids}, { 24}: 3714–3739


\bibitem {Jenabidehkordi`et`2020}
Jenabidehkordi A, Abadi R, Rabczuk T. (2020)
Computational modeling of meso-scale fracture in polymer matrix composites employing peridynamics {\tenit Composite Structures},
{\tenbf 253}, 112740


\bibitem{Kachanov`S`2018}
Kachanov M, Sevostianov I (2018)
{ \tenit Micromechanics of materials, with applications}.
Springer International. Cham

\bibitem{Kanaun`L`1994} 
Kanaun~SK, Levin~VM (1994) Effective field method on mechanics of matrix composite materials. In: Markov~KZ (ed), {\it Advances in Math Modelling of Composite Materials.} World Scientific, Singapore, 1--58


\bibitem{Kanaun`L`2008}
Kanaun KK, Levin VM (2008) {\tenit Self-Consistent Methods for Composites}. Vol. 1, 2, Springer,
Dordrecht

\bibitem{Kanit`et`2003} 
Kanit~T, Forest~S, Galliet~I, Mounoury~V, Jeulin~D
(2003) Determination of the size of the representative volume
element for random composites: statistical
and numerical approach. {\it Int J Solids Struct,} {\bf 40}:3647--3679


\bibitem{Karniadakis`et`2021}
Karniadakis GE, Kevrekidis IG, Lu L, Perdikaris P, Wang S, Yang L. (2021)
Physics- informed machine learning.
{\tenit Nature Reviews Physics}, https://doi.org/10.1038/
s42254-021-00314-5

\bibitem{Khoroshun`1978} 
Khoroshun, L.P., 1978. Random functions theory in problems on the macroscopic
characteristics of microinhomogeneous media. {\tenit Priklad Mekh}, {\tenbf 14}(2), pp. 3--17
(In Russian. Engl Transl. {\tenit Soviet Appl Mech}, {\tenbf 14}, pp. 113--124)

\bibitem{Khoroshun`1996} 
Khoroshun~L (1996)
On a mathematical model for inhomogeneous deformation of composites.
{\it Priklad Mekh,} {\bf 32}(5):22--29 (In Russian. Engl Transl. {\it Int Appl Mech,} {\bf 
32}:341--348)

\bibitem{Kilic`2008}
Kilic B (2008)
{ \tenit Peridynamic theory for progressive failure prediction in homogeneous and heterogeneous materials}.
{Ph.D. Thesis}, Dep. Mechan. Engng, The University of Arizona, 1--262

\bibitem{Kilic`M`2010}
Kilic B, Madenci E (2010) Peridynamic theory for thermomechanical analysis. {\tenit IEEE Trans Adv Packag}, { 33}: 97--105

{\color{black} \bibitem{Kilic`M`2010b}
Kilic B, Madenci E. (2010) An adaptive dynamic relaxation method for quasi-static simulations
using the peridynamic theory. {\tenit Theor. Applied Fract. Mech.} 53(3),
194–204.}

\bibitem{Kim`L`2024}
Kim D, Lee J. (2024)
A review of physics informed neural networks for multiscale analysis
and inverse problems. 
{\tenit Multiscale Science and Engineering}, {\tenbf 6}: 1–11

\bibitem{Kingma`B`2014}
Kingma DP, Ba J (2014) Adam: A method for stochastic optimization. {\tenit arXiv:1412.6980}

\bibitem{Kouznetsova`et`2001}
Kouznetsova VG, Brekelmans WAM, Baaijens FPT (2001)
An approach to micro–macro modeling of heterogeneous materials, {\tenit Comput. Mech.} {\tenbf 27}: 37-–48

\bibitem{Kouznetsova `et`2004a}
 Kouznetsova V,  Geers M, Brekelmans W.  (2004) 
Size of a representative volume element in a second-order computational homogenization framework.
{\tenit Int. J. Multiscale Comput. Eng.}, {\tenbf 2}:  575--598

\bibitem{Kouznetsova `et`2004b}
 Kouznetsova V,  Geers M, Brekelmans W.  (2004) 
Multi-scale second-order computational homogenization of multi-phase materials: a nested finite element solution strategy.
{\tenbf Comput. Methods Appl. Mech. Eng.},  {\tenbf 193}:  5525--5550


\bibitem{Kroner`1958}
Kr\"oner E. (1967) Elasticity theory of materials with long range cohesive forces. {\tenit Int. J. Solids Struct}. {\tenbf 3}, 731–-742

\bibitem{Kumara`Y`2023}
Kumara H, Yadav N. (2023)
Deep learning algorithms for solving differential equations: a
survey. {\tenit J. Experimental \& Theoret. Artificial Intelligence}: 
https://doi.org/10.1080/0952813X.2023.212356

\bibitem{Lahellec`et`2003}
Lahellec N, Michel J C, Moulinec H., Suquet P (2003) Analysis of inhomogeneous materials at large
strains using fast Fourier transforms. {\tenit IUTAM Symposium on Computational Mechanics of Solid
Materials at Large Strains} (Berlin: Springer) pp. 247–58

\bibitem{Lanthaler`et`2024}
Lanthaler S, Li Z, Stuart AM. (2024) Nonlocal and noblinerity implies universality in operator lerning.
arXiv:2304.13221v2

\bibitem{Laurien`et`2023}
Laurien M, Javili A, Steinmann P.(2023)
Peridynamic modeling of nonlocal degrading interfaces in composites.
{\tenit Forces in Mechanics}, {\tenbf 10}: 100124.

\bibitem{Lax`1952}
Lax M (1952) Multiple scattering of waves II. The effective
fields dense systems. {\tenit Phys. Rev.} { 85}: 621--629.

\bibitem{Le`et`2014}
Le QV, Chan WK, Schwartz J. (2014) A two-dimensional ordinary, state-based peridynamic model for linearly elastic solids.
{\tenit Int. J. Numerical Methods Engng} {\tenbf 98}, 547–561.


\bibitem{Lehoucq`S`2008}
Lehoucq RB, Silling SA (2008) Force flux and the peridynamic stress
tensor. {\tenit J. Mech. Phys. Solids}, {\tenbf 56}: 1566–-1577.

\bibitem{LiF`et`2023}
{\color{black} Li F, Yang X, Gao W, Liu W. (2023)
A single-layer peridynamic model for failure analysis of
composite laminates. {\tenit Materials Today Communications}, {\tenbf 37}, 106988}

\bibitem{Li`et`2022a}
Li, J., Li, S., Lai, X., Liu, L. (2022)
Peridynamic stress is the static first Piola–Kirchhoff Virial stress.
{\tenit Int. J. Solids and Structures}, {\tenbf 241}, 111478.

\bibitem{Li`et`2022b}
Li J, Wang Q, Li X, Ju L, Zhang Y. (2022)
Homogenization of periodic microstructure based on representative volume element using improved bond-based peridynamics
{\tenit Engng Analysis Boundary Elements}, {\tenbf 143}: 152--162

\bibitem{LiM`et`2024}
Li M, Wang B, Hu J, Li G, Ding P, Ji C. (2024)
Artificial neural network-based homogenization model for predicting multiscale thermo-mechanical properties of woven composites
{\tenit Int. J. Solids Struct.}, {\tenbf 301}: 112965

{\color{black} \bibitem{Li`et`2020}
Li S, Jin Y, Huang X, Zhai L. (2020) An extended bond-based peridynamic approach for analysis
on fracture in brittle materials. {\tenit Math. Problems Engineering}, ID 9568015, 1–12. }

{\color{black} \bibitem{Li`et`2022b}
Li X, Gu X, Xia X, Madenci E, Chen X, Zhang Q. (2022)
Effect of water-cement ratio and size on
tensile damage in hardened cement paste: Insight from peridynamic simulations. {\tenit Construction
Building Materials}, {\tenbf 356}: 129256}



\bibitem{Li`et`2003}
Li Z, Kovachki N, Azizzadenesheli K, Liu B, Bhattacharya K, Stuart A, Anandkumar, A. (2003) Neural
operator: Graph kernel network for partial differential equations, arXiv preprint arXiv:2003.03485.

\bibitem{Liang`et`2021}
Liang X, Wang L, Xu J, Wang J. (2021)
The boundary element method of peridynamics
{\tenit Int. J. Numerical Methods Enging}, {\tenbf 122}, 5558--5593.


\bibitem{Littlewood`et`2024}
Littlewood DJ, Parks ML, Foster JT, Mitchell JA. (2024)
The Peridigm meshfree peridynamics code
{\tenit J. Peridynamics Nonlocal Modeling}, {\tenbf 6}, 118--148

\bibitem{Littlewood`et`2015}
Littlewood, D.J., Silling, S.A., Mitchell, J.A., Seleson, P.D., Bond, S.D., Parks, M.L., Turner, D.Z., Burnett, D.J., Ostien, J., Gunzburger, M. (2015)
{\tenit Strong local-nonlocal coupling for integrated fracture modeling}. Technical report SAND2015-7998, Sandia National Laboratories, Albuquerque, NM.


\bibitem{Liu`H`2012}
Liu W, Hong J-W. (2012)
Discretized peridynamics for brittle and ductile solids.
{\tenit Int. J. Numer. Meth. Engng}, {\tenbf 89}: 1028--1046

\bibitem{Liu`et`2011}
Liu YL, Mukherjee S, Nishimura N, Schanz M, Ye W, Sutradhar A, Pan E, Dumont NA, Frangi A, Saez A. (2011)
Recent advances and emerging applications of the boundary
element method. {\tenit Applied Mechanics Reviews}, {\tenbf 64}, 031001 (38 pages)

{\color{black} \bibitem{Lu`N`2022}
Lu J, Nie Y. (2022) A reduced-order fast reproducing kernel collocation method for nonlocal models
with inhomogeneous volume constraints. {\tenit Computers \& Mathematics Applications}. {\tenbf 121}: 52–61.}

{\color{black} \bibitem{Lu`et`2022}
Lu J, Yang M, Nie Y. (2022) Convergence analysis of Jacobi spectral collocation methods for
weakly singular nonlocal diffusion equations with volume constraints. {\tenit Applied Mathematics
Computation}. {\tenbf 431}: 127345.}

\bibitem{Luciano`W`2001} 
Luciano~R, Willis~JR (2001) 
Non-local effective relations for fibre-reinforced composites loaded by 
configuration-dependent body forces. {\it J Mech Phys Solids,} {\bf 49}:2705--2717

\bibitem{Macek`S`2007}
Macek RW, Silling SA (2007) Peridynamics via finite element analysis
{\tenit Finite Elements in Analysis and Design}, {\tenbf 43}: 1169--1178


\bibitem{Madenci`et`2016}
Madenci E, Barut A, Futch M. (2016) Peridynamic differential operator and its applications. {\tenit Comput. Methods Appl. Mech. Engrg.}
{\tenbf 304}, 408–451

\bibitem{Madenci`et`2017}
Madenci E, Barut A, Phan ND (2017) Peridynamic unit cell homogenization, 58th
{\tenit AIAA/ASCE/AHS/ASC Structures, Structural Dynamics, and Materials Conference, AIAA}
SciTech Forum, (AIAA 2017-1138)

\bibitem{Madenci`et`2018}
Madenci E, Barut A, Phan N. (2018) Peridynamic unit cell homogenization for thermoelastic properties of heterogenous microstructures with defects. {\tenit Composite Structures}, {\tenbf 188}: 104-115.


\bibitem{Madenci`et`2019}
Madenci E, Dorduncu M, Gu X. (2019) Peridynamic least squares minimization. {\tenit Comput. Methods Appl. Mech. Engrg.},
{\tenbf 348}, 846–874.

\bibitem{Madenci`G`2015}
{\color{black} Madenci~E, Guven~I. (2015)
{\tenit The Finite Element Method and Applications
in Engineering Using ANSYS}. Springer, NY}

\bibitem{Madenci`O`2014}
Madenci~E, Oterkus~E (2014) {\tenit Peridynamic Theory and Its Applications.} Springer, NY

\bibitem{Madenci`O`2016}
Madenci E, Oterkus S (2016) Ordinary state-based peridynamics for plastic deformation according to von Mises yield criteria with isotropic hardening.
{\tenit J. Mech. Phys. Solids}, { 86}: 192–219

\bibitem{Madenci`et`2021}
Madenci E, Yaghoobi A, Barut A, Phan N (2021) Peridynamic modeling of
compression after impact damage in composite laminates. {\tenit J Peridyn Nonlocal Model},
{ 3}: 327–347

\bibitem{Madenci`et`2023}
Madenci E, Yaghoobi A, Barut A, Phan N (2023)
Peridynamics for failure prediction in variable angle tow composites
{\tenit Archive of Applied Mechanics}, {\tenbf 93}: 93–107

\bibitem{Malyarenko`O`2019}
Malyarenko~A, Ostoja-Starzewski~M. (2019) {\tenit Tensor-Valued Random Fields for Continuum Physics}, Cambridge University Press,
Cambridge, UK


\bibitem{Matous`et`2017}
Matouš K, Geers MGD, Kouznetsova VG, Gillman A (2017) A review of predictive
nonlinear theories for multiscale modeling of heterogeneous materials.
{ \tenit J. Comput. Physics}, {\tenbf 330}: 192–220

\bibitem{Maxwell`1873} 
Maxwell~JC (1873) A Treatise on Electricity and Magnetism,
Dover, New York (1954). (Republication of 3rd edition of
1892.)

\bibitem{Mehrmashhadi `et`2019}
{\color{black} Mehrmashhadi J, Chen Z, Zhao J, Bobaru F. (2019)
A stochastically homogenized peridynamic model for intraply fracture in fiber-reinforced composites
{\tenit Composites Science Technology}, {\tenbf 182}: 107770}

\bibitem{Mengesha`D`2014}
Mengesha T, Du Q. (2014) The bond-based peridynamic system with Dirichlet-type volume constraint.
{\tenit Proc. R. Soc. Edinburgh}, {\tenbf A 144}: 161-–186

\bibitem{Mikata`2012}
Mikata Y (2012) Analytical solutions of peristatic and peridynamic problems for a 1D
infinite rod. {\tenit Int. J. Solids and Structures}, {\tenbf 49}: 2887--2897

\bibitem{Mikata`2023}
MikataY(2023) Analytical solutions of peristatics and peridynamics
for 3D isotropic materials. {\tenit Eur J Mech A/Solids}, {\tenbf 101}: 104978

\bibitem{Miehe`K`2002}
Miehe C, Koch A (2002) Computational micro-to-macro transition of discretized microstructures undergoing small strain.
{\tenit Arch. Appl. Mech.}{\tenbf 72}: 300–-317

\bibitem{Moes`B`2002}
Mo$\ddot{\rm o}$es N., Belytschko T. (2002)
Extended finite element method for cohesive crack growth
{\tenit Enginng Fracture Mechanics}, {\tenbf 69}: 813--833


\bibitem{Mori`T`1973}
{Mori T}, {Tanaka K.} (1973) Average stress in matrix and average elastic energy of materials with misfitting inclusions. {\tenit Acta Metall}. { 21}: 571—574


\bibitem{Mousavi`et`2021}
Mousavi F, Jafarzadeh S, Bobaru F (2021)
An ordinary state-based peridynamic elastoplastic 2D model consistent with J2 plasticity.
{\tenit Int. J. Solids Structures}, {\tenbf 229}, 111146

\bibitem{Moumen`et`2021}  
Moumen~AE,  Kanit~T, Imad~A (2021)
Numerical evaluation of the representative volume element for
random composites. {\it European Journal of Mechanics / A Solids}, {\bf 86}:104181


\bibitem{Mura`1987}
Mura T. (1987) {\tenit Micromechanics of Defects in Solids (Mechanics of Elastic and Inelastic Solids)} 2nd edn
Berlin: Springer


\bibitem{Nemat-Nasser`H`1993}
Nemat-Nasser S, Hori M (1993) {\tenit Micromechanics: Overall Properties of Heterogeneous
Materials}. Elsevier, North-Holland.

\bibitem{Ning`et`2023}
Ning L, Cai Z, Dong H, Liu Y, Wang W. (2023)
A peridynamic-informed neural network for continuum elastic
displacement characterization
{\tenit Comput. Methods Appl. Mech. Engrg.}, {\tenbf 407}: 115909

\bibitem{Nguyen`et`2021}
Nguyen CT, Oterkus S, Oterkus E (2021)
An energy-based peridynamic model for fatigue cracking
{\tenit Engineering Fracture Mechanics}, {\tenbf 241}: 107373


{\color{black} \bibitem{Nowak`et`2023}
Nowak M, Mulewska K, Azarov A, Ustrzycka A, {\tenit et al.}
(2023) A peridynamic elasto-plastic damage
model for ion-irradiated materials. {\tenit Int. J. Mechanical Sciences} {\tenbf 237}:
107806.}



\bibitem{Ongaro`et`2022}
Ongaro G, Bertani R, Galvanetto U, Pontefisso A, Zaccariotto M. (2022)
A multiscale peridynamic framework for modelling mechanical
properties of polymer-based nanocomposites.
{\it Engng Fract. Mechanics}, {\tenbf 274}: 108751

\bibitem{Ongaro`et`2021}
Ongaro G, Seleson P, Galvanetto U, Ni T, Zaccariotto M. (2021)
Overall equilibrium in the coupling of peridynamics and classical continuum mechanics
{\tenit Comput. Meth. Appl. Mech. Engng}, {\tenbf 381}: 113515

\bibitem{Ongaro`et`2023}
Ongaro G, Shojaei A, Mossaiby f, Hermann A, Cyron CJ, Trovalusci P. (2023)
Multi-adaptive spatial discretization of bond-based peridynamics.
{\tenit Int J Fract}, {\tenbf 244}: 1--24

\bibitem{Ostoja`et`2016}
Ostoja-Starzewski M, Kale S, Karimi P, Malyarenko A, Raghavan B, Ranganathan SI,
Zhang J (2016) Scaling to RVE in random media. {tenit Adv. Appl. Mech.}, {\tenbf 49}:111–211

\bibitem{Oterkus`O`2024}
Oterkus E, Oterkus S. (2024)
Recent advances in peridynamic theory: A review
{\tenit AIMS Materials Science}, {\tenbf 11}: 515--546

\bibitem{Oterqus`et`2014}
Oterkus, S., Madenci, E., Agwai, A. (2014). Fully coupled peridynamic thermomechanics.
{\tenit J. Mech. Phys. Solids}, {\tenbf 64}: 1–23. 


\bibitem{Pan`et`2024}
Pan Y, Wu P, Fan S, Peng X, Chen Z. (2024)
Peridynamic simulation of fatigue crack growth in
porous materials. {\tenit Engng Fracture Mech.}, {\tenbf 300}: 109984.

\bibitem{Parks`et`2011}
Parks ML, Seleson P, Plimpton SJ, Silling SA, Lehoucq RB. (2011)
{\tenit Peridynamics with LAMMPS: A user guide v0.3 beta, SAND Report 2011–8523}, Sandia
National Laboratories, Albuquerque, NM, and Livermore, CA

\bibitem{Parnell`2016}
Parnell WJ. (2016)
The Eshelby, Hill, moment and concentration tensors for ellipsoidal inhomogeneities in the Newtonian potential problem and linear elastostatics
{\tenit J Elast}, {\tenbf 125}: 231--294.

\bibitem{Paszke`et`2019}
Paszke A, Gross S, Massa F et al (2019) PyTorch: an imperative
style, high-performance deep learning library. {\tenit Adv Neural Inf Process
Syst}, {\tenbf 32}:8024--8035



\bibitem{Qi`et`2024}
Qi J, Li C, Tie Y, Zheng Y, Cui Z, Duan Y. (2024)
A peridynamic-based homogenization method to compute effective properties of periodic microstructure
{\tenit Computational Particle Mechanics.} https://doi.org/10.1007/s40571-023-00698-4


\bibitem{Raissi`et`2019}
Raissi M, Perdikaris P, Karniadakis GE. (2019)
Physics-informed neural networks: A deep learning framework for solving forward and
inverse problems involving nonlinear partial differential equations. {\tenit J. Comput. Phys.}, {\tenbf 378}: 686–707


\bibitem{Rayleigh`1892}
Rayleigh L. (1892) On the influence of obstacles arranged in rectangular order upon the properties of a medium.
{\tenit Philosophical Magazine}, {\tenbf 34}: 481--502.

\bibitem{Ren`et`2017}
Ren, B., Wu, C., Askari, E. (2017) A 3D discontinuous Galerkin finite element method with the bond-based peridynamics model for dynamic brittle failure analysis. {\tenit International Journal of Impact Engineering}, {\tenbf 99}: 14--25

\bibitem{Ren`et`2022}
Ren B, Wu CT, Seleson S, Zeng D, Nishi M, Pasetto M (2022)
An FEM-Based Peridynamic Model for Failure Analysis of Unidirectional Fiber-Reinforced Laminates
{\tenit J.
Peridynamics and Nonlocal Modeling}, {\tenbf 4}: 139—158

\bibitem{Ren`et`2017}
Ren H, Zhuang X, Rabczuk T. (2017)
Dual-horizon peridynamics: A stable solution to varying horizons
{\tenit Comput. Methods Appl. Mech. Engrg}, {\tenbf. 318}, 762--782

\bibitem{Ren`L`2024}
Ren X, Lyu X. (2024)
Mixed form based physics-informed neural networks for performance
evaluation of two-phase random materials
{\tenit Engng Appl. Artificial Intelligence}, {\tenbf 127}: 107250


\bibitem{Sab`N`2005}
Sab~K, Nedjar~B (2005)
Periodization of random media and representative
volume element size for linear composites. 
{\it C R Mecanique}, {\bf 333}:187--195

\bibitem{Sarego`et`2016}
Sarego G, Le QV, Bobaru F, Zaccariotto M, Galvanetto U. (2016)
Linearized state-based peridynamics for 2-D problems
{\tenit Int. J. Numer. Meth. Engng}, {\tenbf 108}: 1174–1197

\bibitem{Scabbia`et`2023}
Scabbia F, Zaccariotto M, Galvanetto U. (2023)
A new surface node method to accurately model
the mechanical behavior of the boundary in 3D state-based
peridynamics.
{\tenit J. Peridyn. Nonloc. Modeling}, {\tenbf 5}: 521–555

\bibitem{Scabbia`et`2023b}
Scabbia F, Zaccariotto M, Galvanetto U. (2023)
Accurate computation of partial volumes in 3D peridynamics
{\tenit Engng with Computers}, {\tenbf 39}: 959--991

\bibitem{Scabbia`et`2024}
Scabbia F, Zaccariotto M, Galvanetto U.
(2024)
A general ordinary state-based peridynamic formulation for
anisotropic materials.
{\tenit Comput. Methods Appl. Mech. Engrg}. {\tenbf 427}: 117059

\bibitem{Scott`M`2020}
Scott JM, Mengesha T. (2020)
Asymptotic analysis of a coupled system of nonlocal equations with oscillatory coefficients
{\tenit Multiscale Modeling \& Simulation}, {\tenbf 18}: 1137/19M1288085

\bibitem{Sejnoha`Z`2013}
Sejnoha, M, Zeman, J.
{\tenit Micromechanics in Practice.} Southampton, UK: WIT Press, 2013.

\bibitem{Seleson`et`2016}
Seleson P, Du Q, Parks M. (2016) On the consistency between nearest-neighbor peridynamic discretizations
and discretized classical elasticity models. {\tenit Comp. Meth. Applied Mech. Engrg.}, {\tenbf 11}, 698--722.
2016.

\bibitem{Seleson`et`2013}
Seleson P, Gunzburger M, Parks ML (2013) Interface problems in nonlocal diffusion and sharp transitions
between local and nonlocal domains. {\tenit Comput. Methods Appl. Mech. Engrg.}, {\tenbf 266}: 185-204

\bibitem{Seleson`L`2016}
Seleson P, Littlewood DJ. (2016)
Convergence studies in mesh-free peridynamic simulations.
{\tenit Comput. Mathematics with Applications}, {\tenbf 71}: 2432--2448

\bibitem{Seleson`P`2011}
Seleson P, Parks ML. (2011) On the role of the influence function in the peridynamic theory.
{\tenit Int. J. Multiscale Comput. Eng.}, {\tenbf 9}: 689-–706

\bibitem{Selezon`et`2024}
Seleson P, Pasetto M, John J, Trageser J, Reeve ST. (2024)
PDMATLAB2D: A peridynamics MATLAB two dimensional code.
{\tenit J. Peridyn. Nonl. Modeling}, {\tenbf 6}:149–205

\bibitem{Selvaraj`S`2023}
Selvaraj J, Said BE. (2023)
Multiscale modelling of strongly heterogeneous materials using geometry
informed clustering
{\tenit Int. J. Solids Struct.}, {\tenbf 280}: 112369

\bibitem{Sevostianov`et`2019}
Sevostianov I, Mogilevskaya SC, Kushch VI (2019) Maxwell’s methodology of estimating
effective properties Alive and well. {\tenit Int. J. Engineering Science}, {\tenbf 140}:35–88

\bibitem{Shermergor`1977}
Shermergor TD. (1977) {\tenit The Theory of Elasticity of Microinhomogeneous Media.} Nauka,
Moscow (In Russian)

\bibitem{Silling`2000}
Silling S. (2000) Reformulation of elasticity theory for discontinuities
and long-range forces. { \tenit J. Mech.
Physics of Solids} { \tenbf 48}: 175--209

\bibitem{Silling`2010}
Silling S. (2010) Linearized theory of peridynamic states. {\tenit Journal of Elasticity,} { \tenbf 99}: 85–111

\bibitem{Silling`2011}
Silling .S (2011) A coarsening method for linear peridynamics. {\tenit Int. J. Multiscale Computational
Engng}, { \tenbf 9}: 609–622


\bibitem{Silling`2014}
Silling, S.A. 2014. Origin and effect of nonlocality in a composite. {\tenit J. Mechanics of Materials
and Structures}, {\tenbf 9}, 245--258.

\bibitem{Silling`2020}
Silling S. (2020) Propagation of a stress pulse in a heterogeneous elastic bar. {\tenit Sandia Report
SAND2020-8197}, Sandia National Laboratories.

\bibitem{Silling`A`2005}
Silling SA, Askari E. (2005) A meshfree method based on the peridynamic model of solid mechanics.
{\tenit Comput. Struct}. { \tenbf 83}: 1526–-153

\bibitem{Silling`et`2007}
Silling SA, Epton M, Weckner O, Xu J, Askari E (2007)
Peridynamic states and constitutive modeling.
{\tenit J. Elasticity}, { \tenbf 88}: 151–-184

{\color{black} \bibitem{Silling`et`2023} 
Silling SA, D'Elia M, Yu Y, You H, Fermen-Coker M. (2023) Peridynamic model for single-layer
graphene obtained from coarse-grained bond forces. {\tenit J Perid. Nonlocal
Modeling}, {\tenbf 5}: 183–204.}

\bibitem{Silling`et`2024}
Silling SA, Jafarzadeh S, Yu Y. (2024)
Peridynamic models for random media found by coarse graining
{\tenit J. Peridynamics and Nonlocal Modeling}, {\tenbf 6}

\bibitem{Silling`L`2008}
Silling SA, Lehoucq RB. (2008)
Convergence of peridynamics to classical elasticity theory,
{\tenit J. Elasticity}, {\tenbf 93}: 13–-37.

\bibitem{Silling`L`2010}
Silling SA, Lehoucq RB. (2010)
Peridynamic theory of solid mechanics. { \it Adv. Appl.Mech.},
{ \tenbf 44}: 73–-168

\bibitem{Silling`et`2003}
Silling SA, Zimmermann M, Abeyaratne R (2003) Deformation of a peridynamic bar.
{\tenit J. Elasticity}, { \tenbf 73}: 173--190.

\bibitem{Smyshlyaev`C`2000} 
Smyshlyaev VP, Cherednichenko KD (2000) A rigorous derivation of 
strain gradient effects in the overall behavior of periodic heterogeneous
media. {\it J Mech Phys Solids,} {\bf 48}:1325--1357


\bibitem{Song`et`2023}
Song Y, Li S, Li Y. (2023)
Peridynamic modeling and simulation of thermo mechanical fracture
in inhomogeneous ice.
{\tenit Engineering with Computers}, {\tenbf 39}: 575–606

\bibitem{Sun`F`2021}
Sun W, Fish J (2021)
Coupling of non-ordinary state-based peridynamics and finite element method for fracture
propagation in saturated porous media. {\tenit Int J Numer Anal Methods Geomech}., { \tenbf 45}: 1260--1281


\bibitem{Sun`et`2020}
Sun W, Fish J, Zhang G (2020)
Superposition of non-ordinary state-based peridynamics
and finite element method for material failure simulations. {\tenit Meccanica}, { \tenbf 55}: 681--699


\bibitem{Talamadupula`et`2020}
Talamadupula KK, Povolny SJ, Prakash N, Seidel GD. (2020)
Mesoscale strain and damage sensing in nanocomposite bonded energetic materials under low velocity impact with frictional heating via peridynamics. {\it Modelling Simul. Mater. Sci. Eng.} {\tenbf 28}: 085011


\bibitem{Terada`K`2001}
Terada K, Kikuchi N (2001) A class of general algorithms for multi-scale analyses of heterogeneous media, {\tenit Comput. Methods Appl. Mech. Eng.} { \tenbf 190}: 5247–-5464


\bibitem{Tian`2024}
Tian H. (2024)
Tensor-involved peridynamics: A unified framework for isotropic and
anisotropic materials.
{\tenit arXiv:2410.10175v2}

\bibitem{Tian`D`2015}
Tian, X., Du, Q. (2015) Nonconforming discontinuous Galerkin methods for nonlocal variational problems. {\tenit SIAM Journal on Numerical Analysis}, {\tenbf 53}, 762-781.


\bibitem{Torquato`2002}
Torquato, S. 2002. {\tenit Random Heterogeneous Materials: Microstucture and Macroscopic
Properties.} Springer-Verlag, New York, Berlin.



\bibitem{Wang`et`2024a}
Wang H, Wu L, Guo J, Yu C, Li Y, Wu Y. (2024)
Three-dimensional modeling and analysis of anisotropic materials with quasi-static deformation and dynamic
fracture in non-ordinary state-based peridynamics. {\tenit Appl. Math. Model.}, {\tenbf 125}: 625–648

\bibitem{Wang`et`2024}
Wang H, Wu L, Huang D, Chen J, Guo J, Yu C, Li Y, Wu Y. (2024)
A machine-learning-based peridynamic surrogate model for
characterizing deformation and failure of materials and structures
{\tenit Engineering with Computers}, 
https://doi.org/10.1007/s00366-024-02014-x

\bibitem{Wang`et`2017}
Wang L, Xu J,Wang J. (2017) Static and dynamic Green’s functions in peridynamics. {\tenit J. Elasticity}, {\tenbf 126}: 95–125


\bibitem{Wang`et`2020}
Wang L, Xu J, Wang J, Karihaloo BL. (2020)
Nonlocal thermo-elastic constitutive relation of fibre-reinforced composites.
{\tenit Acta Mechanica Sinica}, {\tenbf 36}: 176–187

\bibitem{WangQ`et`2024}
Wang QZ, Hu YL, Yu Y, Wu D, Yao ZY. (2024)
A peridynamic differential operator modeling approach for ceramic matrix
composites microstructure with uniform or non-uniform discretization.
{\tenit Compos. Struct.} {\tenbf 344}: 118282

\bibitem{Wang`Y`2024}
Wang X, Yin Z-Y. (2024) 
Interpretable physics-encoded finite element network to handle
concentration features and multi-material heterogeneity
in hyperelasticity.
{\tenit Comput. Meth. Applied Mech. Engng}, {tenit 431}: 117268

\bibitem {Wang`et`2018}
{\color{black} Wang Y, Zhou X, Wang Y, Shou Y. (2018)
A 3-D conjugated bond-pair-based peridynamic formulation for initiation and propagation of cracks in brittle solids
{\tenit Int. J. Solids Structures}, {\tenbf 134}: 89–115}


\bibitem{Weckner`A`2005}
Weckner O, Abeyaratne R (2005) The effect of long-range forces on the dynamics of a bar. { \tenit J. Mech. Physics of Solids} { \tenbf 53}: 705--728

\bibitem{Weckner`et`2009}
Weckner O, Brunk G, Epton MA, Silling SA, Askari E. (2009) Green'ss functions in nonlocal three-dimensional linear elasticity. {\tenit Proc. R. Soc.}, {\tenbf A 465}: 3463–3487

\bibitem{Weckner`E`2005}
Weckner O, Emmrich E. 2005 Numerical simulation of the dynamics of a nonlocal,
inhomogeneous, infinite bar. {\tenit J. Comp. Appl. Mech.}, {\tenbf 6}, 311--319

\bibitem{Wen `et`2023}
Wen Z, Hou C, Zhao M, Wan X. (2023)
A peridynamic model for non-Fourier heat transfer in orthotropic plate with uninsulated cracks.
{\tenit Applied Mathematical Modelling}, {\tenbf 115}: 706--723

\bibitem{Weng`1992}
Weng GJ (1992) Explicit evaluation of Willis' bounds with ellipsoidal inclusions
{\tenit Int. J. Engng Science}, {\tenbf 30}: 83--92


\bibitem{Wildman`et`2017}
Wildman RA, O’Grady JT, Gazonas GA. (2017)
A hybrid multiscale finite element/peridynamics method.
{\tenit Int J Fract.}, {\tenbf 207}: 41--53

\bibitem{Willis`1977}
Willis JR (1977) Variational and related methods for the overall properties and selfconsistent
estimates for the overall properties. {\tenit J Mech Phys Solids}, {\tenbf 25}, 85–202

\bibitem{Willis`1980}
Willis JR. (1980) A polarization approach to the scattering of elastic waves – I. Scattering by a
single inclusion. II. Multiple scattering from inclusions. {\tenit J. Mech. Physics
of Solids}, {\tenbf 28}, 287–327.

\bibitem{Willis`1981}
Willis JR. (1981)
Variational and related methods for the overall properties
of composites. {\tenit Advances in Applied Mechanics},
{ \tenbf 21}: 1--78

{\color{black}
\bibitem{Wu`C`2023}
Wu P, Chen Z. (2023)
Peridynamic electromechanical modeling of damaging and cracking in
conductive composites: A stochastically homogenized approach
{\tenit Composite Structures}, {\tenbf 305}: 116528

\bibitem{Wu`et`2021}
Wu P, Yang F, Chen Z, Bobaru F. (2021) Stochastically homogenized peridynamic model
for dynamic fracture analysis of concrete. {\tenit Eng Fract Mech}, 107863.}


\bibitem{Xia`et`2019}
Xia, W., Galadima, Y.K., Oterkus, E. and Oterkus, S., 2019, Representative volume element homogenization of a composite material by using bond-based peridynamics
{\tenit J. Compos. Biodegradable Polymers}, {\tenbf 7}: 51-56.

\bibitem{Xia`et`2020}
Xia W, Oterkus E, Oterkus S. (2020) Peridynamic modeling of periodic microstructured
materials. {\tenit Procedia Structural Integrity}, { \tenbf 28}: 820–828

\bibitem{Xia`et`2021a}
Xia W, Oterkus E, Oterkus S. (2021) 3-dimensional bond-based peridynamic representative volume element homogenization.
{\tenit Physical Mesomechanics}, {\tenbf 24}: 45-51.

\bibitem{Xia`et`2021b}
Xia W, Oterkus E, Oterkus S. (2021) Ordinary state-based peridynamic homogenization of periodic micro-structured materials
{\tenit Theoret. Applied Fract. Mech.}, {\tenbf 113}, 102960.


\bibitem{Xu`et`2008}
Xu J, Askari A, Weckner O, Silling SA (2008) Peridynamic analysis of impact
damage in composite laminates. {\tenit J. Aerospace Engineering}, { \tenbf 21}: 187–194

\bibitem{Xu`et`2024}
Xu J, Yang Z, Wang Z, Wang X, Li Y, Zhang J. (2024)
Peridynamic simulation for the damage patterns of glass considering the influence of rate-dependence and pre-defects.
{\tenit Engng Fracture Mechanics}, {\tenbf 291}: 109539



\bibitem{Xu`et`2021}
Xu X, D’Elia M, Foster JT (2021) A machine-learning framework for peridynamic material
models with physical constraints. 
{\tenit Computer Meth Appl. Mech. Engng} {\tenbf 386}: 114062

\bibitem{Xu`F`2020}
{\color{black} Xu X, Foster JT. (2020) Deriving peridynamic influence functions for
one-dimensional elastic materials with periodic
microstructure. {\tenit J. Peridyn. Nonlocal Modeling}, {\tenbf 2}: 337–351}

\bibitem{YangC`et`2024}
Yang C, Zhu F, Zhao J. (2024)
A multi-horizon fully coupled thermo-mechanical peridynamics
{\tenit J. Mech. Phys. Solids}, {\tenbf 191}: 105758


\bibitem{Yang`et`2024}
Yang X, Gao W, Liu W, Zhang X. (2024)
Coupling four-parameters bond-based peridynamic and finite elements for
damage analysis of composite structures
{\tenit Theor. Appl. Fracture Mech.}, {\tenbf 129}: 104230


\bibitem{Yang`et`2019} 
Yang Y, Ragnvaldsen O, Bai Y, Yi M, Xu BX. (2019) 3D non-isothermal phase-field simulation of microstructure evolution during selective laser sintering. {\tenit Npj Comput Mater}, {\tenbf 5}: 81 (12 pages).


\bibitem{Yang`et`2024}
Yang Z, Shen S, Guan X, He X, Cui J. (2024)
Multiscale analysis-based peridynamic simulation of fracture in porous media.
{\tenit Front. Struct. Civ. Eng}. {\tenbf 18}: 1–13

\bibitem{Yang`et`2023}
{\color{black} Yang Z, Zheng S, Han F, Cui, J. (2023)
An efficient peridynamics-based statistical multiscale method for fracture in
composite structures.
{\tenit Int. J. Mech. Sciences}, {\tenbf 259}: 108611}

\bibitem{Yang`et`2023b}
Yang Z, Zheng S, Han F, Guand X, Zhange J.
(2023) An adaptive coupling approach of local and non-local micromechanics
{\tenit J. Comput. Physics}, {\tenbf 489}, 112277


\bibitem{Yilbas`2013}
Yilbas, B.S. (2013) {\tenit Laser Drilling-Practical Applications}. Springer: Heidelberg, Germany.


\bibitem{You`et`2023}
You H, Xu X, Yu Y, Silling S, D’Elia M, Foster J. (2023) Towards a unified nonlocal, peridynamics
framework for the coarse-graining of molecular dynamics data with fractures. {\tenit Applied Mathematics
Mechanics}, {\tenbf 44}: 1125–1150.

\bibitem{You`et`2020}
You H, Yu Y, Silling S, D’Elia M (2020) Data-driven learning of nonlocal models: from
high-fidelity simulations to constitutive laws. { \tenit arXiv:2012.04157}

\bibitem{You`et`2022}
You H, Yu Y, Silling S, D'Eliac M. (2022) A data-driven peridynamic continuum model
for upscaling molecular dynamics. {\tenit Comput. Meth. Appl. Mechanics Engng.}, {\tenbf 389}: 114400.

\bibitem{You`et`2024}
You H, Yu Y, Silling S, D'Eliac M. (2024)
Nonlocal operator learning for homogenized models:
from high-fidelity simulations to constitutive laws.
{\tenit J. Peridynamics Nonlocal Modeling},
https://doi.org/10.1007/s42102-024-00119-x

\bibitem{You`et`2021}
You H, Yu Y, Trask N, Gulian M, D’Elia M (2021) Data-driven learning of robust nonlocal
physics from high-fidelity synthetic data. {\tenit Computer Methods Applied Mech. Engineering}, {\tenbf 374}: 113553

\bibitem{You`et`2022b}
You H, Zhang Q, Ross C, Lee C-H, Hsu M-C, Yu Y. (2022)
A physics-guided neural operator learning approach to model biological tissues from digital image correlation measurements. {\tenit J. Biomechanical Engng}, {\tenbf 144}, 121012


\bibitem{Yu`Z`2024}
Yu X-L, Zhou X-P. (2024)
A nonlocal energy-informed neural network for peridynamic
correspondence material models. {\tenit Engng Anal. Boundary Elements}, {\tenbf 160}, 273--297

\bibitem{Yu`Z`2024b}
Yu X-L, Zhou X-P. (2024)
A nonlocal energy-informed neural network based on peridynamics
for elastic solids with discontinuities {\tenit Computational Mechanics}, {\tenbf 73}: 233--255

{\color{black} \bibitem{Yu`et`2018}
Yu Y, Bargos FF, You H, Parks ML, Bittencourt ML, Karniadakis GE. (2018)
A partitioned
coupling framework for peridynamics and classical theory: analysis and simulations. {\tenit Comput.
Meth. Appl. Mech. Engineering}, {\tenbf 340}: 905–931}

{\color{black} \bibitem{Zaccariotto`et`2018}
Zaccariotto M, Mudric T, Tomasi D, Shojaei A, Galvanetto U. (2018)
Coupling of FEM meshes
with peridynamic grids. {\tenit Comput. Meth. Applied Mech. Engineering}, {\tenbf 330}:
471–497.}

\bibitem{Zaoui`2002}
Zaoui A. (2002) Continuum micromechanics: survey. {\tenit J. Engng Mech., ASCE}, 
{\tenbf 128}: 808--816

\bibitem{Zhan`et`2021}
Zhan JM, Yao XH, Zhang XQ. (2021)
Study on predicting the mechanical properties and fracturing behaviors of particle reinforced metal matrix composites by non-local approach. {\tenit Mech. Materials}, {\tenbf 155}: 103790


{\color{black} \bibitem{ZhangJ`et`2023}
Zhang J, Han F, Yang Z, Cui J. (2023)
Coupling of an atomistic model and bond-based peridynamic
model using an extended Arlequin framework. {\tenit Comput.
Meth. Appl. Mech. Engineering}, {\tenbf 403}: 115663.}


{\color{black} \bibitem{Zhang`N`2023}
Zhang S, Nie Y. (2023) Localized Chebyshev and MLS collocation methods for solving 2D steady
state nonlocal diffusion and peridynamic equations. {\tenit Math. Computers in Simulation}. {\tenbf 206}: 264–285}


\bibitem{Zhang`et`2016a}
Zhang X, Gunzburger M, Ju L. (2016a) Nodal-type collocation methods for hypersingular integral equations
and nonlocal diffusion problems. {\tenit Comput. Meth. Appl. Mech. Engrg.}, {\tenbf 299}: 401--420

\bibitem{Zhang`et`2016b}
Zhang X, Gunzburger M, Ju L. (2016b) Quadrature rules for finite element approximations of 1D nonlocal
problems. {\tenit J. Comp. Phys.}, {\tenbf 310}: 213-–236.

\bibitem{Zhang`Q`2021}
Zhang Y, Qiao P. (2021) A fully-discrete peridynamic modeling approach for tensile fracture of fiber-reinforced cementitious composites.
{\tenit Eng Fract Mech}, {\tenbf 242}: 107454

\bibitem{Zhao`et`2024}
Zhao S, Zhang Q, Miao Y, Zhang W, Zhao X, Xu W. (2024)
Sub-homogeneous peridynamic model for fracture and failure analysis
of roadway surrounding rock. {\tenit Computer Modeling Engineering \& Sciences}, {\tenbf 139}: no. 3

\bibitem{Zhong`et`2024}
Zhong J, Han F, Du Z, Guo X. (2024)
Multi-time-step coupling of peridynamics and classical continuum mechanics for dynamic brittle fracture
{\tenit Engng Fract. Mechanics}, {\tenbf 307}: 110264

\bibitem{Zhou`et`2013}
Zhou K, Hoh HJ, Wang X, Keer LM, Pang JHL, Song B, Wang QJ. (2013)
A review of recent works on inclusions
{\tenit Mechanics of Materials}, {\tenbf 60}: 144--158.

\bibitem {Zhou`W`2021}
{\color{black} Zhou XP, Wang YT. (2021)
State-of-the-art review on the progressive failure characteristics of geomaterials in peridynamic theory.
{\tenit J. Eng. Mech.}, {\tenbf 147}: 03120001}

\bibitem{Zhou`Y`2024}
Zhou X-P, Yu X-L. (2024)
Transfer learning enhanced nonlocal energy-informed neural
network for quasi-static fracture in rock-like materials.
{\tenit Comput. Methods Appl. Mech. Engrg.} https://doi.org/10.1016/j.cma.2024.117226


\bibitem{Zohdi`W`2008}
Zohdi TI, Wriggers P. (2008)
{\tenit Introduction to Computational Micromechanics.}
Berlin: Springer.





} }



\end{thebibliography}
\end{document}